\documentclass[12pt,english]{article}
\usepackage{mathptmx}

\usepackage[T1]{fontenc}
\usepackage[latin9]{inputenc}
\usepackage{geometry}
\usepackage{fancyhdr}
\usepackage{babel}
\usepackage{amsmath}
\usepackage{amssymb}
\usepackage{setspace}
\usepackage{wasysym}
\usepackage{esint}
\usepackage[unicode=true,pdfusetitle,
 bookmarks=true,bookmarksnumbered=false,bookmarksopen=false,
 breaklinks=false,pdfborder={0 0 1},backref=false,colorlinks=false]
 {hyperref}

\makeatletter
\newcommand{\lyxaddress}[1]{
	\par {\raggedright #1
	\vspace{1.4em}
	\noindent\par}
}
\newenvironment{lyxcode}
	{\par\begin{list}{}{
		\setlength{\rightmargin}{\leftmargin}
		\setlength{\listparindent}{0pt}
		\raggedright
		\setlength{\itemsep}{0pt}
		\setlength{\parsep}{0pt}
		\normalfont\ttfamily}%
	 \item[]}
	{\end{list}}

\date{}
\usepackage{tikz}
\usetikzlibrary{shapes,graphs,arrows,calc,decorations.pathmorphing,backgrounds,positioning,fit,petri}
\usepackage{xstring}
\numberwithin{equation}{section}
\usepackage{bbm}

\makeatother

\begin{document}

\title{The action of long strings in supersymmetric field theories}

\author{Tomer Solberg and Mykhailo Yutushui}
\maketitle

\lyxaddress{Department of Particle Physics and Astrophysics, Weizmann Institute
of Science, Rehovot 76100, Israel}
\begin{lyxcode}
E-mails:~Tomer.Solberg@weizmann.ac.il

Mykhailo.Yutushui@weizmann.ac.il~
\end{lyxcode}
\begin{abstract}
Long strings emerge in many Quantum Field Theories, for example as
vortices in Abelian Higgs theories, or flux tubes in Yang-Mills theories.
The actions of such objects can be expanded in the number of derivatives,
around a long straight string solution. This corresponds to the expansion
of energy levels in powers of $1/L$, with $L$ the length of the
string. Doing so reveals that the first few terms in the expansions
are universal, and only from a certain term do they become dependent
on the originating field theory. Such classifications have been made
before for bosonic strings. In this work we expand upon that and classify
also strings with fermionic degrees of freedom, where the string breaks
$D=4\ N=1$ SUSY completely. An example is the confining string in
$N=1$ SYM theory. We find a general method for generating supersymmetric
action terms from their bosonic counterparts, as well as new fermionic
terms which do not exist in the non-supersymmetric case. These terms
lead to energy corrections at a lower order in $1/L$ than in the
bosonic case.
\end{abstract}
\global\long\def\bra#1{\left\langle #1\right|}
\global\long\def\ket#1{\left|#1\right\rangle }
\global\long\def\braket#1#2{\left\langle #1|#2\right\rangle }
\global\long\def\ketbra#1#2{\left|#1\right\rangle \left\langle #2\right|}

{\tiny{}}\global\long\def\b#1{\Bra{#1}}
{\tiny{}}\global\long\def\k#1{\Ket{#1}}
{\tiny{}}\global\long\def\set#1{\Set{#1}}
{\tiny{}}\global\long\def\bk#1{\Braket{#1}}
{\tiny{}}\global\long\def\norm#1{\left\Vert #1\right\Vert }
{\tiny\par}

\global\long\def\broket#1#2#3{\bra{#1}#2\ket{#3}}

\global\long\def\clop#1{\left[#1\right)}
\global\long\def\opcl#1{\left(#1\right]}

\global\long\def\beC#1#2{\begC#2{#1}}
\global\long\def\enC#1#2{\endC#2{#1}}

{\tiny{}}\global\long\def\at#1{\left.#1\right|}
{\tiny\par}

\global\long\def\paren#1{\left(#1\right.}
\global\long\def\thesis#1{\left.#1\right)}

\global\long\def\del{\nabla}
\global\long\def\cross{\times}
\global\long\def\div{\del\cdot}
\global\long\def\rot{\del\cross}
{\tiny{}}\global\long\def\lap{\nabla^{2}}
{\tiny{} }{\tiny\par}

\global\long\def\divv{\vec{\nabla}\cdot}
\global\long\def\delv{\vec{\nabla}}
\global\long\def\rotv{\vec{\nabla}\cross}

\global\long\def\diff#1#2{\frac{\partial#1}{\partial#2}}
\global\long\def\Diff#1#2{\frac{d#1}{d#2}}

\global\long\def\up{\uparrow}
\global\long\def\down{\downarrow}
\global\long\def\tg{\prime}
\global\long\def\dag{\dagger}

\global\long\def\sinc{\mbox{sinc}}
\global\long\def\Tr{\mbox{Tr}}
\global\long\def\const{\mbox{const}}
{\tiny{}}\global\long\def\trace{{\rm trace}}
{\tiny\par}

\global\long\def\defined{\equiv}
\global\long\def\r{\rightarrow}

\global\long\def\matt#1#2#3#4{\left(\begin{array}{cc}
 #1  &  #2\\
 #3  &  #4 
\end{array}\right)}
\global\long\def\colt#1#2{\left(\begin{array}{c}
 #1\\
 #2 
\end{array}\right)}
\global\long\def\rowt#1#2{\left(\begin{array}{cc}
 #1  &  #2\end{array}\right)}

\global\long\def\identity{\mathbbm{1}}
\global\long\def\paulix{\begin{pmatrix}  &  1\\
 1 
\end{pmatrix}}
\global\long\def\pauliy{\begin{pmatrix}  &  -i\\
 i 
\end{pmatrix}}
\global\long\def\pauliz{\begin{pmatrix}1\\
  &  -1 
\end{pmatrix}}

{\tiny{}}\global\long\def\vecpro#1#2#3#4#5#6{\left|\begin{array}{ccc}
 \hat{x}  &  \hat{y}  &  \hat{z}\\
 #1  &  #2  &  #3\\
 #4  &  #5  &  #6 
\end{array}\right|}
{\tiny\par}

\global\long\def\undercom#1#2{\underset{_{#2}}{\underbrace{#1}}}
{\tiny{}}\global\long\def\explain#1#2{\underset{\mathclap{\overset{\uparrow}{#2}}}{#1}}
{\tiny{}}\global\long\def\iexplain#1#2{\underset{\mathclap{\overset{\big{\uparrow}}{#2}}}{#1}}
{\tiny{}}\global\long\def\bracedown#1#2{\underbrace{#1}_{\mathclap{#2}}}
{\tiny\par}

{\tiny{}}\global\long\def\explainup#1#2{\overset{\mathclap{\underset{\downarrow}{#2}}}{#1}}
{\tiny{}}\global\long\def\iexplainup#1#2{\overset{\mathclap{\underset{\big{\downarrow}}{#2}}}{#1}}
{\tiny{}}\global\long\def\braceup#1#2{\overbrace{#1}^{\mathclap{#2}}}
{\tiny\par}

{\tiny{}}\global\long\def\rmto#1#2{\cancelto{#2}{#1}}
{\tiny{}}\global\long\def\rmpart#1{\cancel{#1}}
\global\long\def\conv#1#2{\underset{_{#1\rightarrow#2}}{\longrightarrow}}

\global\long\def\MA{\mathcal{A}}
\global\long\def\MB{\mathcal{B}}
\global\long\def\MC{\mathcal{C}}
\global\long\def\MD{\mathcal{D}}
\global\long\def\ME{\mathcal{E}}
\global\long\def\MF{\mathcal{F}}
\global\long\def\MG{\mathcal{G}}
\global\long\def\MH{\mathcal{H}}
\global\long\def\MI{\mathcal{I}}
\global\long\def\MJ{\mathcal{J}}
\global\long\def\MK{\mathcal{K}}
\global\long\def\ML{\mathcal{L}}
\global\long\def\MM{\mathcal{M}}
\global\long\def\MN{\mathcal{N}}
\global\long\def\MO{\mathcal{O}}
\global\long\def\MP{\mathcal{P}}
\global\long\def\MQ{\mathcal{Q}}
\global\long\def\MR{\mathcal{R}}
\global\long\def\MS{\mathcal{S}}
\global\long\def\MT{\mathcal{T}}
\global\long\def\MU{\mathcal{U}}
\global\long\def\MV{\mathcal{V}}
\global\long\def\MW{\mathcal{W}}
\global\long\def\MX{\mathcal{X}}
\global\long\def\MY{\mathcal{Y}}
\global\long\def\MZ{\mathcal{Z}}
\global\long\def\BA{\mathbb{A}}
\global\long\def\BB{\mathbb{B}}
\global\long\def\BC{\mathbb{C}}
\global\long\def\BD{\mathbb{D}}
\global\long\def\BE{\mathbb{E}}
\global\long\def\BF{\mathbb{F}}
\global\long\def\BG{\mathbb{G}}
\global\long\def\BH{\mathbb{H}}
\global\long\def\BI{\mathbb{I}}
\global\long\def\BJ{\mathbb{J}}
\global\long\def\BK{\mathbb{K}}
\global\long\def\BL{\mathbb{L}}
\global\long\def\BM{\mathbb{M}}
\global\long\def\BN{\mathbb{N}}
\global\long\def\BO{\mathbb{O}}
\global\long\def\BP{\mathbb{P}}
\global\long\def\BQ{\mathbb{Q}}
\global\long\def\BR{\mathbb{R}}
\global\long\def\BS{\mathbb{S}}
\global\long\def\BT{\mathbb{T}}
\global\long\def\BU{\mathbb{U}}
\global\long\def\BV{\mathbb{V}}
\global\long\def\BW{\mathbb{W}}
\global\long\def\BX{\mathbb{X}}
\global\long\def\BY{\mathbb{Y}}
\global\long\def\BZ{\mathbb{Z}}

{\tiny{}}\global\long\def\eps{\varepsilon}
\global\long\def\ph{\varphi}
\global\long\def\th{\theta}
\global\long\def\d{\partial}
\global\long\def\h{\hbar}
{\tiny{}}\global\long\def\angstrom{\overset{\circ}{\mathrm{A}}}
{\tiny\par}

{\tiny{}}\global\long\def\dbar{{\mathchar'26\mkern-12mu  {\rm d} }}
{\tiny\par}

\global\long\def\ind#1#2#3{#1_{\hphantom{#2}#3}^{#2}}
\global\long\def\rind#1#2#3{#1_{#2}^{\hphantom{#2}#3}}

\global\long\def\mom#1#2{\momentum{#1}{#2}}
\global\long\def\momb#1#2{\momentum[bot]{#1}{#2}}

\global\long\def\vlup#1{\vertexlabel^{{#1}}}
\global\long\def\vldown#1{\vertexlabel_{{#1}}}

{\tiny{}}\global\long\def\wcal#1#2#3{\contraction{}{#1}{#2}{#3}#1#2#3}
{\tiny{}}\global\long\def\wcah#1#2#3{\contraction[2ex]{}{#1}{#2}{#3}#1#2#3}
{\tiny\par}

{\tiny{}}\global\long\def\wcbl#1#2#3{\bcontraction{}{#1}{#2}{#3}#1#2#3}
{\tiny{}}\global\long\def\wcbh#1#2#3{\bcontraction[2ex]{}{#1}{#2}{#3}#1#2#3}
{\tiny\par}

{\tiny{}}\global\long\def\doublewick#1#2#3#4#5#6#7{\contraction{}{#1}{#2#3#4}{#5}\contraction[2ex]{#1#2}{#3}{#4#5#6}{#7}#1#2#3#4#5#6#7}
{\tiny\par}

{\tiny{}}\global\long\def\vec#1{\mathbf{#1}}
{\tiny{}}\global\long\def\hatt#1#2{\hat{#1}_{#2}}
{\tiny\par}

\global\long\def\limm#1#2{\lim\limits _{#1\rightarrow#2}}

\global\long\def\sig#1#2{\sum\limits _{#1}^{#2}}
\global\long\def\intt#1#2{\int\limits _{#1}^{#2}}
 {\tiny{}}\global\long\def\inftop{\intop_{-\infty}^{\infty}}
{\tiny\par}

{\tiny{}}\global\long\def\ra{\quad\Rightarrow\quad}
{\tiny{}}\global\long\def\sp{\hfill}
{\tiny{}}\global\long\def\comm{\quad,\quad}
{\tiny\par}

{\tiny{}}\global\long\def\e#1{\cdot10^{#1}}
{\tiny\par}

\global\long\def\ofr{\left(\overrightarrow{r}\right)}
\global\long\def\kb{k_{B}}
\global\long\def\kbt{k_{B}T}

\global\long\def\fermionr{\mathcal{\vartriangleright}}
\global\long\def\fermionl{\mathcal{\vartriangleleft}}
\global\long\def\susyr{\mathcal{\blacktriangleright}}
\global\long\def\susyl{\blacktriangleleft}
\global\long\def\boson{\bigcirc}
\global\long\def\lorentz{\CIRCLE}
\global\long\def\susy{\CIRCLE}
\global\long\def\worldsheet{\overline{\qquad}}
\global\long\def\transverse{\photon}
\global\long\def\spinor{\gluon}
\global\long\def\scaleone{\mathcal{\lozenge}}

\tikzset {
worldsheet/.style = thick,
transverse/.style = {thick,decorate,decoration=snake,segment length=1.5mm},
spinor/.style = {thick,decorate,decoration={coil,aspect=1,segment length=1.2mm}},
boson/.style = {circle,inner sep=0.25em,draw},
fermion/.style = {isosceles triangle,inner sep=0.25em,draw},
fermionr/.style = {isosceles triangle,shape border rotate=180,inner sep=0.25em,draw},
dfdf/.style = {diamond,aspect=0.5,inner sep=0.25em,draw},
lorentz/.style = {circle,inner sep=0.25em,draw,fill=black},
comb/.style = {inner sep=0.25em,draw},
susy/.style = {isosceles triangle,inner sep=0.25em,draw,fill=black},
susyr/.style = {isosceles triangle,shape border rotate=180,inner sep=0.25em,draw,fill=black},
symb/.style = {inner sep=0pt,scale=1.2},
large/.style = {inner sep=0pt,scale=1.5},
head/.style = {inner sep=0pt},
graphs/every graph/.style = {grow right sep = 7mm,branch up = 7mm}
}

\global\long\def\gammavertexa{\tikz\graph{/--[transverse]"\ensuremath{\gamma}"[symb]--[spinor]/};}

\global\long\def\gammavertexb{\tikz\graph{/--[worldsheet]"\ensuremath{\gamma}"[symb]--[spinor]/};}

\global\long\def\dvertex{\tikz\graph{/--[spinor]"\ensuremath{\partial}"[symb]--[worldsheet]/};}

\global\long\def\bosonvertex{\tikz\graph{/--[transverse]A/[boson]--[worldsheet]/};}

\global\long\def\fermionvertexa{\tikz\graph{/--[transverse]A/[fermion]--[worldsheet]/};}

\global\long\def\fermionvertexb{\tikz\graph{/--[worldsheet]A/[fermion]--[worldsheet]/};}

\global\long\def\fermionvertexc{\tikz\graph{/--[worldsheet]A/[fermionr]--[worldsheet]/};}

\global\long\def\combinationvertex{\tikz\graph{/--[worldsheet]A/[comb]--[worldsheet]/};}

\global\long\def\dfdfvertex{\tikz\graph{/--[worldsheet]A/[dfdf]--[worldsheet]/};}

\global\long\def\fullfermiona{\tikz\graph{/--[transverse]"\ensuremath{\gamma}"[symb]--[spinor]"\ensuremath{\partial}"[symb]--[worldsheet]/};}

\global\long\def\fullfermionb{\tikz\graph{/--[worldsheet]"\ensuremath{\gamma}"[symb]--[spinor]"\ensuremath{\partial}"[symb]--[worldsheet]/};}

\global\long\def\fermiontransaa{\tikz\graph{/--[transverse]A/[susy]--[worldsheet]/};}

\global\long\def\fermiontransab{\tikz\graph{/--[transverse]A/[fermion]--[worldsheet]B/[susy]--[worldsheet]/};}

\global\long\def\fermiontransac{\tikz\graph{/--[transverse]A/[lorentz]--[worldsheet]B/[fermion]--[worldsheet]/};}

\global\long\def\fermiontransad{\tikz\graph{/--[transverse]A/[fermion]--[worldsheet]C/[lorentz]--[transverse]B/[boson]--[worldsheet]/};}

\global\long\def\fermiontransba{\tikz\graph{/--[worldsheet]A/[susy]--[worldsheet]/};}

\global\long\def\fermiontransbb{\tikz\graph{/--[worldsheet]A/[fermion]--[worldsheet]B/[susy]--[worldsheet]/};}

\global\long\def\fermiontransbc{\tikz\graph{/--[worldsheet]A/[lorentz]--[transverse]B/[fermion]--[worldsheet]/};}

\global\long\def\fermiontransbd{\tikz\graph{/--[worldsheet]A/[fermion]--[worldsheet]C/[lorentz]--[transverse]B/[boson]--[worldsheet]/};}

\global\long\def\fulltransa{\tikz\graph{/--[transverse]A/[lorentz]--[spinor]"\ensuremath{\partial}"[symb]--[worldsheet]/};}

\global\long\def\fulltransb{\tikz\graph{/--[worldsheet]A/[lorentz]--[spinor]"\ensuremath{\partial}"[symb]--[worldsheet]/};}

\global\long\def\gammatransaa{\tikz\graph{/--[transverse]A/[lorentz]--[spinor]/};}

\global\long\def\gammatransab{\tikz\graph{/--[transverse]A/[lorentz]--[worldsheet]"\ensuremath{\gamma}"[symb]--[spinor]/};}

\global\long\def\gammatumoraa{\tikz\graph{/--[transverse]A/"\ensuremath{\gamma}"[symb]--[spinor]/;(A.north)--[spinor]"\ensuremath{\partial}"[symb]--[worldsheet]B/[lorentz]--[spinor]"\ensuremath{\psi}"[symb]};}

\global\long\def\gammatumorab{\tikz\graph{/--[transverse]A/"\ensuremath{\gamma}"[symb]--[spinor]/;(A.north)--[spinor]"\ensuremath{\partial}"[symb]--[worldsheet]B/[lorentz]--[transverse]"\ensuremath{X}"[symb]};}

\global\long\def\gammatransba{\tikz\graph{/--[worldsheet]A/[lorentz]--[spinor]/};}

\global\long\def\gammatransbb{\tikz\graph{/--[worldsheet]A/[lorentz]--[transverse]"\ensuremath{\gamma}"[symb]--[spinor]/};}

\global\long\def\gammatumorba{\tikz\graph{/--[worldsheet]A/"\ensuremath{\gamma}"[symb]--[spinor]/;(A.north)--[spinor]"\ensuremath{\partial}"[symb]--[worldsheet]B/[lorentz]--[spinor]"\ensuremath{\psi}"[symb]};}

\global\long\def\gammatumorbb{\tikz\graph{/--[worldsheet]A/"\ensuremath{\gamma}"[symb]--[spinor]/;(A.north)--[spinor]"\ensuremath{\partial}"[symb]--[worldsheet]B/[lorentz]--[transverse]"\ensuremath{X}"[symb]};}

\global\long\def\dtransa{\tikz\graph{/--[spinor]B/"\ensuremath{\partial}"[symb]--[worldsheet]A/[lorentz]--[spinor]C/"\ensuremath{\partial}"[symb]--[worldsheet]/};}

\global\long\def\dtransb{\tikz\graph{/--[spinor]B/"\ensuremath{\partial}"[symb]--[worldsheet]A/[lorentz]--[transverse]C/[boson]--[worldsheet]/};}

\global\long\def\dtumora{\tikz\graph{/--[spinor]A/"\ensuremath{\partial}"[symb]--[worldsheet]/;(A.north)--[worldsheet]B/[lorentz]--[spinor]"\ensuremath{\psi}"[symb]};}

\global\long\def\dtumorb{\tikz\graph{/--[spinor]A/"\ensuremath{\partial}"[symb]--[worldsheet]/;(A.north)--[worldsheet]B/[lorentz]--[transverse]"\ensuremath{X}"[symb]};}

\global\long\def\bosontransa{\tikz\graph{/--[transverse]A/[lorentz]--[worldsheet]/};}

\global\long\def\bosontransb{\tikz\graph{/--[transverse]A/[boson]--[worldsheet]B/[lorentz]--[transverse]C/[boson]--[worldsheet]/};}

\global\long\def\bosontransc{\tikz\graph{/--[transverse]A/[lorentz]--[spinor]B/"\ensuremath{\partial}"[symb]--[worldsheet]/};}

\global\long\def\bosontransd{\tikz\graph{/--[transverse]C/[boson]--[worldsheet]A/[lorentz]--[spinor]B/"\ensuremath{\partial}"[symb]--[worldsheet]/};}

\global\long\def\bosontranse{\tikz\graph{/--[transverse]C/[boson]--[worldsheet]A/[susy]--[worldsheet]/};}

\global\long\def\bosontumora{\tikz\graph{/--[transverse]A/[boson]--[worldsheet]/;(A.north)--[worldsheet]B/[lorentz]--[spinor]"\ensuremath{\psi}"[symb]};}

\global\long\def\bosontumorb{\tikz\graph{/--[transverse]A/[boson]--[worldsheet]/;(A.north)--[worldsheet]B/[lorentz]--[transverse]"\ensuremath{X}"[symb]};}

\global\long\def\doubleboson{\tikz\graph{/--[worldsheet]B/[boson]--[transverse]C/[boson]--[worldsheet]/};}

\global\long\def\doublebosontransa{\tikz\graph{/--[worldsheet]B/[boson]--[transverse]C/[lorentz]--[worldsheet]/};}

\global\long\def\doublebosontransb{\tikz\graph{/--[worldsheet]B/[boson]--[transverse]C/[lorentz]--[worldsheet]D/[boson]--[transverse]E/[boson]--[worldsheet]/};}

\global\long\def\doublebosontransc{\tikz\graph{/--[worldsheet]B/[boson]--[transverse]C/[susy]--[worldsheet]/};}

\global\long\def\doublebosontransd{\tikz\graph{/--[worldsheet]B/[boson]--[transverse]D/[boson]--[worldsheet]C/[susy]--[worldsheet]/};}

\global\long\def\bosonfermion{\tikz\graph{/--[worldsheet]B/[boson]--[transverse]C/[fermion]--[worldsheet]/};}

\global\long\def\fermionboson{\tikz\graph{/--[worldsheet]B/[fermionr]--[transverse]C/[boson]--[worldsheet]/};}

\global\long\def\bosonfermiontransa{\tikz\graph{/--[worldsheet]B/[boson]--[transverse]C/[susy]--[worldsheet]/};}

\global\long\def\bosonfermiontransb{\tikz\graph{/--[worldsheet]B/[lorentz]--[transverse]C/[fermion]--[worldsheet]/};}

\global\long\def\bosonfermiontransc{\tikz\graph{/--[worldsheet]B/[boson]--[transverse]D/[fermion]--[worldsheet]C/[susy]--[worldsheet]/};}

\global\long\def\bosonfermiontransd{\tikz\graph{/--[worldsheet]B/[susyr]--[transverse]C/[fermion]--[worldsheet]/};}

\global\long\def\bosonfermiontranse{\tikz\graph{/--[worldsheet]B/[boson]--[transverse]D/[lorentz]--[worldsheet]C/[fermion]--[worldsheet]/};}

\global\long\def\bosonfermiontransf{\tikz\graph{/--[worldsheet]B/[susyr]--[worldsheet]D/[boson]--[transverse]C/[fermion]--[worldsheet]/};}

\global\long\def\bosonfermiontransg{\tikz\graph{/--[worldsheet]B/[boson]--[transverse]C/[fermion]--[worldsheet]D/[lorentz]--[transverse]E/[boson]--[worldsheet]/};}

\global\long\def\bosonfermiontransh{\tikz\graph{/--[worldsheet]B/[boson]--[transverse]C/[lorentz]--[worldsheet]D/[boson]--[transverse]E/[fermion]--[worldsheet]/};}

\global\long\def\doublefermiona{\tikz\graph{/--[worldsheet]B/[fermionr]--[transverse]C/[fermion]--[worldsheet]/};}

\global\long\def\doublefermionatransa{\tikz\graph{/--[worldsheet]B/[susyr]--[transverse]C/[fermion]--[worldsheet]/};}

\global\long\def\doublefermionatransb{\tikz\graph{/--[worldsheet]B/[susyr]--[worldsheet]D/[fermionr]--[transverse]C/[fermion]--[worldsheet]/};}

\global\long\def\doublefermionatransc{\tikz\graph{/--[worldsheet]B/[fermionr]--[worldsheet]D/[lorentz]--[transverse]C/[fermion]--[worldsheet]/};}

\global\long\def\doublefermionatransd{\tikz\graph{/--[worldsheet]B/[fermionr]--[transverse]C/[fermion]--[worldsheet]D/[lorentz]--[transverse]E/[boson]--[worldsheet]/};}

\global\long\def\doublefermionb{\tikz\graph{/--[worldsheet]B/[fermionr]--[worldsheet]C/[fermion]--[worldsheet]/};}

\global\long\def\doublefermionbtransa{\tikz\graph{/--[worldsheet]B/[fermionr]--[worldsheet]C/[susy]--[worldsheet]/};}

\global\long\def\doublefermionbtransb{\tikz\graph{/--[worldsheet]A/[fermionr]--[worldsheet]D/[fermion]--[worldsheet]B/[fermionr]--[worldsheet]C/[susy]--[worldsheet]/};}

\global\long\def\doublefermionbtransc{\tikz\graph{/--[worldsheet]B/[fermionr]--[worldsheet]D/[lorentz]--[transverse]C/[fermion]--[worldsheet]/};}

\global\long\def\doublefermionbtransd{\tikz\graph{/--[worldsheet]A/[fermionr]--[worldsheet]D/[fermion]--[worldsheet]B/[lorentz]--[worldsheet]C/[boson]--[worldsheet]/};}

\global\long\def\dfdftransa{\tikz\graph{/--[worldsheet]B/[dfdf]--[worldsheet]C/[susy]--[worldsheet]/};}

\global\long\def\dfdftransb{\tikz\graph{/--[worldsheet]B/[susyr]--[worldsheet]C/[dfdf]--[worldsheet]/};}

\global\long\def\dfdftransc{\tikz\graph{/--[worldsheet]B/[dfdf]--[worldsheet]A/[lorentz]--[transverse]C/[boson]--[worldsheet]/};}

\global\long\def\dfdftransd{\tikz\graph{/--[worldsheet]B/[boson]--[transverse]A/[lorentz]--[worldsheet]C/[dfdf]--[worldsheet]/};}

\global\long\def\scaleonechaina{\tikz[baseline={(0,0)}]\graph{/--[transverse]A/\ScaleOneHead{bos}[head]};}

\global\long\def\scaleonechainb{\tikz[baseline={(0,0)}]\graph{/--[transverse]A/\ScaleOneHead{fer}[head]};}

\global\long\def\scaleonechainc{\tikz[baseline={(0,0)}]\graph{/--[transverse]B/[boson]--[worldsheet]A/\ScaleOneHead{fer}[head]};}

\global\long\def\scaleonechaind{\tikz[baseline={(0,0)}]\graph{/--[transverse]B/[fermion]--[worldsheet]A/\ScaleOneHead{fer}[head]};}

\global\long\def\scaleonechaine{\ScaleOneVertex{fern}{bos}}

\global\long\def\scaleonechainf{\ScaleOneVertex{fern}{fer}}

\newcommand{\Aring}[1]{
	\IfEqCase{#1}{
		{2}{\tikz [every new --/.style = {bend left=90},baseline={(0,-0.25)}] {\node (b0) [boson] at (90:0.65) {}; \node (b1) [boson] at (90+360*2/4:0.65) {}; \node (n) at (0:0) {$A_{2}$}; \graph {(b0) --[worldsheet] (b1) --[transverse]  (b0)};}}
		{4}{\tikz [every new --/.style = {bend left},baseline={(0,-0.25)}] {\node (b0) [boson] at (90:0.65) {}; \node (b1) [boson] at (90+360*3/4:0.65) {}; \node (b2) [boson] at (90+360*2/4:0.65) {}; \node (b3) [boson] at (90+360*1/4:0.65) {}; \node (n) at (0:0) {$A_{4}$}; \graph {(b0) --[worldsheet] (b1) --[transverse] (b2) --[worldsheet] (b3) --[transverse] (b0)};}}
		{6}{\tikz [every new --/.style = {bend left},baseline={(0,-0.25)}] {\node (b0) [boson] at (90:0.65) {}; \node (b1) [boson] at (90+360*5/6:0.65) {}; \node (b2) [boson] at (90+360*4/6:0.65) {}; \node (b3) [boson] at (90+360*3/6:0.65) {}; \node (b4) [boson] at (90+360*2/6:0.65) {}; \node (b5) [boson] at (90+360*1/6:0.65) {}; \node (n) at (0:0) {$A_{6}$}; \graph {(b0) --[worldsheet] (b1) --[transverse] (b2) --[worldsheet] (b3) --[transverse] (b4) --[worldsheet] (b5) --[transverse] (b0)};}}
		{2n}{\tikz [every new --/.style = {bend left},baseline={(0,-0.25)}] {\node (b0) [boson] at (90:0.65) {}; \node (b1) [boson] at (90+360*6/7:0.65) {}; \node (b2) [boson] at (90+360*5/7:0.65) {}; \node (b3) [boson] at (90+360*4/7:0.65) {}; \node (b4) [boson] at (90+360*3/7:0.65) {}; \node (p1) [coordinate] at (90+360*2/7:0.65) {}; \node (p2) [coordinate] at (90+360*1/7:0.65) {}; \node (n) at (0:0) {$A_{2n}$}; \graph {(b0) --[worldsheet] (b1) --[transverse] (b2) --[worldsheet] (b3) --[transverse] (b4) --[worldsheet] (p1) --[loosely dotted, edge label=2n] (p2) --[transverse] (b0)};}}
		{2l}{\tikz [every new --/.style = {bend left=90},baseline={(0,-0.25)}] {\node (b0) [lorentz] at (90:0.65) {}; \node (b1) [boson] at (90+360*2/4:0.65) {}; \graph {(b0) --[worldsheet] (b1) --[transverse]  (b0)};}}
		{4l}{\tikz [every new --/.style = {bend left},baseline={(0,-0.25)}] {\node (b0) [lorentz] at (90:0.65) {}; \node (b1) [boson] at (90+360*3/4:0.65) {}; \node (b2) [boson] at (90+360*2/4:0.65) {}; \node (b3) [boson] at (90+360*1/4:0.65) {}; \graph {(b0) --[worldsheet] (b1) --[transverse] (b2) --[worldsheet] (b3) --[transverse] (b0)};}}
		{6l}{\tikz [every new --/.style = {bend left},baseline={(0,-0.25)}] {\node (b0) [lorentz] at (90:0.65) {}; \node (b1) [boson] at (90+360*5/6:0.65) {}; \node (b2) [boson] at (90+360*4/6:0.65) {}; \node (b3) [boson] at (90+360*3/6:0.65) {}; \node (b4) [boson] at (90+360*2/6:0.65) {}; \node (b5) [boson] at (90+360*1/6:0.65) {}; \graph {(b0) --[worldsheet] (b1) --[transverse] (b2) --[worldsheet] (b3) --[transverse] (b4) --[worldsheet] (b5) --[transverse] (b0)};}}
		{2nl}{\tikz [every new --/.style = {bend left},baseline={(0,-0.25)}] {\node (b0) [lorentz] at (90:0.65) {}; \node (b1) [boson] at (90+360*6/7:0.65) {}; \node (b2) [boson] at (90+360*5/7:0.65) {}; \node (b3) [boson] at (90+360*4/7:0.65) {}; \node (b4) [boson] at (90+360*3/7:0.65) {}; \node (p1) [coordinate] at (90+360*2/7:0.65) {}; \node (p2) [coordinate] at (90+360*1/7:0.65) {}; \graph {(b0) --[worldsheet] (b1) --[transverse] (b2) --[worldsheet] (b3) --[transverse] (b4) --[worldsheet] (p1) --[loosely dotted, edge label'=2n] (p2) --[transverse] (b0)};}}
		{2n2l}{\tikz [every new --/.style = {bend left},baseline={(0,-0.25)}] {\node (b0) [lorentz] at (90:0.65) {}; \node (b1) [boson] at (90+360*6/7:0.65) {}; \node (b2) [boson] at (90+360*5/7:0.65) {}; \node (b3) [boson] at (90+360*4/7:0.65) {}; \node (b4) [boson] at (90+360*3/7:0.65) {}; \node (p1) [coordinate] at (90+360*2/7:0.65) {}; \node (p2) [coordinate] at (90+360*1/7:0.65) {}; \graph {(b0) --[worldsheet] (b1) --[transverse] (b2) --[worldsheet] (b3) --[transverse] (b4) --[worldsheet] (p1) --[loosely dotted, edge label'=2n+2] (p2) --[transverse] (b0)};}}
		{tumor}{\tikz [every new --/.style = {bend left},baseline={(0,-0.25)}] {\node (b0) [boson] at (90:0.65) {}; \node (b1) [boson] at (90+360*6/7:0.65) {}; \node (b2) [boson] at (90+360*5/7:0.65) {}; \node (b3) [boson] at (90+360*4/7:0.65) {}; \node (b4) [boson] at (90+360*3/7:0.65) {}; \node (p1) [coordinate] at (90+360*2/7:0.65) {}; \node (p2) [coordinate] at (90+360*1/7:0.65) {}; \node (l0) [lorentz] at (90:1.2) {}; \node (s0) [symb] at (0.65,1.2) {$X$}; \graph {(s0) --[transverse,bend left=0] (l0) --[worldsheet,bend left=0] (b0) --[worldsheet] (b1) --[transverse] (b2) --[worldsheet] (b3) --[transverse] (b4) --[worldsheet] (p1) --[loosely dotted, edge label'=2n] (p2) --[transverse] (b0)};}}
	}[\PackageError{Aring}{Undefined option to A ring: #1}{}]
}

\newcommand{\Bring}[1]{
	\IfEqCase{#1}{
		{0}{\tikz [every new --/.style = {bend left=90},baseline={(0,-0.25)}] {\node (f1) [fermion] at (90:0.65) {}; \node (b1) [coordinate] at (90+360*2/4:0.65) {}; \node (n) at (0:0) {$B_{0}$}; \graph {(f1.east) --[worldsheet] (b1) --[worldsheet]  (f1.west)};}}
		{1}{\tikz [every new --/.style = {bend left=90},baseline={(0,-0.25)}] {\node (f1) [fermion] at (90:0.65) {}; \node (b1) [boson] at (90+360*2/4:0.65) {}; \node (n) at (0:0) {$B_{1}$}; \graph {(f1.east) --[worldsheet] (b1) --[transverse]  (f1.west)};}}
		{2}{\tikz [every new --/.style = {bend left},baseline={(0,-0.25)}] {\node (f) [fermion] at (90:0.65) {}; \node (b1) [boson] at (90+360*2/3:0.65) {}; \node (b2) [boson] at (90+360/3:0.65) {}; \node (n) at (0:0) {$B_{2}$}; \graph {(f.east) --[worldsheet] (b1) --[transverse] (b2) --[worldsheet] (f.west)};}}
		{3}{\tikz [every new --/.style = {bend left},baseline={(0,-0.25)}] {\node (f) [fermion] at (90:0.65) {}; \node (b1) [boson] at (90+360*3/4:0.65) {}; \node (b2) [boson] at (90+360*2/4:0.65) {}; \node (b3) [boson] at (90+360*1/4:0.65) {}; \node (n) at (0:0) {$B_{3}$}; \graph {(f.east) --[worldsheet] (b1) --[transverse] (b2) --[worldsheet] (b3) --[transverse] (f.west)};}}
		{n}{\tikz [every new --/.style = {bend left},baseline={(0,-0.25)}] {\node (f) [fermion] at (90:0.65) {}; \node (b1) [boson] at (90+360*6/7:0.65) {}; \node (b2) [boson] at (90+360*5/7:0.65) {}; \node (b3) [boson] at (90+360*4/7:0.65) {}; \node (b4) [boson] at (90+360*3/7:0.65) {}; \node (p1) [coordinate] at (90+360*2/7:0.65) {}; \node (p2) [coordinate] at (90+360*1/7:0.65) {}; \node (n) at (0:0) {$B_{n}$}; \graph {(f.east) --[worldsheet] (b1) --[transverse] (b2) --[worldsheet] (b3) --[transverse] (b4) --[worldsheet] (p1) --[loosely dotted, edge label=n] (p2) --[transverse] (f.west);(p2) --[worldsheet] (f.west)};}}
		{0s}{\tikz [every new --/.style = {bend left=90},baseline={(0,-0.25)}] {\node (f1) [susy] at (90:0.65) {}; \node (b1) [coordinate] at (90+360*2/4:0.65) {}; \graph {(f1.east) --[worldsheet] (b1) --[worldsheet]  (f1.west)};}}
		{1s}{\tikz [every new --/.style = {bend left=90},baseline={(0,-0.25)}] {\node (f1) [susy] at (90:0.65) {}; \node (b1) [boson] at (90+360*2/4:0.65) {}; \graph {(f1.east) --[worldsheet] (b1) --[transverse]  (f1.west)};}}
		{2s}{\tikz [every new --/.style = {bend left},baseline={(0,-0.25)}] {\node (f) [susy] at (90:0.65) {}; \node (b1) [boson] at (90+360*2/3:0.65) {}; \node (b2) [boson] at (90+360/3:0.65) {}; \graph {(f.east) --[worldsheet] (b1) --[transverse] (b2) --[worldsheet] (f.west)};}}
		{3s}{\tikz [every new --/.style = {bend left},baseline={(0,-0.25)}] {\node (f) [susy] at (90:0.65) {}; \node (b1) [boson] at (90+360*3/4:0.65) {}; \node (b2) [boson] at (90+360*2/4:0.65) {}; \node (b3) [boson] at (90+360*1/4:0.65) {}; \graph {(f.east) --[worldsheet] (b1) --[transverse] (b2) --[worldsheet] (b3) --[transverse] (f.west)};}}
		{1l}{\tikz [every new --/.style = {bend left=90},baseline={(0,-0.25)}] {\node (f1) [fermion] at (90:0.65) {}; \node (b1) [lorentz] at (90+360*2/4:0.65) {}; \graph {(f1.east) --[worldsheet] (b1) --[transverse]  (f1.west)};}}
		{2l1}{\tikz [every new --/.style = {bend left},baseline={(0,-0.25)}] {\node (f) [fermion] at (90:0.65) {}; \node (b1) [lorentz] at (90+360*2/3:0.65) {}; \node (b2) [boson] at (90+360/3:0.65) {}; \graph {(f.east) --[worldsheet] (b1) --[transverse] (b2) --[worldsheet] (f.west)};}}
		{2l2}{\tikz [every new --/.style = {bend left},baseline={(0,-0.25)}] {\node (f) [fermion] at (90:0.65) {}; \node (b1) [boson] at (90+360*2/3:0.65) {}; \node (b2) [lorentz] at (90+360/3:0.65) {}; \graph {(f.east) --[worldsheet] (b1) --[transverse] (b2) --[worldsheet] (f.west)};}}
		{3l1}{\tikz [every new --/.style = {bend left},baseline={(0,-0.25)}] {\node (f) [fermion] at (90:0.65) {}; \node (b1) [lorentz] at (90+360*3/4:0.65) {}; \node (b2) [boson] at (90+360*2/4:0.65) {}; \node (b3) [boson] at (90+360*1/4:0.65) {};  \graph {(f.east) --[worldsheet] (b1) --[transverse] (b2) --[worldsheet] (b3) --[transverse] (f.west)};}}
		{3l2}{\tikz [every new --/.style = {bend left},baseline={(0,-0.25)}] {\node (f) [fermion] at (90:0.65) {}; \node (b1) [boson] at (90+360*3/4:0.65) {}; \node (b2) [lorentz] at (90+360*2/4:0.65) {}; \node (b3) [boson] at (90+360*1/4:0.65) {};  \graph {(f.east) --[worldsheet] (b1) --[transverse] (b2) --[worldsheet] (b3) --[transverse] (f.west)};}}
		{3l3}{\tikz [every new --/.style = {bend left},baseline={(0,-0.25)}] {\node (f) [fermion] at (90:0.65) {}; \node (b1) [boson] at (90+360*3/4:0.65) {}; \node (b2) [boson] at (90+360*2/4:0.65) {}; \node (b3) [lorentz] at (90+360*1/4:0.65) {};  \graph {(f.east) --[worldsheet] (b1) --[transverse] (b2) --[worldsheet] (b3) --[transverse] (f.west)};}}
	}[\PackageError{Bring}{Undefined option to B ring: #1}{}]
}

\newcommand{\Cring}[2]{
	\IfEqCase{#1}{
		{}{\IfEqCase{#2}{
					{00}{\tikz [every new --/.style = {bend left},baseline={(0,-0.25)}] {\node (f1) [fermion] at (90:0.65) {}; \node (p1) [coordinate] at (90+360*3/4:0.65) {}; \node (f2) [fermion, rotate=180] at (90+360*2/4:0.65) {}; \node (p2) [coordinate] at (90+360*1/4:0.65) {}; \node (n) at (0:0) {$C_{0,0}$}; \graph {(f1.east) --[worldsheet] (p1) --[worldsheet] (f2.west); (f2.east) --[worldsheet] (p2) --[worldsheet] (f1.west)};}}
					{01}{\tikz [every new --/.style = {bend left},baseline={(0,-0.25)}] {\node (f1) [fermion] at (90:0.65) {}; \node (b1) [boson] at (90+360*3/4:0.65) {}; \node (f2) [fermion, rotate=180] at (90+360*2/4:0.65) {}; \node (p2) [coordinate] at (90+360*1/4:0.65) {}; \node (n) at (0:0) {$C_{1,0}$}; \graph {(f1.east) --[worldsheet] (b1) --[transverse] (f2.west); (f2.east) --[worldsheet] (p2) --[worldsheet] (f1.west)};}}
					{nm}{\tikz [every new --/.style = {bend left},baseline={(0,-0.25)}] {\node (f1) [fermion] at (90:0.65) {}; \node (b1) [boson] at (90+360*7/8:0.65) {}; \node (p1) [coordinate] at (90+360*6/8:0.65) {}; \node (p2) [coordinate] at (90+360*5/8:0.65) {}; \node (f2) [fermion, rotate=180] at (90+360*4/8:0.65) {}; \node (b2) [boson] at (90+360*3/8:0.65) {}; \node (p3) [coordinate] at (90+360*2/8:0.65) {}; \node (p4) [coordinate] at (90+360*1/8:0.65) {}; \node (n) at (0:0) {$C_{n,m}$}; \graph {(f1.east) --[worldsheet] (b1) --[transverse] (p1) --[loosely dotted, edge label=n] (p2) --[transverse] (f2.west); (p2) --[worldsheet] (f2.west); (f2.east) --[worldsheet] (b2) --[transverse] (p3) --[loosely dotted, edge label=m] (p4) --[transverse] (f1.west); (p4) --[worldsheet] (f1.west)};}}
					{00s}{\tikz [every new --/.style = {bend left},baseline={(0,-0.25)}] {\node (f1) [susy] at (90:0.65) {}; \node (p1) [coordinate] at (90+360*3/4:0.65) {}; \node (f2) [fermion, rotate=180] at (90+360*2/4:0.65) {}; \node (p2) [coordinate] at (90+360*1/4:0.65) {}; \graph {(f1.east) --[worldsheet] (p1) --[worldsheet] (f2.west); (f2.east) --[worldsheet] (p2) --[worldsheet] (f1.west)};}}
					{01s1}{\tikz [every new --/.style = {bend left},baseline={(0,-0.25)}] {\node (f1) [susy] at (90:0.65) {}; \node (b1) [boson] at (90+360*3/4:0.65) {}; \node (f2) [fermion, rotate=180] at (90+360*2/4:0.65) {}; \node (p2) [coordinate] at (90+360*1/4:0.65) {}; \graph {(f1.east) --[worldsheet] (b1) --[transverse] (f2.west); (f2.east) --[worldsheet] (p2) --[worldsheet] (f1.west)};}}
					{01s2}{\tikz [every new --/.style = {bend left},baseline={(0,-0.25)}] {\node (f1) [fermion] at (90:0.65) {}; \node (b1) [boson] at (90+360*3/4:0.65) {}; \node (f2) [susy, rotate=180] at (90+360*2/4:0.65) {}; \node (p2) [coordinate] at (90+360*1/4:0.65) {}; \graph {(f1.east) --[worldsheet] (b1) --[transverse] (f2.west); (f2.east) --[worldsheet] (p2) --[worldsheet] (f1.west)};}}
				}[\PackageError{Cring}{Undefined option to C ring: #2}{}]};

		{p}{\IfEqCase{#2}{
				{00}{\tikz [every new --/.style = {bend left},baseline={(0,-0.25)}] {\node (f1) [fermion] at (90:0.65) {}; \node (p1) [coordinate] at (90+360*3/4:0.65) {}; \node (f2) [fermion] at (90+360*2/4:0.65) {}; \node (p2) [coordinate] at (90+360*1/4:0.65) {}; \node (n) at (0:0) {$C'_{0,0}$}; \graph {(f1.east) --[worldsheet] (p1) --[worldsheet] (f2.east); (f2.west) --[worldsheet] (p2) --[worldsheet] (f1.west)};}}
				{nm}{\tikz [every new --/.style = {bend left},baseline={(0,-0.25)}] {\node (f1) [fermion] at (90:0.65) {}; \node (b1) [boson] at (90+360*7/8:0.65) {}; \node (p1) [coordinate] at (90+360*6/8:0.65) {}; \node (p2) [coordinate] at (90+360*5/8:0.65) {}; \node (f2) [fermion] at (90+360*4/8:0.65) {}; \node (b2) [boson] at (90+360*3/8:0.65) {}; \node (p3) [coordinate] at (90+360*2/8:0.65) {}; \node (p4) [coordinate] at (90+360*1/8:0.65) {}; \node (n) at (0:0) {$C'_{2n,m}$}; \graph {(f1.east) --[worldsheet] (b1) --[transverse] (p1) --[loosely dotted, edge label=2n] (p2) --[worldsheet] (f2.east); (f2.west) --[worldsheet] (b2) --[transverse] (p3) --[loosely dotted, edge label=m] (p4) --[transverse] (f1.west); (p4) --[worldsheet] (f1.west)};}}
				{01}{\tikz [every new --/.style = {bend left},baseline={(0,-0.25)}] {\node (f1) [fermion] at (90:0.65) {}; \node (p1) [coordinate] at (90+360*3/4:0.65) {}; \node (f2) [fermion] at (90+360*2/4:0.65) {}; \node (b1) [boson] at (90+360*1/4:0.65) {}; \node (n) at (0:0) {$C'_{0,1}$}; \graph {(f1.east) --[worldsheet] (p1) --[worldsheet] (f2.east); (f2.west) --[transverse] (b1) --[worldsheet] (f1.west)};}}
				{02}{\tikz [every new --/.style = {bend left},baseline={(0,-0.25)}] {\node (f1) [fermion] at (90:0.65) {}; \node (p1) [coordinate] at (90+360*3/4:0.65) {}; \node (f2) [fermion] at (90+360*2/4:0.65) {}; \node (b1) [boson] at (90+360*2/6:0.65) {}; \node (b2) [boson] at (90+360*1/6:0.65) {}; \node (n) at (0:0) {$C'_{0,2}$}; \graph {(f1.east) --[worldsheet] (p1) --[worldsheet] (f2.east); (f2.west) --[worldsheet] (b1) --[transverse] (b2) --[worldsheet] (f1.west)};}}
				{20}{\tikz [every new --/.style = {bend left},baseline={(0,-0.25)}] {\node (f1) [fermion] at (90:0.65) {}; \node (b1) [boson] at (90+360*5/6:0.65) {}; \node (b2) [boson] at (90+360*4/6:0.65) {}; \node (f2) [fermion] at (90+360*2/4:0.65) {}; \node (p2) [coordinate] at (90+360*1/4:0.65) {}; \node (n) at (0:0) {$C'_{2,0}$}; \graph {(f1.east) --[worldsheet] (b1) --[transverse] (b2) --[worldsheet] (f2.east); (f2.west) --[worldsheet] (p2) --[worldsheet] (f1.west)};}}
				{21}{\tikz [every new --/.style = {bend left},baseline={(0,-0.25)}] {\node (f1) [fermion] at (90:0.65) {}; \node (b1) [boson] at (90+360*5/6:0.65) {}; \node (b2) [boson] at (90+360*4/6:0.65) {}; \node (f2) [fermion] at (90+360*2/4:0.65) {}; \node (b3) [boson] at (90+360*1/4:0.65) {}; \node (n) at (0:0) {$C'_{2,1}$}; \graph {(f1.east) --[worldsheet] (b1) --[transverse] (b2) --[worldsheet] (f2.east); (f2.west) --[transverse] (b3) --[worldsheet] (f1.west)};}}
				{00s}{\tikz [every new --/.style = {bend left},baseline={(0,-0.25)}] {\node (f1) [susy] at (90:0.65) {}; \node (p1) [coordinate] at (90+360*3/4:0.65) {}; \node (f2) [fermion] at (90+360*2/4:0.65) {}; \node (p2) [coordinate] at (90+360*1/4:0.65) {}; \graph {(f1.east) --[worldsheet] (p1) --[worldsheet] (f2.east); (f2.west) --[worldsheet] (p2) --[worldsheet] (f1.west)};}}
				{01s1}{\tikz [every new --/.style = {bend left},baseline={(0,-0.25)}] {\node (f1) [susy] at (90:0.65) {}; \node (p1) [coordinate] at (90+360*3/4:0.65) {}; \node (f2) [fermion] at (90+360*2/4:0.65) {}; \node (b1) [boson] at (90+360*1/4:0.65) {}; \graph {(f1.east) --[worldsheet] (p1) --[worldsheet] (f2.east); (f2.west) --[transverse] (b1) --[worldsheet] (f1.west)};}}
				{01s2}{\tikz [every new --/.style = {bend left},baseline={(0,-0.25)}] {\node (f1) [fermion] at (90:0.65) {}; \node (p1) [coordinate] at (90+360*3/4:0.65) {}; \node (f2) [susy] at (90+360*2/4:0.65) {}; \node (b1) [boson] at (90+360*1/4:0.65) {}; \graph {(f1.east) --[worldsheet] (p1) --[worldsheet] (f2.east); (f2.west) --[transverse] (b1) --[worldsheet] (f1.west)};}}
			}[\PackageError{Cring}{Undefined option to C ring: #2}{}]};

		{pp}{\IfEqCase{#2}{
					{00}{\tikz [every new --/.style = {bend left},baseline={(0,-0.25)}] {\node (f1) [fermion] at (90:0.65) {}; \node (p1) [coordinate] at (90+360*3/4:0.65) {}; \node (f2) [fermion] at (90+360*2/4:0.65) {}; \node (p2) [coordinate] at (90+360*1/4:0.65) {}; \node (n) at (0:0) {$C''_{0,0}$}; \graph {(f1.east) --[worldsheet] (p1) --[worldsheet] (f2.east); (f2.west) --[transverse] (p2) --[transverse] (f1.west)};}}
					{nm}{\tikz [every new --/.style = {bend left},baseline={(0,-0.25)}] {\node (f1) [fermion] at (90:0.65) {}; \node (b1) [boson] at (90+360*7/8:0.65) {}; \node (p1) [coordinate] at (90+360*6/8:0.65) {}; \node (p2) [coordinate] at (90+360*5/8:0.65) {}; \node (f2) [fermion] at (90+360*4/8:0.65) {}; \node (b2) [boson] at (90+360*3/8:0.65) {}; \node (p3) [coordinate] at (90+360*2/8:0.65) {}; \node (p4) [coordinate] at (90+360*1/8:0.65) {}; \node (n) at (0:0) {$C''_{2n,2m}$}; \graph {(f1.east) --[worldsheet] (b1) --[transverse] (p1) --[loosely dotted, edge label=2n] (p2) --[worldsheet] (f2.east); (f2.west) --[transverse] (b2) --[worldsheet] (p3) --[loosely dotted, edge label=2m] (p4) --[transverse] (f1.west)};}}
					{20}{\tikz [every new --/.style = {bend left},baseline={(0,-0.25)}] {\node (f1) [fermion] at (90:0.65) {}; \node (b1) [boson] at (90+360*5/6:0.65) {}; \node (b2) [boson] at (90+360*4/6:0.65) {}; \node (f2) [fermion] at (90+360*2/4:0.65) {}; \node (p2) [coordinate] at (90+360*1/4:0.65) {}; \node (n) at (0:0) {$C''_{2,0}$}; \graph {(f1.east) --[worldsheet] (b1) --[transverse] (b2) --[worldsheet] (f2.east); (f2.west) --[transverse] (p2) --[transverse] (f1.west)};}}
					{00s}{\tikz [every new --/.style = {bend left},baseline={(0,-0.25)}] {\node (f1) [susy] at (90:0.65) {}; \node (p1) [coordinate] at (90+360*3/4:0.65) {}; \node (f2) [fermion] at (90+360*2/4:0.65) {}; \node (p2) [coordinate] at (90+360*1/4:0.65) {}; \graph {(f1.east) --[worldsheet] (p1) --[worldsheet] (f2.east); (f2.west) --[transverse] (p2) --[transverse] (f1.west)};}}
				}[\PackageError{Cring}{Undefined option to C ring: #2}{}]}

	}[\PackageError{Cring}{Undefined option to C ring: #1}{}]
}

\newcommand{\Dring}[2]{
	\IfEqCase{#1}{
		{}{\IfEqCase{#2}{
					{000}{\tikz [every new --/.style = {bend left},baseline={(0,-0.25)}] {\node (f1) [fermion] at (90:0.65) {}; \node (f2) [fermion, rotate=240] at (90+360*6/9:0.65) {}; \node (f3) [fermion, rotate=120] at (90+360*3/9:0.65) {}; \node (n) at (0:0) {$D_{000}$}; \graph {(f1.east) --[worldsheet] (f2.west); (f2.east) --[worldsheet] (f3.west); (f3.east) --[worldsheet] (f1.west)};}}
					{nmk}{\tikz [every new --/.style = {bend left},baseline={(0,-0.25)}] {\node (f1) [fermion] at (90:0.65) {};  \node (p1) [coordinate] at (90+360*8/9:0.65) {}; \node (p2) [coordinate] at (90+360*7/9:0.65) {}; \node (f2) [fermion, rotate=240] at (90+360*6/9:0.65) {};\node (p3) [coordinate] at (90+360*5/9:0.65) {}; \node (p4) [coordinate] at (90+360*4/9:0.65) {}; \node (f3) [fermion, rotate=120] at (90+360*3/9:0.65) {};\node (p5) [coordinate] at (90+360*2/9:0.65) {}; \node (p6) [coordinate] at (90+360*1/9:0.65) {}; \node (n) at (0:0) {$D_{n,m,k}$}; \graph {(f1.east) --[worldsheet] (p1) --[loosely dotted, edge label=n] (p2) --[worldsheet] (f2.west); (p2) --[transverse] (f2.west); (f2.east) --[worldsheet] (p3) --[loosely dotted, edge label=m] (p4) --[worldsheet] (f3.west); (p4) --[transverse] (f3.west); (f3.east) --[worldsheet] (p5) --[loosely dotted, edge label=k] (p6) --[worldsheet] (f1.west); (p6) --[transverse] (f1.west)};}}
					{000s}{\tikz [every new --/.style = {bend left},baseline={(0,-0.25)}] {\node (f1) [susy] at (90:0.65) {}; \node (f2) [fermion, rotate=240] at (90+360*6/9:0.65) {}; \node (f3) [fermion, rotate=120] at (90+360*3/9:0.65) {}; \graph {(f1.east) --[worldsheet] (f2.west); (f2.east) --[worldsheet] (f3.west); (f3.east) --[worldsheet] (f1.west)};}}
				}[\PackageError{Dring}{Undefined option to D ring: #2}{}]};
		
		{p}{\IfEqCase{#2}{
				{000}{\tikz [every new --/.style = {bend left},baseline={(0,-0.25)}] {\node (f1) [fermion] at (90:0.65) {}; \node (f2) [fermion, rotate=60] at (90+360*6/9:0.65) {}; \node (f3) [fermion, rotate=120] at (90+360*3/9:0.65) {}; \node (n) at (0:0) {$D'_{000}$}; \graph {(f1.east) --[worldsheet] (f2.east); (f2.west) --[worldsheet] (f3.west); (f3.east) --[worldsheet]  (f1.west)};}}
				{nmk}{\tikz [every new --/.style = {bend left},baseline={(0,-0.25)}] {\node (f1) [fermion] at (90:0.65) {};  \node (p1) [coordinate] at (90+360*8/9:0.65) {}; \node (p2) [coordinate] at (90+360*7/9:0.65) {}; \node (f2) [fermion, rotate=60] at (90+360*6/9:0.65) {};\node (p3) [coordinate] at (90+360*5/9:0.65) {}; \node (p4) [coordinate] at (90+360*4/9:0.65) {}; \node (f3) [fermion, rotate=120] at (90+360*3/9:0.65) {};\node (p5) [coordinate] at (90+360*2/9:0.65) {}; \node (p6) [coordinate] at (90+360*1/9:0.65) {}; \node (n) at (0:0) {$D'_{2n,m,k}$}; \graph {(f1.east) --[worldsheet] (p1) --[loosely dotted, edge label=2n] (p2) --[worldsheet] (f2.east); (f2.west) --[worldsheet] (p3) --[loosely dotted, edge label=m] (p4) --[worldsheet] (f3.west); (p4) --[transverse] (f3.west); (f3.east) --[worldsheet] (p5) --[loosely dotted, edge label=k] (p6) --[worldsheet] (f1.west); (p6) --[transverse] (f1.west)};}}
				{000s1}{\tikz [every new --/.style = {bend left},baseline={(0,-0.25)}] {\node (f1) [susy] at (90:0.65) {}; \node (f2) [fermion, rotate=60] at (90+360*6/9:0.65) {}; \node (f3) [fermion, rotate=120] at (90+360*3/9:0.65) {}; \graph {(f1.east) --[worldsheet] (f2.east); (f2.west) --[worldsheet] (f3.west); (f3.east) --[worldsheet]  (f1.west)};}}
				{000s2}{\tikz [every new --/.style = {bend left},baseline={(0,-0.25)}] {\node (f1) [fermion] at (90:0.65) {}; \node (f2) [susy, rotate=60] at (90+360*6/9:0.65) {}; \node (f3) [fermion, rotate=120] at (90+360*3/9:0.65) {}; \graph {(f1.east) --[worldsheet] (f2.east); (f2.west) --[worldsheet] (f3.west); (f3.east) --[worldsheet]  (f1.west)};}}
				{000s3}{\tikz [every new --/.style = {bend left},baseline={(0,-0.25)}] {\node (f1) [fermion] at (90:0.65) {}; \node (f2) [fermion, rotate=60] at (90+360*6/9:0.65) {}; \node (f3) [susy, rotate=120] at (90+360*3/9:0.65) {}; \graph {(f1.east) --[worldsheet] (f2.east); (f2.west) --[worldsheet] (f3.west); (f3.east) --[worldsheet]  (f1.west)};}}
			}[\PackageError{Dring}{Undefined option to D ring: #2}{}]};
		
		{pp}{\IfEqCase{#2}{
					{000}{\tikz [every new --/.style = {bend left},baseline={(0,-0.25)}] {\node (f1) [fermion] at (90:0.65) {}; \node (f2) [fermion, rotate=60] at (90+360*6/9:0.65) {}; \node (f3) [fermion, rotate=120] at (90+360*3/9:0.65) {}; \node (n) at (0:0) {$D''_{000}$}; \graph {(f1.east) --[worldsheet] (f2.east); (f2.west) --[transverse] (f3.west); (f3.east) --[worldsheet]  (f1.west)};}}
					{nmk}{\tikz [every new --/.style = {bend left},baseline={(0,-0.25)}] {\node (f1) [fermion] at (90:0.65) {};  \node (p1) [coordinate] at (90+360*8/9:0.65) {}; \node (p2) [coordinate] at (90+360*7/9:0.65) {}; \node (f2) [fermion, rotate=60] at (90+360*6/9:0.65) {};\node (p3) [coordinate] at (90+360*5/9:0.65) {}; \node (p4) [coordinate] at (90+360*4/9:0.65) {}; \node (f3) [fermion, rotate=120] at (90+360*3/9:0.65) {};\node (p5) [coordinate] at (90+360*2/9:0.65) {}; \node (p6) [coordinate] at (90+360*1/9:0.65) {}; \node (n) at (0:0) {$D''_{2n,m,k}$}; \graph {(f1.east) --[worldsheet] (p1) --[loosely dotted, edge label=2n] (p2) --[worldsheet] (f2.east); (f2.west) --[transverse] (p3) --[loosely dotted, edge label=m] (p4) --[worldsheet] (f3.west); (p4) --[transverse] (f3.west); (f3.east) --[worldsheet] (p5) --[loosely dotted, edge label=k] (p6) --[worldsheet] (f1.west); (p6) --[transverse] (f1.west)};}}
					{000s1}{\tikz [every new --/.style = {bend left},baseline={(0,-0.25)}] {\node (f1) [susy] at (90:0.65) {}; \node (f2) [fermion, rotate=60] at (90+360*6/9:0.65) {}; \node (f3) [fermion, rotate=120] at (90+360*3/9:0.65) {}; \graph {(f1.east) --[worldsheet] (f2.east); (f2.west) --[transverse] (f3.west); (f3.east) --[worldsheet]  (f1.west)};}}
					{000s2}{\tikz [every new --/.style = {bend left},baseline={(0,-0.25)}] {\node (f1) [fermion] at (90:0.65) {}; \node (f2) [susy, rotate=60] at (90+360*6/9:0.65) {}; \node (f3) [fermion, rotate=120] at (90+360*3/9:0.65) {}; \graph {(f1.east) --[worldsheet] (f2.east); (f2.west) --[transverse] (f3.west); (f3.east) --[worldsheet]  (f1.west)};}}
					{000s3}{\tikz [every new --/.style = {bend left},baseline={(0,-0.25)}] {\node (f1) [fermion] at (90:0.65) {}; \node (f2) [fermion, rotate=60] at (90+360*6/9:0.65) {}; \node (f3) [susy, rotate=120] at (90+360*3/9:0.65) {}; \graph {(f1.east) --[worldsheet] (f2.east); (f2.west) --[transverse] (f3.west); (f3.east) --[worldsheet]  (f1.west)};}}
				}[\PackageError{Dring}{Undefined option to D ring: #2}{}]}
		
	}[\PackageError{Dring}{Undefined option to D ring: #1}{}]
}

\newcommand{\Ering}[2]{
	\IfEqCase{#1}{
		{}{\IfEqCase{#2}{
				{0000}{\tikz [every new --/.style = {bend left},baseline={(0,-0.25)}] {\node (f1) [fermion] at (90:0.65) {}; \node (f2) [fermion, rotate=270] at (90+360*9/12:0.65) {}; \node (f3) [fermion, rotate=180] at (90+360*6/12:0.65) {}; \node (f4) [fermion, rotate=90] at (90+360*3/12:0.65) {}; \node (n) at (0:0) {$E_{0000}$}; \graph {(f1.east) --[worldsheet] (f2.west); (f2.east) --[worldsheet] (f3.west); (f3.east) --[worldsheet] (f4.west); (f4.east) --[worldsheet] (f1.west)};}}
				{nmkl}{\tikz [baseline={(0,-0.25)}] {\node (f1) [fermion] at (90:1) {};  \node (p1) [coordinate] at (90+360*11/12:1*1.1547) {}; \node (p2) [coordinate] at (90+360*10/12:1*1.1547) {}; \node (f2) [fermion, rotate=270] at (90+360*9/12:1) {};\node (p3) [coordinate] at (90+360*8/12:1*1.1547) {}; \node (p4) [coordinate] at (90+360*7/12:1*1.1547) {}; \node (f3) [fermion, rotate=180] at (90+360*6/12:1) {};\node (p5) [coordinate] at (90+360*5/12:1*1.1547) {}; \node (p6) [coordinate] at (90+360*4/12:1*1.1547) {}; \node (f4) [fermion, rotate=90] at (90+360*3/12:1) {};\node (p7) [coordinate] at (90+360*2/12:1*1.1547) {}; \node (p8) [coordinate] at (90+360*1/12:1*1.1547) {}; \node (n) at (0:0) {$E_{n,m,k,\ell}$}; \graph {(f1.east) --[worldsheet] (p1) --[loosely dotted, edge label=n] (p2) --[worldsheet] (f2.west); (p2) --[transverse] (f2.west); (f2.east) --[worldsheet] (p3) --[loosely dotted, edge label=m] (p4) --[worldsheet] (f3.west); (p4) --[transverse] (f3.west); (f3.east) --[worldsheet] (p5) --[loosely dotted, edge label=k] (p6) --[worldsheet] (f4.west); (p6) --[transverse] (f4.west); (f4.east) --[worldsheet] (p7) --[loosely dotted, edge label=$\ell$] (p8) --[worldsheet] (f1.west); (p8) --[transverse] (f1.west)};}}
				{0000s}{\tikz [every new --/.style = {bend left},baseline={(0,-0.25)}] {\node (f1) [susy] at (90:0.65) {}; \node (f2) [fermion, rotate=270] at (90+360*9/12:0.65) {}; \node (f3) [fermion, rotate=180] at (90+360*6/12:0.65) {}; \node (f4) [fermion, rotate=90] at (90+360*3/12:0.65) {}; \graph {(f1.east) --[worldsheet] (f2.west); (f2.east) --[worldsheet] (f3.west); (f3.east) --[worldsheet] (f4.west); (f4.east) --[worldsheet] (f1.west)};}}
			}[\PackageError{Ering}{Undefined option to E ring: #2}{}]};
		
		{p}{\IfEqCase{#2}{
				{0000}{\tikz [every new --/.style = {bend left},baseline={(0,-0.25)}] {\node (f1) [fermion] at (90:0.65) {}; \node (f2) [fermion, rotate=90] at (90+360*9/12:0.65) {}; \node (f3) [fermion, rotate=180] at (90+360*6/12:0.65) {}; \node (f4) [fermion, rotate=90] at (90+360*3/12:0.65) {}; \node (n) at (0:0) {$E'_{0000}$}; \graph {(f1.east) --[worldsheet] (f2.east); (f2.west) --[worldsheet] (f3.west); (f3.east) --[worldsheet] (f4.west); (f4.east) --[worldsheet] (f1.west)};}}
				{nmkl}{\tikz [baseline={(0,-0.25)}] {\node (f1) [fermion] at (90:1) {};  \node (p1) [coordinate] at (90+360*11/12:1*1.1547) {}; \node (p2) [coordinate] at (90+360*10/12:1*1.1547) {}; \node (f2) [fermion, rotate=90] at (90+360*9/12:1) {};\node (p3) [coordinate] at (90+360*8/12:1*1.1547) {}; \node (p4) [coordinate] at (90+360*7/12:1*1.1547) {}; \node (f3) [fermion, rotate=180] at (90+360*6/12:1) {};\node (p5) [coordinate] at (90+360*5/12:1*1.1547) {}; \node (p6) [coordinate] at (90+360*4/12:1*1.1547) {}; \node (f4) [fermion, rotate=90] at (90+360*3/12:1) {};\node (p7) [coordinate] at (90+360*2/12:1*1.1547) {}; \node (p8) [coordinate] at (90+360*1/12:1*1.1547) {}; \node (n) at (0:0) {$E'_{2n,m,k,\ell}$}; \graph {(f1.east) --[worldsheet] (p1) --[loosely dotted, edge label=2n] (p2) --[worldsheet] (f2.east); (f2.west) --[worldsheet] (p3) --[loosely dotted, edge label=m] (p4) --[worldsheet] (f3.west); (p4) --[transverse] (f3.west); (f3.east) --[worldsheet] (p5) --[loosely dotted, edge label=k] (p6) --[worldsheet] (f4.west); (p6) --[transverse] (f4.west); (f4.east) --[worldsheet] (p7) --[loosely dotted, edge label=$\ell$] (p8) --[worldsheet] (f1.west); (p8) --[transverse] (f1.west)};}}
				{0000s1}{\tikz [every new --/.style = {bend left},baseline={(0,-0.25)}] {\node (f1) [susy] at (90:0.65) {}; \node (f2) [fermion, rotate=90] at (90+360*9/12:0.65) {}; \node (f3) [fermion, rotate=180] at (90+360*6/12:0.65) {}; \node (f4) [fermion, rotate=90] at (90+360*3/12:0.65) {}; \graph {(f1.east) --[worldsheet] (f2.east); (f2.west) --[worldsheet] (f3.west); (f3.east) --[worldsheet] (f4.west); (f4.east) --[worldsheet] (f1.west)};}}
				{0000s2}{\tikz [every new --/.style = {bend left},baseline={(0,-0.25)}] {\node (f1) [fermion] at (90:0.65) {}; \node (f2) [susy, rotate=90] at (90+360*9/12:0.65) {}; \node (f3) [fermion, rotate=180] at (90+360*6/12:0.65) {}; \node (f4) [fermion, rotate=90] at (90+360*3/12:0.65) {}; \graph {(f1.east) --[worldsheet] (f2.east); (f2.west) --[worldsheet] (f3.west); (f3.east) --[worldsheet] (f4.west); (f4.east) --[worldsheet] (f1.west)};}}
				{0000s3}{\tikz [every new --/.style = {bend left},baseline={(0,-0.25)}] {\node (f1) [fermion] at (90:0.65) {}; \node (f2) [fermion, rotate=90] at (90+360*9/12:0.65) {}; \node (f3) [susy, rotate=180] at (90+360*6/12:0.65) {}; \node (f4) [fermion, rotate=90] at (90+360*3/12:0.65) {}; \graph {(f1.east) --[worldsheet] (f2.east); (f2.west) --[worldsheet] (f3.west); (f3.east) --[worldsheet] (f4.west); (f4.east) --[worldsheet] (f1.west)};}}
				{0000s4}{\tikz [every new --/.style = {bend left},baseline={(0,-0.25)}] {\node (f1) [fermion] at (90:0.65) {}; \node (f2) [fermion, rotate=90] at (90+360*9/12:0.65) {}; \node (f3) [fermion, rotate=180] at (90+360*6/12:0.65) {}; \node (f4) [susy, rotate=90] at (90+360*3/12:0.65) {}; \graph {(f1.east) --[worldsheet] (f2.east); (f2.west) --[worldsheet] (f3.west); (f3.east) --[worldsheet] (f4.west); (f4.east) --[worldsheet] (f1.west)};}}
			}[\PackageError{Ering}{Undefined option to E ring: #2}{}]};
		
		{pp}{\IfEqCase{#2}{
				{0000}{\tikz [every new --/.style = {bend left},baseline={(0,-0.25)}] {\node (f1) [fermion] at (90:0.65) {}; \node (f2) [fermion, rotate=90] at (90+360*9/12:0.65) {}; \node (f3) [fermion, rotate=180] at (90+360*6/12:0.65) {}; \node (f4) [fermion, rotate=90] at (90+360*3/12:0.65) {}; \node (n) at (0:0) {$E''_{0000}$}; \graph {(f1.east) --[worldsheet] (f2.east); (f2.west) --[transverse] (f3.west); (f3.east) --[worldsheet] (f4.west); (f4.east) --[worldsheet] (f1.west)};}}
				{nmkl}{\tikz [baseline={(0,-0.25)}] {\node (f1) [fermion] at (90:1) {};  \node (p1) [coordinate] at (90+360*11/12:1*1.1547) {}; \node (p2) [coordinate] at (90+360*10/12:1*1.1547) {}; \node (f2) [fermion, rotate=90] at (90+360*9/12:1) {};\node (p3) [coordinate] at (90+360*8/12:1*1.1547) {}; \node (p4) [coordinate] at (90+360*7/12:1*1.1547) {}; \node (f3) [fermion, rotate=180] at (90+360*6/12:1) {};\node (p5) [coordinate] at (90+360*5/12:1*1.1547) {}; \node (p6) [coordinate] at (90+360*4/12:1*1.1547) {}; \node (f4) [fermion, rotate=90] at (90+360*3/12:1) {};\node (p7) [coordinate] at (90+360*2/12:1*1.1547) {}; \node (p8) [coordinate] at (90+360*1/12:1*1.1547) {}; \node (n) at (0:0) {$E''_{2n,m,k,\ell}$}; \graph {(f1.east) --[worldsheet] (p1) --[loosely dotted, edge label=2n] (p2) --[worldsheet] (f2.east); (f2.west) --[transverse] (p3) --[loosely dotted, edge label=m] (p4) --[worldsheet] (f3.west); (p4) --[transverse] (f3.west); (f3.east) --[worldsheet] (p5) --[loosely dotted, edge label=k] (p6) --[worldsheet] (f4.west); (p6) --[transverse] (f4.west); (f4.east) --[worldsheet] (p7) --[loosely dotted, edge label=$\ell$] (p8) --[worldsheet] (f1.west); (p8) --[transverse] (f1.west)};}}
				{0000s1}{\tikz [every new --/.style = {bend left},baseline={(0,-0.25)}] {\node (f1) [susy] at (90:0.65) {}; \node (f2) [fermion, rotate=90] at (90+360*9/12:0.65) {}; \node (f3) [fermion, rotate=180] at (90+360*6/12:0.65) {}; \node (f4) [fermion, rotate=90] at (90+360*3/12:0.65) {}; \graph {(f1.east) --[worldsheet] (f2.east); (f2.west) --[transverse] (f3.west); (f3.east) --[worldsheet] (f4.west); (f4.east) --[worldsheet] (f1.west)};}}
				{0000s2}{\tikz [every new --/.style = {bend left},baseline={(0,-0.25)}] {\node (f1) [fermion] at (90:0.65) {}; \node (f2) [susy, rotate=90] at (90+360*9/12:0.65) {}; \node (f3) [fermion, rotate=180] at (90+360*6/12:0.65) {}; \node (f4) [fermion, rotate=90] at (90+360*3/12:0.65) {}; \graph {(f1.east) --[worldsheet] (f2.east); (f2.west) --[transverse] (f3.west); (f3.east) --[worldsheet] (f4.west); (f4.east) --[worldsheet] (f1.west)};}}
				{0000s3}{\tikz [every new --/.style = {bend left},baseline={(0,-0.25)}] {\node (f1) [fermion] at (90:0.65) {}; \node (f2) [fermion, rotate=90] at (90+360*9/12:0.65) {}; \node (f3) [susy, rotate=180] at (90+360*6/12:0.65) {}; \node (f4) [fermion, rotate=90] at (90+360*3/12:0.65) {}; \graph {(f1.east) --[worldsheet] (f2.east); (f2.west) --[transverse] (f3.west); (f3.east) --[worldsheet] (f4.west); (f4.east) --[worldsheet] (f1.west)};}}
				{0000s4}{\tikz [every new --/.style = {bend left},baseline={(0,-0.25)}] {\node (f1) [fermion] at (90:0.65) {}; \node (f2) [fermion, rotate=90] at (90+360*9/12:0.65) {}; \node (f3) [fermion, rotate=180] at (90+360*6/12:0.65) {}; \node (f4) [susy, rotate=90] at (90+360*3/12:0.65) {}; \graph {(f1.east) --[worldsheet] (f2.east); (f2.west) --[transverse] (f3.west); (f3.east) --[worldsheet] (f4.west); (f4.east) --[worldsheet] (f1.west)};}}
			}[\PackageError{Ering}{Undefined option to E ring: #2}{}]};
		
		{b}{\IfEqCase{#2}{
				{0000}{\tikz [every new --/.style = {bend left},baseline={(0,-0.25)}] {\node (f1) [fermion] at (90:0.65) {}; \node (f2) [fermion, rotate=90] at (90+360*9/12:0.65) {}; \node (f3) [fermion, rotate=180] at (90+360*6/12:0.65) {}; \node (f4) [fermion, rotate=270] at (90+360*3/12:0.65) {}; \node (n) at (0:0) {$\overline{E}_{0000}$}; \graph {(f1.east) --[worldsheet] (f2.east); (f2.west) --[worldsheet] (f3.west); (f3.east) --[worldsheet] (f4.east); (f4.west) --[worldsheet] (f1.west)};}}
				{nmkl}{\tikz [baseline={(0,-0.25)}] {\node (f1) [fermion] at (90:1) {};  \node (p1) [coordinate] at (90+360*11/12:1*1.1547) {}; \node (p2) [coordinate] at (90+360*10/12:1*1.1547) {}; \node (f2) [fermion, rotate=90] at (90+360*9/12:1) {};\node (p3) [coordinate] at (90+360*8/12:1*1.1547) {}; \node (p4) [coordinate] at (90+360*7/12:1*1.1547) {}; \node (f3) [fermion, rotate=180] at (90+360*6/12:1) {};\node (p5) [coordinate] at (90+360*5/12:1*1.1547) {}; \node (p6) [coordinate] at (90+360*4/12:1*1.1547) {}; \node (f4) [fermion, rotate=270] at (90+360*3/12:1) {};\node (p7) [coordinate] at (90+360*2/12:1*1.1547) {}; \node (p8) [coordinate] at (90+360*1/12:1*1.1547) {}; \node (n) at (0:0) {$\overline{E}_{2n,m,2k,\ell}$}; \graph {(f1.east) --[worldsheet] (p1) --[loosely dotted, edge label=2n] (p2) --[worldsheet] (f2.east); (f2.west) --[worldsheet] (p3) --[loosely dotted, edge label=m] (p4) --[worldsheet] (f3.west); (p4) --[transverse] (f3.west); (f3.east) --[worldsheet] (p5) --[loosely dotted, edge label=2k] (p6) --[worldsheet] (f4.east); (f4.west) --[worldsheet] (p7); (f4.west) --[transverse] (p7) --[loosely dotted, edge label=$\ell$] (p8) --[worldsheet] (f1.west)};}}
				{0000s}{\tikz [every new --/.style = {bend left},baseline={(0,-0.25)}] {\node (f1) [susy] at (90:0.65) {}; \node (f2) [fermion, rotate=90] at (90+360*9/12:0.65) {}; \node (f3) [fermion, rotate=180] at (90+360*6/12:0.65) {}; \node (f4) [fermion, rotate=270] at (90+360*3/12:0.65) {}; \graph {(f1.east) --[worldsheet] (f2.east); (f2.west) --[worldsheet] (f3.west); (f3.east) --[worldsheet] (f4.east); (f4.west) --[worldsheet] (f1.west)};}}
			}[\PackageError{Ering}{Undefined option to E ring: #2}{}]};
		
		{bb}{\IfEqCase{#2}{
				{0000}{\tikz [every new --/.style = {bend left},baseline={(0,-0.25)}] {\node (f1) [fermion] at (90:0.65) {}; \node (f2) [fermion, rotate=90] at (90+360*9/12:0.65) {}; \node (f3) [fermion, rotate=180] at (90+360*6/12:0.65) {}; \node (f4) [fermion, rotate=270] at (90+360*3/12:0.65) {}; \node (n) at (0:0) {$\overline{\overline{E}}_{0000}$}; \graph {(f1.east) --[worldsheet] (f2.east); (f2.west) --[transverse] (f3.west); (f3.east) --[worldsheet] (f4.east); (f4.west) --[transverse] (f1.west)};}}
				{nmkl}{\tikz [baseline={(0,-0.25)}] {\node (f1) [fermion] at (90:1) {};  \node (p1) [coordinate] at (90+360*11/12:1*1.1547) {}; \node (p2) [coordinate] at (90+360*10/12:1*1.1547) {}; \node (f2) [fermion, rotate=90] at (90+360*9/12:1) {};\node (p3) [coordinate] at (90+360*8/12:1*1.1547) {}; \node (p4) [coordinate] at (90+360*7/12:1*1.1547) {}; \node (f3) [fermion, rotate=180] at (90+360*6/12:1) {};\node (p5) [coordinate] at (90+360*5/12:1*1.1547) {}; \node (p6) [coordinate] at (90+360*4/12:1*1.1547) {}; \node (f4) [fermion, rotate=270] at (90+360*3/12:1) {};\node (p7) [coordinate] at (90+360*2/12:1*1.1547) {}; \node (p8) [coordinate] at (90+360*1/12:1*1.1547) {}; \node (n) at (0:0) {$\overline{\overline{E}}_{2n,m,2k,\ell}$}; \graph {(f1.east) --[worldsheet] (p1) --[loosely dotted, edge label=2n] (p2) --[worldsheet] (f2.east); (f2.west) --[transverse] (p3) --[loosely dotted, edge label=m] (p4) --[worldsheet] (f3.west); (p4) --[transverse] (f3.west); (f3.east) --[worldsheet] (p5) --[loosely dotted, edge label=2k] (p6) --[worldsheet] (f4.east); (f4.west) --[transverse] (p7) --[loosely dotted, edge label=$\ell$] (p8) --[worldsheet] (f1.west); (p8) --[transverse] (f1.west)};}}
				{0000s}{\tikz [every new --/.style = {bend left},baseline={(0,-0.25)}] {\node (f1) [susy] at (90:0.65) {}; \node (f2) [fermion, rotate=90] at (90+360*9/12:0.65) {}; \node (f3) [fermion, rotate=180] at (90+360*6/12:0.65) {}; \node (f4) [fermion, rotate=270] at (90+360*3/12:0.65) {}; \graph {(f1.east) --[worldsheet] (f2.east); (f2.west) --[transverse] (f3.west); (f3.east) --[worldsheet] (f4.east); (f4.west) --[transverse] (f1.west)};}}
			}[\PackageError{Ering}{Undefined option to E ring: #2}{}]};
		
		{bbb}{\IfEqCase{#2}{
				{0000}{\tikz [every new --/.style = {bend left},baseline={(0,-0.25)}] {\node (f1) [fermion] at (90:0.65) {}; \node (f2) [fermion, rotate=90] at (90+360*9/12:0.65) {}; \node (f3) [fermion, rotate=180] at (90+360*6/12:0.65) {}; \node (f4) [fermion, rotate=270] at (90+360*3/12:0.65) {}; \node (n) at (0:0) {$\overline{\overline{\overline{E}}}_{0000}$}; \graph {(f1.east) --[worldsheet] (f2.east); (f2.west) --[transverse] (f3.west); (f3.east) --[worldsheet] (f4.east); (f4.west) --[worldsheet] (f1.west)};}}
				{nmkl}{\tikz [baseline={(0,-0.25)}] {\node (f1) [fermion] at (90:1) {};  \node (p1) [coordinate] at (90+360*11/12:1*1.1547) {}; \node (p2) [coordinate] at (90+360*10/12:1*1.1547) {}; \node (f2) [fermion, rotate=90] at (90+360*9/12:1) {};\node (p3) [coordinate] at (90+360*8/12:1*1.1547) {}; \node (p4) [coordinate] at (90+360*7/12:1*1.1547) {}; \node (f3) [fermion, rotate=180] at (90+360*6/12:1) {};\node (p5) [coordinate] at (90+360*5/12:1*1.1547) {}; \node (p6) [coordinate] at (90+360*4/12:1*1.1547) {}; \node (f4) [fermion, rotate=270] at (90+360*3/12:1) {};\node (p7) [coordinate] at (90+360*2/12:1*1.1547) {}; \node (p8) [coordinate] at (90+360*1/12:1*1.1547) {}; \node (n) at (0:0) {$\overline{\overline{\overline{E}}}_{2n,2m,2k,2\ell}$}; \graph {(f1.east) --[worldsheet] (p1) --[loosely dotted, edge label=2n] (p2) --[worldsheet] (f2.east); (f2.west) --[transverse] (p3) --[loosely dotted, edge label=2m] (p4) --[transverse] (f3.west); (f3.east) --[worldsheet] (p5) --[loosely dotted, edge label=2k] (p6) --[worldsheet] (f4.east); (f4.west) --[worldsheet] (p7) --[loosely dotted, edge label=$2\ell$] (p8) --[worldsheet] (f1.west)};}}
				{0000s1}{\tikz [every new --/.style = {bend left},baseline={(0,-0.25)}] {\node (f1) [susy] at (90:0.65) {}; \node (f2) [fermion, rotate=90] at (90+360*9/12:0.65) {}; \node (f3) [fermion, rotate=180] at (90+360*6/12:0.65) {}; \node (f4) [fermion, rotate=270] at (90+360*3/12:0.65) {}; \graph {(f1.east) --[worldsheet] (f2.east); (f2.west) --[transverse] (f3.west); (f3.east) --[worldsheet] (f4.east); (f4.west) --[worldsheet] (f1.west)};}}
				{0000s2}{\tikz [every new --/.style = {bend left},baseline={(0,-0.25)}] {\node (f1) [fermion] at (90:0.65) {}; \node (f2) [susy, rotate=90] at (90+360*9/12:0.65) {}; \node (f3) [fermion, rotate=180] at (90+360*6/12:0.65) {}; \node (f4) [fermion, rotate=270] at (90+360*3/12:0.65) {}; \graph {(f1.east) --[worldsheet] (f2.east); (f2.west) --[transverse] (f3.west); (f3.east) --[worldsheet] (f4.east); (f4.west) --[worldsheet] (f1.west)};}}
			}[\PackageError{Ering}{Undefined option to E ring: #2}{}]};
		
		{t}{\IfEqCase{#2}{
				{0000}{\tikz [every new --/.style = {bend left},baseline={(0,-0.25)}] {\node (f1) [fermion] at (90:0.65) {}; \node (f2) [fermion, rotate=90] at (90+360*9/12:0.65) {}; \node (f3) [fermion] at (90+360*6/12:0.65) {}; \node (f4) [fermion, rotate=90] at (90+360*3/12:0.65) {}; \node (n) at (0:0) {$\tilde{E}_{0000}$}; \graph {(f1.east) --[worldsheet] (f2.east); (f2.west) --[worldsheet] (f3.east); (f3.west) --[worldsheet] (f4.west); (f4.east) --[worldsheet] (f1.west)};}}
				{nmkl}{\tikz [baseline={(0,-0.25)}] {\node (f1) [fermion] at (90:1) {};  \node (p1) [coordinate] at (90+360*11/12:1*1.1547) {}; \node (p2) [coordinate] at (90+360*10/12:1*1.1547) {}; \node (f2) [fermion, rotate=90] at (90+360*9/12:1) {};\node (p3) [coordinate] at (90+360*8/12:1*1.1547) {}; \node (p4) [coordinate] at (90+360*7/12:1*1.1547) {}; \node (f3) [fermion] at (90+360*6/12:1) {};\node (p5) [coordinate] at (90+360*5/12:1*1.1547) {}; \node (p6) [coordinate] at (90+360*4/12:1*1.1547) {}; \node (f4) [fermion, rotate=90] at (90+360*3/12:1) {};\node (p7) [coordinate] at (90+360*2/12:1*1.1547) {}; \node (p8) [coordinate] at (90+360*1/12:1*1.1547) {}; \node (n) at (0:0) {$\tilde{E}_{2n,m,k,\ell}$}; \graph {(f1.east) --[worldsheet] (p1) --[loosely dotted, edge label=2n] (p2) --[worldsheet] (f2.east); (f2.west) --[transverse] (p3); (f2.west) --[worldsheet] (p3) --[loosely dotted, edge label=m] (p4) --[worldsheet] (f3.east); (f3.west) --[worldsheet] (p5) --[loosely dotted, edge label=k] (p6) --[worldsheet] (f4.west); (p6) --[transverse] (f4.west); (f4.east) --[worldsheet] (p7) --[loosely dotted, edge label=$\ell$] (p8) --[worldsheet] (f1.west); (p8) --[transverse] (f1.west)};}}
				{0000s1}{\tikz [every new --/.style = {bend left},baseline={(0,-0.25)}] {\node (f1) [susy] at (90:0.65) {}; \node (f2) [fermion, rotate=90] at (90+360*9/12:0.65) {}; \node (f3) [fermion] at (90+360*6/12:0.65) {}; \node (f4) [fermion, rotate=90] at (90+360*3/12:0.65) {}; \graph {(f1.east) --[worldsheet] (f2.east); (f2.west) --[worldsheet] (f3.east); (f3.west) --[worldsheet] (f4.west); (f4.east) --[worldsheet] (f1.west)};}}
				{0000s2}{\tikz [every new --/.style = {bend left},baseline={(0,-0.25)}] {\node (f1) [fermion] at (90:0.65) {}; \node (f2) [fermion, rotate=90] at (90+360*9/12:0.65) {}; \node (f3) [fermion] at (90+360*6/12:0.65) {}; \node (f4) [susy, rotate=90] at (90+360*3/12:0.65) {}; \graph {(f1.east) --[worldsheet] (f2.east); (f2.west) --[worldsheet] (f3.east); (f3.west) --[worldsheet] (f4.west); (f4.east) --[worldsheet] (f1.west)};}}
			}[\PackageError{Ering}{Undefined option to E ring: #2}{}]};
		
		{tt}{\IfEqCase{#2}{
				{0000}{\tikz [every new --/.style = {bend left},baseline={(0,-0.25)}] {\node (f1) [fermion] at (90:0.65) {}; \node (f2) [fermion, rotate=90] at (90+360*9/12:0.65) {}; \node (f3) [fermion] at (90+360*6/12:0.65) {}; \node (f4) [fermion, rotate=90] at (90+360*3/12:0.65) {}; \node (n) at (0:0) {$\tilde{\tilde{E}}_{0000}$}; \graph {(f1.east) --[worldsheet] (f2.east); (f2.west) --[worldsheet] (f3.east); (f3.west) --[transverse] (f4.west); (f4.east) --[worldsheet] (f1.west)};}}
				{nmkl}{\tikz [baseline={(0,-0.25)}] {\node (f1) [fermion] at (90:1) {};  \node (p1) [coordinate] at (90+360*11/12:1*1.1547) {}; \node (p2) [coordinate] at (90+360*10/12:1*1.1547) {}; \node (f2) [fermion, rotate=90] at (90+360*9/12:1) {};\node (p3) [coordinate] at (90+360*8/12:1*1.1547) {}; \node (p4) [coordinate] at (90+360*7/12:1*1.1547) {}; \node (f3) [fermion] at (90+360*6/12:1) {};\node (p5) [coordinate] at (90+360*5/12:1*1.1547) {}; \node (p6) [coordinate] at (90+360*4/12:1*1.1547) {}; \node (f4) [fermion, rotate=90] at (90+360*3/12:1) {};\node (p7) [coordinate] at (90+360*2/12:1*1.1547) {}; \node (p8) [coordinate] at (90+360*1/12:1*1.1547) {}; \node (n) at (0:0) {$\tilde{\tilde{E}}_{2n,m,2k,\ell}$}; \graph {(f1.east) --[worldsheet] (p1) --[loosely dotted, edge label=2n] (p2) --[worldsheet] (f2.east); (f2.west) --[transverse] (p3); (f2.west) --[worldsheet] (p3) --[loosely dotted, edge label=m] (p4) --[worldsheet] (f3.east); (f3.west) --[transverse] (p5) --[loosely dotted, edge label=2k] (p6) --[transverse] (f4.west); (f4.east) --[worldsheet] (p7) --[loosely dotted, edge label=$\ell$] (p8) --[worldsheet] (f1.west); (p8) --[transverse] (f1.west)};}}
				{0000s1}{\tikz [every new --/.style = {bend left},baseline={(0,-0.25)}] {\node (f1) [susy] at (90:0.65) {}; \node (f2) [fermion, rotate=90] at (90+360*9/12:0.65) {}; \node (f3) [fermion] at (90+360*6/12:0.65) {}; \node (f4) [fermion, rotate=90] at (90+360*3/12:0.65) {}; \graph {(f1.east) --[worldsheet] (f2.east); (f2.west) --[worldsheet] (f3.east); (f3.west) --[transverse] (f4.west); (f4.east) --[worldsheet] (f1.west)};}}
				{0000s2}{\tikz [every new --/.style = {bend left},baseline={(0,-0.25)}] {\node (f1) [fermion] at (90:0.65) {}; \node (f2) [fermion, rotate=90] at (90+360*9/12:0.65) {}; \node (f3) [fermion] at (90+360*6/12:0.65) {}; \node (f4) [susy, rotate=90] at (90+360*3/12:0.65) {}; \graph {(f1.east) --[worldsheet] (f2.east); (f2.west) --[worldsheet] (f3.east); (f3.west) --[transverse] (f4.west); (f4.east) --[worldsheet] (f1.west)};}}
			}[\PackageError{Ering}{Undefined option to E ring: #2}{}]}
		
	}[\PackageError{Ering}{Undefined option to E ring: #1}{}]
}

\newcommand{\Scaleoneringkl}{
	\tikz [every new --/.style = {bend left},baseline={(0,-0.25)}] {\node (f1) [dfdf] at (90:0.65) {}; \node (b1) [comb,rotate=-45] at (90+360*7/8:0.65) {}; \node (p1) [coordinate] at (90+360*6/8:0.65) {}; \node (p2) [coordinate] at (90+360*5/8:0.65) {}; \node (f2) [dfdf, rotate=180] at (90+360*4/8:0.65) {}; \node (b2) [comb,rotate=135] at (90+360*3/8:0.65) {}; \node (p3) [coordinate] at (90+360*2/8:0.65) {}; \node (p4) [coordinate] at (90+360*1/8:0.65) {}; \node (n) at (0:0) {$B_{k\ell}$}; \graph {(f1.east) --[worldsheet] (b1.west); (b1.east) --[worldsheet] (p1) --[loosely dotted, edge label=k] (p2) --[worldsheet] (f2.west); (f2.east) --[worldsheet] (b2.west); (b2.east) --[worldsheet] (p3) --[loosely dotted, edge label=$\ell$] (p4) --[worldsheet] (f1.west)};}
}

\newcommand{\ScaleTwoSeed}[1]{
	\IfEqCase{#1}{
		{Ring}{\tikz [every new --/.style = {bend left=90},baseline={(0,-0.25)}] {\node (b0) [dfdf] at (90:0.65) {}; \node (b1) [dfdf] at (90+360*2/4:0.65) {}; \graph {(b0) --[worldsheet] (b1) --[worldsheet]  (b0)};}}
		{Theta1}{\tikz [every new --/.style = {bend left=90},baseline={(0,-0.25)}] {\node (b0) [boson] at (180:0.65) {}; \node (b1) [boson] at (180+360*2/4:0.65) {}; \graph {(b0) --[worldsheet,out=135,in=45,looseness=1.6] (b1) --[transverse,looseness=0.6]  (b0);(b0) --[worldsheet,out=45,in=135,looseness=0.6] (b1)};}}
		{Theta2}{\tikz [every new --/.style = {bend left=90},baseline={(0,-0.25)}] {\node (b0) [boson] at (180:0.65) {}; \node (b1) [boson] at (180+360*3/4:0.65) {}; \node (b2) [boson] at (180+360*2/4:0.65) {}; \node (b3) [boson] at (180+360*1/4:0.65) {}; \graph {(b0) --[worldsheet,in=135,out=90,looseness=1.6] (b1) --[transverse,in=135,out=45,looseness=1.6] (b2) --[worldsheet,in=135,out=90,looseness=1.6] (b3) --[transverse,in=135,out=45,looseness=1.6]  (b0);(b0) --[worldsheet,out=45,in=225,looseness=1] (b2)};}}
		{Theta3}{\tikz [every new --/.style = {bend left=90},baseline={(0,-0.25)}] {\node (f0) [fermion, rotate=90] at (180:0.65) {}; \node (f1) [fermion, rotate=90] at (180+360*2/4:0.65) {}; \graph {(f0) --[worldsheet,out=135,in=45,looseness=1.6] (f1) --[worldsheet,looseness=0.6]  (f0);(f0) --[worldsheet,out=45,in=135,looseness=0.6] (f1)};}}
		{Theta4}{\tikz [every new --/.style = {bend left=90},baseline={(0,-0.25)}] {\node (f0) [fermion, rotate=90] at (180:0.65) {}; \node (f1) [fermion, rotate=270] at (180+360*2/4:0.65) {}; \graph {(f0) --[worldsheet,out=135,in=90,looseness=1.6] (f1) --[worldsheet,out=135,in=90,looseness=1.6]  (f0);(f0) --[worldsheet,out=45,in=225,looseness=1] (f1)};}}
		{Theta5}{\tikz [every new --/.style = {bend left=90},baseline={(0,-0.25)}] {\node (f0) [fermion, rotate=90] at (180:0.65) {}; \node (b0) [boson] at (180+360*2/4:0.65) {}; \node (b1) [boson] at (180+360*1/4:0.45) {}; \graph {(f0) --[worldsheet,out=135,in=45,looseness=1.6] (b0) --[transverse,out=45,in=135,looseness=0.6]  (b1) --[worldsheet,out=45,in=135,looseness=0.6]  (f0);(f0) --[worldsheet,out=45,in=135,looseness=0.6] (b0)};}}
		{Theta6}{\tikz [every new --/.style = {bend left=90},baseline={(0,-0.25)}] {\node (f0) [fermion, rotate=90] at (180:0.65) {}; \node (b0) [boson] at (180+360*3/4:0.65) {}; \node (b1) [boson] at (180+360*2/4:0.65) {}; \graph {(f0) --[worldsheet,out=90,in=135,looseness=1.6] (b0) --[transverse,out=45,in=135,looseness=1.6] (b1)  --[worldsheet,out=135,in=90,looseness=1.6]  (f0);(f0) --[worldsheet,out=45,in=225,looseness=1] (b1)};}}
		{Dumbbell1}{\tikz [every new --/.style = {bend left=90},baseline={(0,-0.25)}] {\node (b0) [boson] at (0,0) {}; \node (b1) [boson] at (1,0) {}; \node (p0) [coordinate] at (-0.65,0) {}; \node (p1) [coordinate] at (1+0.65,0) {}; \graph {(b0) --[transverse,bend left=0] (b1);(b0) --[worldsheet,out=45,in=90,looseness=1.6] (p0) --[worldsheet,out=90,in=135,looseness=1.6] (b0);(b1) --[worldsheet,out=45,in=90,looseness=1.6] (p1) --[worldsheet,out=90,in=135,looseness=1.6] (b1)};}}
		{Dumbbell2}{\tikz [every new --/.style = {bend left=90},baseline={(0,-0.25)}] {\node (b0) [boson] at (0,0) {}; \node (b1) [boson] at (1,0) {}; \node (b2) [boson] at (-1,0) {}; \node (b3) [boson] at (2,0) {}; \graph {(b0) --[worldsheet,out=45,in=135,looseness=0.6] (b1);(b0) --[transverse,out=90,in=90,looseness=0.6] (b2) --[worldsheet,out=90,in=135,looseness=1] (b0);(b1) --[worldsheet,out=45,in=90,looseness=1] (b3) --[transverse,out=90,in=90,looseness=0.6] (b1)};}}
		{Dumbbell3}{\tikz [every new --/.style = {bend left=90},baseline={(0,-0.25)}] {\node (b0) [boson] at (0,0) {}; \node (b1) [boson] at (1,0) {}; \node (p0) [coordinate] at (-0.65,0) {}; \node (b2) [boson] at (2,0) {}; \node (b3) [boson] at (3,0) {}; \graph {(b0) --[transverse,bend left=0] (b1);(b0) --[worldsheet,out=45,in=90,looseness=1.6] (p0) --[worldsheet,out=90,in=135,looseness=1.6] (b0);(b1) --[worldsheet,bend left=45] (b2);(b2) --[worldsheet,out=45,in=90,looseness=1] (b3) --[transverse,out=90,in=90,looseness=0.6] (b2)};}}
		{Dumbbell4}{\tikz [every new --/.style = {bend left=90},baseline={(0,-0.25)}] {\node (f0) [fermion,rotate=180] at (0,0) {}; \node (f1) [fermion] at (1,0) {}; \node (p0) [coordinate] at (-0.65,0) {}; \node (p1) [coordinate] at (1+0.65,0) {}; \graph {(f0) --[worldsheet,bend left=0] (f1);(f0) --[worldsheet,out=45,in=90,looseness=1.6] (p0) --[worldsheet,out=90,in=135,looseness=1.6] (f0);(f1) --[worldsheet,out=45,in=90,looseness=1.6] (p1) --[worldsheet,out=90,in=135,looseness=1.6] (f1)};}}
		{Dumbbell5}{\tikz [every new --/.style = {bend left=90},baseline={(0,-0.25)}] {\node (f0) [fermion,rotate=90] at (0,0) {}; \node (f1) [fermion,rotate=270] at (1,0) {}; \node (p0) [coordinate] at (-0.65,0) {}; \node (p1) [coordinate] at (1+0.65,0) {}; \graph {(f0) --[worldsheet,out=45,in=225,looseness=1] (f1);(f0) --[worldsheet,out=90,in=90,looseness=1.6] (p0) --[worldsheet,out=90,in=135,looseness=1.6] (f0);(f1) --[worldsheet,out=90,in=90,looseness=1.6] (p1) --[worldsheet,out=90,in=135,looseness=1.6] (f1)};}}
		{Dumbbell6}{\tikz [every new --/.style = {bend left=90},baseline={(0,-0.25)}] {\node (f0) [fermion,rotate=180] at (0,0) {}; \node (f1) [fermion,rotate=270] at (1,0) {}; \node (p0) [coordinate] at (-0.65,0) {}; \node (p1) [coordinate] at (1+0.65,0) {}; \graph {(f0) --[worldsheet,out=0,in=225,looseness=1] (f1);(f0) --[worldsheet,out=45,in=90,looseness=1.6] (p0) --[worldsheet,out=90,in=135,looseness=1.6] (f0);(f1) --[worldsheet,out=90,in=90,looseness=1.6] (p1) --[worldsheet,out=90,in=135,looseness=1.6] (f1)};}}
		{Dumbbell7}{\tikz [every new --/.style = {bend left=90},baseline={(0,-0.25)}] {\node (f0) [fermion,rotate=180] at (0,0) {}; \node (p0) [coordinate] at (-0.65,0) {}; \node (b2) [boson] at (1,0) {}; \node (b3) [boson] at (2,0) {}; \graph {(f0) --[worldsheet,bend left=0] (b2);(b0) --[worldsheet,out=45,in=90,looseness=1.6] (p0) --[worldsheet,out=90,in=135,looseness=1.6] (b0);(b2) --[worldsheet,out=45,in=90,looseness=1] (b3) --[transverse,out=90,in=90,looseness=0.6] (b2)};}}
		{Dumbbell8}{\tikz [every new --/.style = {bend left=90},baseline={(0,-0.25)}] {\node (f0) [fermion,rotate=180] at (0,0) {}; \node (b0) [boson] at (1,0) {}; \node (p0) [coordinate] at (-0.65,0) {}; \node (b1) [boson] at (2,0) {}; \node (p1) [coordinate] at (2+0.65,0) {}; \graph {(f0) --[worldsheet,bend left=0] (b0) --[transverse,bend left=0] (b1);(f0) --[worldsheet,out=45,in=90,looseness=1.6] (p0) --[worldsheet,out=90,in=135,looseness=1.6] (f0);(b1) --[worldsheet,out=45,in=90,looseness=1.6] (p1) --[worldsheet,out=90,in=135,looseness=1.6] (b1)};}}
		{Dumbbell9}{\tikz [every new --/.style = {bend left=90},baseline={(0,-0.25)}] {\node (f0) [fermion,rotate=90] at (0,0) {}; \node (p0) [coordinate] at (-0.65,0) {}; \node (b2) [boson] at (1,0) {}; \node (b3) [boson] at (2,0) {}; \graph {(f0) --[worldsheet,out=90,in=90,looseness=1.6] (p0) --[worldsheet,out=90,in=135,looseness=1.6] (f0);(f0) --[worldsheet,bend left=45] (b2);(b2) --[worldsheet,out=45,in=90,looseness=1] (b3) --[transverse,out=90,in=90,looseness=0.6] (b2)};}}
		{Dumbbell10}{\tikz [every new --/.style = {bend left=90},baseline={(0,-0.25)}] {\node (f0) [fermion,rotate=90] at (0,0) {}; \node (b0) [boson] at (1,0) {}; \node (p0) [coordinate] at (-0.65,0) {}; \node (b1) [boson] at (2,0) {}; \node (p1) [coordinate] at (2+0.65,0) {}; \graph {(f0) --[worldsheet,out=45,in=225,looseness=1] (b0) --[transverse,bend left=0] (b1);(f0) --[worldsheet,out=90,in=90,looseness=1.6] (p0) --[worldsheet,out=90,in=135,looseness=1.6] (f0);(b1) --[worldsheet,out=45,in=90,looseness=1.6] (p1) --[worldsheet,out=90,in=135,looseness=1.6] (b1)};}}
		{forbidden1}{\tikz [every new --/.style = {bend left=90},baseline={(0,-0.25)}] {\node (b0) [boson] at (180:0.65) {}; \node (b1) [lorentz] at (180+360*3/4:0.65) {}; \node (b2) [boson] at (180+360*2/4:0.65) {}; \node (b3) [boson] at (180+360*1/4:0.65) {}; \graph {(b0) --[worldsheet,in=135,out=90,looseness=1.6] (b1) --[transverse,in=135,out=45,looseness=1.6] (b2) --[worldsheet,in=135,out=90,looseness=1.6] (b3) --[transverse,in=135,out=45,looseness=1.6]  (b0);(b0) --[worldsheet,out=45,in=225,looseness=1] (b2)};}}
		{forbidden2}{\tikz [every new --/.style = {bend left=90},baseline={(0,-0.25)}] {\node (f0) [susy,rotate=180] at (0,0) {}; \node (f1) [fermion] at (1,0) {}; \node (p0) [coordinate] at (-0.65,0) {}; \node (p1) [coordinate] at (1+0.65,0) {}; \graph {(f0) --[worldsheet,bend left=0] (f1);(f0) --[worldsheet,out=45,in=90,looseness=1.6] (p0) --[worldsheet,out=90,in=135,looseness=1.6] (f0);(f1) --[worldsheet,out=45,in=90,looseness=1.6] (p1) --[worldsheet,out=90,in=135,looseness=1.6] (f1)};}}
		{Eight1}{\tikz [every new --/.style = {bend left=90},baseline={(0,-0.25)}] {\node (b0) [boson] at (0,0) {}; \node (p1) [coordinate] at (1,0) {}; \node (b2) [boson] at (-1,0) {}; \graph {(b0) --[worldsheet,out=45,in=90,looseness=1] (p1) --[worldsheet,out=90,in=135,looseness=1] (b0);(b0) --[transverse,out=45,in=90,looseness=0.6] (b2) --[worldsheet,out=90,in=135,looseness=1] (b0)};}}
		{Eight2}{\tikz [every new --/.style = {bend left=90},baseline={(0,-0.25)}] {\node (f0) [fermion,rotate=90] at (0,0) {}; \node (p0) [coordinate] at (-0.65,0) {}; \node (p1) [coordinate] at (0.65,0.5) {}; \graph {(f0) --[worldsheet,out=90,in=90,looseness=1.6] (p0) --[worldsheet,out=90,in=135,looseness=1.6] (f0);(f0.east) --[worldsheet] (p1) --[worldsheet] (f0.south east)};}}
		{forbidden3}{\tikz [every new --/.style = {bend left=90},baseline={(0,-0.25)}] {\node (b0) [boson] at (0,0) {}; \node (p1) [coordinate] at (1,0) {}; \node (b2) [lorentz] at (-1,0) {}; \graph {(b0) --[worldsheet,out=45,in=90,looseness=1] (p1) --[worldsheet,out=90,in=135,looseness=1] (b0);(b0) --[transverse,out=45,in=90,looseness=0.6] (b2) --[worldsheet,out=90,in=135,looseness=1] (b0)};}}
		{forbidden4}{\tikz [every new --/.style = {bend left=90},baseline={(0,-0.25)}] {\node (f0) [susy,rotate=90] at (0,0) {}; \node (p0) [coordinate] at (-0.65,0) {}; \node (p1) [coordinate] at (0.65,0.5) {}; \graph {(f0) --[worldsheet,out=90,in=90,looseness=1.6] (p0) --[worldsheet,out=90,in=135,looseness=1.6] (f0);(f0.east) --[worldsheet] (p1) --[worldsheet] (f0.south east)};}}
	}[\PackageError{ScaleTwoSeed}{Undefined option to ScaleTwoSeed: #1}{}]
}

\newcommand{\ScaleOneVertex}[2]{
	\IfEqCase{#1}{
		{bos}{\IfEqCase{#2}{
				{}{\tikz [baseline={(0,0)}] {\node (p1) [coordinate] at (0,0) {}; \node (b1) [boson] at (7mm,0) {}; \node (p2) [coordinate] at (14mm,5mm) {}; \node (p3) [coordinate] at (14mm,-5mm) {}; \graph {(p1) --[transverse] (b1) --[worldsheet] (p2); (b1) --[worldsheet] (p3)};}}
				{trans1}{\tikz [baseline={(0,0)}] {\node (p1) [coordinate] at (0,0) {}; \node (b1) [boson] at (7mm,0) {}; \node (b2) [lorentz] at (14mm,0) {}; \node (b3) [boson] at (21mm,0) {}; \node (p2) [coordinate] at (28mm,5mm) {}; \node (p3) [coordinate] at (28mm,-5mm) {}; \graph {(p1) --[transverse] (b1) --[worldsheet] (b2) --[transverse] (b3) --[worldsheet] (p2); (b3) --[worldsheet] (p3)};}}
				{trans2}{\tikz [baseline={(0,0)}] {\node (p1) [coordinate] at (0,0) {}; \node (b1) [boson] at (7mm,0) {}; \node (b2) [lorentz] at (14mm,5mm) {}; \node (p3) [coordinate] at (14mm,-5mm) {}; \node (b3) [boson] at (21mm,5mm) {}; \node (p2) [coordinate] at (28mm,5mm) {}; \graph {(p1) --[transverse] (b1) --[worldsheet] (b2) --[transverse] (b3) --[worldsheet] (p2); (b1) --[worldsheet] (p3)};}}
				{trans3}{\tikz [baseline={(0,0)}] {\node (p1) [coordinate] at (0,0) {}; \node (b1) [boson] at (7mm,0) {}; \node (b2) [lorentz] at (14mm,-5mm) {}; \node (p3) [coordinate] at (14mm,5mm) {}; \node (b3) [boson] at (21mm,-5mm) {}; \node (p2) [coordinate] at (28mm,-5mm) {}; \graph {(p1) --[transverse] (b1) --[worldsheet] (b2) --[transverse] (b3) --[worldsheet] (p2); (b1) --[worldsheet] (p3)};}}
				{trans4}{\tikz [baseline={(0,0)}] {\node (p1) [coordinate] at (0,0) {}; \node (b1) [susy] at (7mm,0) {}; \node (p2) [coordinate] at (14mm,5mm) {}; \node (p3) [coordinate] at (14mm,-5mm) {}; \graph {(p1) --[transverse] (b1) --[worldsheet] (p2); (b1) --[worldsheet] (p3)};}}
				{trans5}{\tikz [baseline={(0,0)}] {\node (p1) [coordinate] at (0,0) {}; \node (b0) [boson] at (7mm,0) {}; \node (b1) [susy] at (14mm,0) {}; \node (p2) [coordinate] at (21mm,5mm) {}; \node (p3) [coordinate] at (21mm,-5mm) {}; \graph {(p1) --[transverse] (b0) --[worldsheet] (b1) --[worldsheet] (p2); (b1) --[worldsheet] (p3)};}}
				{trans6}{\tikz [baseline={(0,0)}] {\node (p1) [coordinate] at (0,0) {}; \node (b1) [boson] at (7mm,0) {}; \node (b2) [susy] at (14mm,5mm) {}; \node (p3) [coordinate] at (14mm,-5mm) {}; \node (p2) [coordinate] at (21mm,5mm) {}; \graph {(p1) --[transverse] (b1) --[worldsheet] (b2) --[worldsheet] (p2); (b1) --[worldsheet] (p3)};}}
				{trans7}{\tikz [baseline={(0,0)}] {\node (p1) [coordinate] at (0,0) {}; \node (b1) [boson] at (7mm,0) {}; \node (b2) [susy] at (14mm,-5mm) {}; \node (p3) [coordinate] at (14mm,5mm) {}; \node (p2) [coordinate] at (21mm,-5mm) {}; \graph {(p1) --[transverse] (b1) --[worldsheet] (b2) --[worldsheet] (p2); (b1) --[worldsheet] (p3)};}}
			}[\PackageError{ScaleOneVertexCring}{Undefined option: #2}{}]};
		
		{fera}{\IfEqCase{#2}{
				{}{\tikz [baseline={(0,0)}] {\node (p1) [coordinate] at (0,0) {}; \node (b1) [fermion] at (7mm,0) {}; \node (p2) [coordinate] at (14mm,5mm) {}; \node (p3) [coordinate] at (14mm,-5mm) {}; \graph {(p1) --[transverse] (b1) --[worldsheet] (p2); (b1) --[worldsheet] (p3)};}}
				{trans1}{\tikz [baseline={(0,0)}] {\node (p1) [coordinate] at (0,0) {}; \node (b1) [fermion] at (7mm,0) {}; \node (b2) [lorentz] at (14mm,0) {}; \node (b3) [boson] at (21mm,0) {}; \node (p2) [coordinate] at (28mm,5mm) {}; \node (p3) [coordinate] at (28mm,-5mm) {}; \graph {(p1) --[transverse] (b1) --[worldsheet] (b2) --[transverse] (b3) --[worldsheet] (p2); (b3) --[worldsheet] (p3)};}}
				{trans2}{\tikz [baseline={(0,0)}] {\node (p1) [coordinate] at (0,0) {}; \node (b1) [fermion] at (7mm,0) {}; \node (b2) [lorentz] at (14mm,5mm) {}; \node (p3) [coordinate] at (14mm,-5mm) {}; \node (b3) [boson] at (21mm,5mm) {}; \node (p2) [coordinate] at (28mm,5mm) {}; \graph {(p1) --[transverse] (b1) --[worldsheet] (b2) --[transverse] (b3) --[worldsheet] (p2); (b1) --[worldsheet] (p3)};}}
				{trans3}{\tikz [baseline={(0,0)}] {\node (p1) [coordinate] at (0,0) {}; \node (b1) [fermion] at (7mm,0) {}; \node (b2) [lorentz] at (14mm,-5mm) {}; \node (p3) [coordinate] at (14mm,5mm) {}; \node (b3) [boson] at (21mm,-5mm) {}; \node (p2) [coordinate] at (28mm,-5mm) {}; \graph {(p1) --[transverse] (b1) --[worldsheet] (b2) --[transverse] (b3) --[worldsheet] (p2); (b1) --[worldsheet] (p3)};}}
				{trans4}{\tikz [baseline={(0,0)}] {\node (p1) [coordinate] at (0,0) {}; \node (b1) [susy] at (7mm,0) {}; \node (p2) [coordinate] at (14mm,5mm) {}; \node (p3) [coordinate] at (14mm,-5mm) {}; \graph {(p1) --[transverse] (b1) --[worldsheet] (p2); (b1) --[worldsheet] (p3)};}}
				{trans5}{\tikz [baseline={(0,0)}] {\node (p1) [coordinate] at (0,0) {}; \node (b0) [fermion] at (7mm,0) {}; \node (b1) [susy] at (14mm,0) {}; \node (p2) [coordinate] at (21mm,5mm) {}; \node (p3) [coordinate] at (21mm,-5mm) {}; \graph {(p1) --[transverse] (b0) --[worldsheet] (b1) --[worldsheet] (p2); (b1) --[worldsheet] (p3)};}}
				{trans6}{\tikz [baseline={(0,0)}] {\node (p1) [coordinate] at (0,0) {}; \node (b1) [fermion] at (7mm,0) {}; \node (b2) [susy] at (14mm,5mm) {}; \node (p3) [coordinate] at (14mm,-5mm) {}; \node (p2) [coordinate] at (21mm,5mm) {}; \graph {(p1) --[transverse] (b1) --[worldsheet] (b2) --[worldsheet] (p2); (b1) --[worldsheet] (p3)};}}
				{trans7}{\tikz [baseline={(0,0)}] {\node (p1) [coordinate] at (0,0) {}; \node (b1) [fermion] at (7mm,0) {}; \node (b2) [susy] at (14mm,-5mm) {}; \node (p3) [coordinate] at (14mm,5mm) {}; \node (p2) [coordinate] at (21mm,-5mm) {}; \graph {(p1) --[transverse] (b1) --[worldsheet] (b2) --[worldsheet] (p2); (b1) --[worldsheet] (p3)};}}
				{trans8}{\tikz [baseline={(0,0)}] {\node (p1) [coordinate] at (0,0) {}; \node (b0) [lorentz] at (7mm,0) {}; \node (b1) [fermion] at (14mm,0) {}; \node (p2) [coordinate] at (21mm,5mm) {}; \node (p3) [coordinate] at (21mm,-5mm) {}; \graph {(p1) --[transverse] (b0) --[worldsheet] (b1) --[worldsheet] (p2); (b1) --[worldsheet] (p3)};}}
			}[\PackageError{ScaleOneVertex}{Undefined option: #2}{}]};
		
		{ferb}{\IfEqCase{#2}{
				{}{\tikz [baseline={(0,0)}] {\node (p1) [coordinate] at (0,0) {}; \node (b1) [fermion] at (7mm,0) {}; \node (p2) [coordinate] at (14mm,5mm) {}; \node (p3) [coordinate] at (14mm,-5mm) {}; \graph {(p1) --[worldsheet] (b1) --[worldsheet] (p2); (b1) --[worldsheet] (p3)};}}
				{trans1}{\tikz [baseline={(0,0)}] {\node (p1) [coordinate] at (0,0) {}; \node (b1) [fermion] at (7mm,0) {}; \node (b2) [lorentz] at (14mm,0) {}; \node (b3) [boson] at (21mm,0) {}; \node (p2) [coordinate] at (28mm,5mm) {}; \node (p3) [coordinate] at (28mm,-5mm) {}; \graph {(p1) --[worldsheet] (b1) --[worldsheet] (b2) --[transverse] (b3) --[worldsheet] (p2); (b3) --[worldsheet] (p3)};}}
				{trans2}{\tikz [baseline={(0,0)}] {\node (p1) [coordinate] at (0,0) {}; \node (b1) [fermion] at (7mm,0) {}; \node (b2) [lorentz] at (14mm,5mm) {}; \node (p3) [coordinate] at (14mm,-5mm) {}; \node (b3) [boson] at (21mm,5mm) {}; \node (p2) [coordinate] at (28mm,5mm) {}; \graph {(p1) --[worldsheet] (b1) --[worldsheet] (b2) --[transverse] (b3) --[worldsheet] (p2); (b1) --[worldsheet] (p3)};}}
				{trans3}{\tikz [baseline={(0,0)}] {\node (p1) [coordinate] at (0,0) {}; \node (b1) [fermion] at (7mm,0) {}; \node (b2) [lorentz] at (14mm,-5mm) {}; \node (p3) [coordinate] at (14mm,5mm) {}; \node (b3) [boson] at (21mm,-5mm) {}; \node (p2) [coordinate] at (28mm,-5mm) {}; \graph {(p1) --[worldsheet] (b1) --[worldsheet] (b2) --[transverse] (b3) --[worldsheet] (p2); (b1) --[worldsheet] (p3)};}}
				{trans4}{\tikz [baseline={(0,0)}] {\node (p1) [coordinate] at (0,0) {}; \node (b1) [susy] at (7mm,0) {}; \node (p2) [coordinate] at (14mm,5mm) {}; \node (p3) [coordinate] at (14mm,-5mm) {}; \graph {(p1) --[worldsheet] (b1) --[worldsheet] (p2); (b1) --[worldsheet] (p3)};}}
				{trans5}{\tikz [baseline={(0,0)}] {\node (p1) [coordinate] at (0,0) {}; \node (b0) [fermion] at (7mm,0) {}; \node (b1) [susy] at (14mm,0) {}; \node (p2) [coordinate] at (21mm,5mm) {}; \node (p3) [coordinate] at (21mm,-5mm) {}; \graph {(p1) --[worldsheet] (b0) --[worldsheet] (b1) --[worldsheet] (p2); (b1) --[worldsheet] (p3)};}}
				{trans6}{\tikz [baseline={(0,0)}] {\node (p1) [coordinate] at (0,0) {}; \node (b1) [fermion] at (7mm,0) {}; \node (b2) [susy] at (14mm,5mm) {}; \node (p3) [coordinate] at (14mm,-5mm) {}; \node (p2) [coordinate] at (21mm,5mm) {}; \graph {(p1) --[worldsheet] (b1) --[worldsheet] (b2) --[worldsheet] (p2); (b1) --[worldsheet] (p3)};}}
				{trans7}{\tikz [baseline={(0,0)}] {\node (p1) [coordinate] at (0,0) {}; \node (b1) [fermion] at (7mm,0) {}; \node (b2) [susy] at (14mm,-5mm) {}; \node (p3) [coordinate] at (14mm,5mm) {}; \node (p2) [coordinate] at (21mm,-5mm) {}; \graph {(p1) --[worldsheet] (b1) --[worldsheet] (b2) --[worldsheet] (p2); (b1) --[worldsheet] (p3)};}}
				{trans8}{\tikz [baseline={(0,0)}] {\node (p1) [coordinate] at (0,0) {}; \node (b0) [lorentz] at (7mm,0) {}; \node (b1) [fermion] at (14mm,0) {}; \node (p2) [coordinate] at (21mm,5mm) {}; \node (p3) [coordinate] at (21mm,-5mm) {}; \graph {(p1) --[worldsheet] (b0) --[transverse] (b1) --[worldsheet] (p2); (b1) --[worldsheet] (p3)};}}
			}[\PackageError{ScaleOneVertex}{Undefined option: #2}{}]};
		
		{ferbos}{\IfEqCase{#2}{
				{}{\tikz [baseline={(0,0)}] {\node (p1) [coordinate] at (-7mm,0) {}; \node (b0) [boson] at (0,0) {}; \node (b1) [fermion] at (7mm,0) {}; \node (p2) [coordinate] at (14mm,5mm) {}; \node (p3) [coordinate] at (14mm,-5mm) {}; \graph {(p1) --[transverse] (b0) --[worldsheet] (b1) --[worldsheet] (p2); (b1) --[worldsheet] (p3)};}}
				{trans1}{\tikz [baseline={(0,0)}] {\node (p1) [coordinate] at (0,0) {}; \node (b1) [fermion] at (7mm,0) {}; \node (b2) [lorentz] at (14mm,0) {}; \node (b3) [boson] at (21mm,0) {}; \node (p2) [coordinate] at (28mm,5mm) {}; \node (p3) [coordinate] at (28mm,-5mm) {}; \graph {(p1) --[worldsheet] (b1) --[worldsheet] (b2) --[transverse] (b3) --[worldsheet] (p2); (b3) --[worldsheet] (p3)};}}
				{trans2}{\tikz [baseline={(0,0)}] {\node (p1) [coordinate] at (0,0) {}; \node (b1) [fermion] at (7mm,0) {}; \node (b2) [lorentz] at (14mm,5mm) {}; \node (p3) [coordinate] at (14mm,-5mm) {}; \node (b3) [boson] at (21mm,5mm) {}; \node (p2) [coordinate] at (28mm,5mm) {}; \graph {(p1) --[worldsheet] (b1) --[worldsheet] (b2) --[transverse] (b3) --[worldsheet] (p2); (b1) --[worldsheet] (p3)};}}
				{trans3}{\tikz [baseline={(0,0)}] {\node (p1) [coordinate] at (0,0) {}; \node (b1) [fermion] at (7mm,0) {}; \node (b2) [lorentz] at (14mm,-5mm) {}; \node (p3) [coordinate] at (14mm,5mm) {}; \node (b3) [boson] at (21mm,-5mm) {}; \node (p2) [coordinate] at (28mm,-5mm) {}; \graph {(p1) --[worldsheet] (b1) --[worldsheet] (b2) --[transverse] (b3) --[worldsheet] (p2); (b1) --[worldsheet] (p3)};}}
				{trans4}{\tikz [baseline={(0,0)}] {\node (p1) [coordinate] at (0,0) {}; \node (b1) [susy] at (7mm,0) {}; \node (p2) [coordinate] at (14mm,5mm) {}; \node (p3) [coordinate] at (14mm,-5mm) {}; \graph {(p1) --[worldsheet] (b1) --[worldsheet] (p2); (b1) --[worldsheet] (p3)};}}
				{trans5}{\tikz [baseline={(0,0)}] {\node (p1) [coordinate] at (0,0) {}; \node (b0) [fermion] at (7mm,0) {}; \node (b1) [susy] at (14mm,0) {}; \node (p2) [coordinate] at (21mm,5mm) {}; \node (p3) [coordinate] at (21mm,-5mm) {}; \graph {(p1) --[worldsheet] (b0) --[worldsheet] (b1) --[worldsheet] (p2); (b1) --[worldsheet] (p3)};}}
				{trans6}{\tikz [baseline={(0,0)}] {\node (p1) [coordinate] at (0,0) {}; \node (b1) [fermion] at (7mm,0) {}; \node (b2) [susy] at (14mm,5mm) {}; \node (p3) [coordinate] at (14mm,-5mm) {}; \node (p2) [coordinate] at (21mm,5mm) {}; \graph {(p1) --[worldsheet] (b1) --[worldsheet] (b2) --[worldsheet] (p2); (b1) --[worldsheet] (p3)};}}
				{trans7}{\tikz [baseline={(0,0)}] {\node (p1) [coordinate] at (0,0) {}; \node (b1) [fermion] at (7mm,0) {}; \node (b2) [susy] at (14mm,-5mm) {}; \node (p3) [coordinate] at (14mm,5mm) {}; \node (p2) [coordinate] at (21mm,-5mm) {}; \graph {(p1) --[worldsheet] (b1) --[worldsheet] (b2) --[worldsheet] (p2); (b1) --[worldsheet] (p3)};}}
				{trans8}{\tikz [baseline={(0,0)}] {\node (p1) [coordinate] at (0,0) {}; \node (b0) [lorentz] at (7mm,0) {}; \node (b1) [fermion] at (14mm,0) {}; \node (p2) [coordinate] at (21mm,5mm) {}; \node (p3) [coordinate] at (21mm,-5mm) {}; \graph {(p1) --[worldsheet] (b0) --[transverse] (b1) --[worldsheet] (p2); (b1) --[worldsheet] (p3)};}}
			}[\PackageError{ScaleOneVertex}{Undefined option: #2}{}]};
		
		{fern}{\IfEqCase{#2}{
				{bos}{\tikz [baseline={(0,0)}] {\node (p1) [coordinate] at (-7mm,0) {}; \node (b0) [boson] at (0,0) {}; \node (f1) [fermion] at (8mm,0) {}; \node (n) [large] at (10.2mm,0) {$ \left( \quad \right)^{n} $}; \node (b1) [fermion] at (19mm,0) {}; \node (p2) [coordinate] at (26mm,5mm) {}; \node (p3) [coordinate] at (26mm,-5mm) {}; \graph {(p1) --[transverse] (b0) --[worldsheet] (f1) --[worldsheet] (b1) --[worldsheet] (p2); (b1) --[worldsheet] (p3)};}}
				{fer}{\tikz [baseline={(0,0)}] {\node (p1) [coordinate] at (-7mm,0) {}; \node (b0) [fermion] at (0,0) {}; \node (f1) [fermion] at (10mm,0) {}; \node (n) [large] at (12.2mm,0) {$ \left( \quad \right)^{n} $}; \node (b1) [fermion] at (21mm,0) {}; \node (p2) [coordinate] at (28mm,5mm) {}; \node (p3) [coordinate] at (28mm,-5mm) {}; \graph {(p1) --[transverse] (b0) --[worldsheet] (f1) --[worldsheet] (b1) --[worldsheet] (p2); (b1) --[worldsheet] (p3)};}}
			}[\PackageError{ScaleOneVertex}{Undefined option: #2}{}]};
		
		{comb}{\IfEqCase{#2}{
				{}{\tikz [baseline={(0,0)}] {\node (p1) [coordinate] at (0,0) {}; \node (b1) [fermion] at (7mm,0) {}; \node (b0) [boson,scale=0.8] at (7mm,0) {}; \node (p2) [coordinate] at (14mm,5mm) {}; \node (p3) [coordinate] at (14mm,-5mm) {}; \graph {(p1) --[transverse] (b1) --[worldsheet] (p2); (b1) --[worldsheet] (p3)};}}
				{fer}{\tikz [baseline={(0,0)}] {\node (p1) [coordinate] at (-7mm,0) {}; \node (b0) [fermion] at (0,0) {}; \node (f1) [fermion] at (10mm,0) {}; \node (n) [large] at (12.2mm,0) {$ \left( \quad \right)^{n} $}; \node (b1) [fermion] at (21mm,0) {}; \node (p2) [coordinate] at (28mm,5mm) {}; \node (p3) [coordinate] at (28mm,-5mm) {}; \graph {(p1) --[transverse] (b0) --[worldsheet] (f1) --[worldsheet] (b1) --[worldsheet] (p2); (b1) --[worldsheet] (p3)};}}
			}[\PackageError{ScaleOneVertex}{Undefined option: #2}{}]}
		
	}[\PackageError{ScaleOneVertex}{Undefined option: #1}{}]
}

\newcommand{\ScaleOneHead}[1]{
	\IfEqCase{#1}{
		{bos}{\tikz [baseline={(0,0)}] {\node (b1) [boson] at (0,0) {}; \node (p2) [coordinate] at (7mm,5mm) {}; \node (p3) [coordinate] at (7mm,-5mm) {}; \graph {(b1) --[worldsheet] (p2); (b1) --[worldsheet] (p3)};}}
		{fer}{\tikz [baseline={(0,0)}] {\node (b1) [fermion] at (0,0) {}; \node (p2) [coordinate] at (7mm,5mm) {}; \node (p3) [coordinate] at (7mm,-5mm) {}; \graph {(b1) --[worldsheet] (p2); (b1) --[worldsheet] (p3)};}}
		{ferv}{\tikz [baseline={(0,0)}] {\node (b1) [susy] at (0,0) {}; \node (p2) [coordinate] at (7mm,5mm) {}; \node (p3) [coordinate] at (7mm,-5mm) {}; \graph {(b1) --[worldsheet] (p2); (b1) --[worldsheet] (p3)};}}
	}[\PackageError{ScaleOneVertex}{Undefined option: #1}{}]
}

\newcommand{\ScaleZeroCancellationA}{
	\tikz [>=stealth,very thick,
	every new ->/.style={shorten >=-1.2mm,shorten <=-1.5mm},
	forward/.style = {out=-75,in=105,looseness=1},
	back/.style = {out=-105,in=75,looseness=1}] {
		\node (B0s) [] at (0,-2.2) {\Bring{0s}};
		\node (B0) [] at (1,0) {\Bring{0}};
		\node (C00s) [] at (2,-2.2) {\Cring{}{00s}};
		\node (C00) [] at (3,0) {\Cring{}{00}};
		\node (D000s) [] at (4,-2.2) {\Dring{}{000s}};
		\node (D000) [] at (5,0) {\Dring{}{000}};
		\node (E0000s) [] at (6,-2.2) {\Ering{}{0000s}};
		\node (E0000) [] at (7,0) {\Ering{}{0000}};
		\node (TD) [] at (1,-1.7*2.2) {Total derivative};
		\graph {
			(B0) -> [back] {(B0s)};
			(B0s) -> [forward] {(TD)};
			(B0) -> [forward] {(C00s)};
			(C00) -> [back, edge label'=2] {(C00s)};
			(C00) -> [forward, edge label=2] {(D000s)};
			(D000) -> [back, edge label'=3] {(D000s)};
			(D000) -> [forward, edge label=3] {(E0000s)};
			(E0000) -> [back, edge label'=4] {(E0000s)};
		};
	}
}

\newcommand{\ScaleZeroCancellationB}{
	\tikz [>=stealth,very thick,
	every new ->/.style={shorten >=-1.2mm,shorten <=-1.5mm}] {
		\node (Cp00s) [] at (0,-1*2.2) {\Cring{p}{00s}};
		\node (Cp00) [] at (1,0) {\Cring{p}{00}};
		\node (Dp000s1) [] at (2,-1*2.2) {\Dring{p}{000s1}};
		\node (Dp000s2) [] at (2,-2*2.2) {\Dring{p}{000s2}};
		\node (Dp000s3) [] at (2,-3*2.2) {\Dring{p}{000s3}};
		\node (Dp000) [] at (4,0) {\Dring{p}{000}};
		\node (Ep0000s1) [] at (6,-1*2.2) {\Ering{p}{0000s1}};
		\node (Ep0000s4) [] at (6,-2*2.2) {\Ering{p}{0000s4}};
		\node (Et0000s2) [] at (6,-3*2.2) {\Ering{t}{0000s2}};
		\node (Ep0000) [] at (8,0) {\Ering{p}{0000}};
		\node (Et0000) [] at (8,-2*2.2) {\Ering{t}{0000}};
		\graph {
			(Cp00) -> [out=-105,in=75, edge label'=2] {(Cp00s)};
			(Cp00) -> [out=-75,in=105, edge label=2] {(Dp000s1)};
			(Dp000) -> [out=-105,in=75] {(Dp000s1)};
			(Dp000) -> [out=-105,in=60] {(Dp000s2)};
			(Dp000) -> [out=-105,in=45] {(Dp000s3)};
			(Dp000) -> [out=-75,in=105] {(Ep0000s1)};
			(Dp000) -> [out=-75,in=120] {(Ep0000s4)};
			(Dp000) -> [out=-75,in=135] {(Et0000s2)};
			(Ep0000) -> [out=-105,in=75] {(Ep0000s1)};
			(Ep0000) -> [out=-105,in=60] {(Ep0000s4)};
			(Et0000) -> [out=-105,in=75] {(Et0000s2)};
		};
	}
}

\newcommand{\ScaleZeroCancellationC}{
	\tikz [>=stealth,very thick,
	every new ->/.style={shorten >=-1.2mm,shorten <=-1.5mm},] {
		\node (A2) [] at (0,0) {\Aring{2}};
		\node (B1s) [] at (1,-1*2.2) {\Bring{1s}};
		\node (B2s) [] at (1,-2*2.2) {\Bring{2s}};
		\node (B1) [] at (2,0) {\Bring{1}};
		\node (B2) [] at (2,-3*2.2) {\Bring{2}};
		\node (Cpp00s) [] at (4,-1*2.2) {\Cring{pp}{00s}};
		\node (C01s1) [] at (4,-2*2.2) {\Cring{}{01s1}};
		\node (C01s2) [] at (4,-4*2.2) {\Cring{}{01s2}};
		\node (Cp01s1) [] at (4,-3*2.2) {\Cring{p}{01s1}};
		\node (Cp01s2) [] at (4,-5*2.2) {\Cring{p}{01s2}};
		\node (Cpp00) [] at (6,0) {\Cring{pp}{00}};
		\node (C01) [] at (6,-2*2.2) {\Cring{}{01}};
		\node (Cp01) [] at (6,-4*2.2) {\Cring{p}{01}};
		\node (Dpp000s1) [] at (8,-1*2.2) {\Dring{pp}{000s1}};
		\node (Dpp000s2) [] at (8,-2*2.2) {\Dring{pp}{000s2}};
		\node (Dpp000s3) [] at (8,-3*2.2) {\Dring{pp}{000s3}};
		\node (Dpp000) [] at (9.5,0) {\Dring{pp}{000}};
		\node (Epp0000s1) [] at (11,-1*2.2) {\Ering{pp}{0000s1}};
		\node (Epp0000s4) [] at (11,-2*2.2) {\Ering{pp}{0000s4}};
		\node (Ett0000s1) [] at (11,-3*2.2) {\Ering{tt}{0000s1}};
		\node (Epp0000) [] at (12.5,0) {\Ering{pp}{0000}};
		\node (Ett0000) [] at (12.5,-4*2.2) {\Ering{tt}{0000}};
		\graph {
			(A2) -> [out=-75,in=105, edge label=2] {(B1s)};
			(A2) -> [out=-90,in=180,looseness=0.6, edge label=2] {(B2s)};
			(B1) -> [out=-105,in=75] {(B1s)};
			(B1) -> [out=-75,in=105] {(Cpp00s)};
			(B1) -> [out=-75,in=120] {(C01s1)};
			(B1) -> [out=-75,in=135,shorten >=-4mm] {(Cp01s1)};
			(B2) -> [in=-75,out=105] {(B2s)};
			(B2) -> [out=-75,in=105,shorten >=-3mm] {(C01s2)};
			(B2) -> [out=-75,in=120] {(Cp01s2)};
			(Cpp00) -> [out=-105,in=75, edge label'=2] {(Cpp00s)};
			(Cpp00) -> [out=-75,in=105, edge label=2] {(Dpp000s1)};
			(C01) -> [] {(C01s1)};
			(C01) -> [out=-105,in=45,shorten >=-4mm] {(C01s2)};
			(C01) -> [] {(Dpp000s2)};
			(Cp01) -> [in=-75,out=105] {(Cp01s1)};
			(Cp01) -> [in=75,out=-105] {(Cp01s2)};
			(Cp01) -> [in=-105,out=75] {(Dpp000s3)};
			(Dpp000) -> [out=-105,in=75] {(Dpp000s1)};
			(Dpp000) -> [out=-105,in=60] {(Dpp000s2)};
			(Dpp000) -> [out=-105,in=45,looseness=0.6,shorten >=-4mm] {(Dpp000s3)};
			(Dpp000) -> [out=-75,in=105] {(Epp0000s1)};
			(Dpp000) -> [out=-75,in=120] {(Epp0000s4)};
			(Dpp000) -> [out=-75,in=135,looseness=0.6,shorten >=-5mm] {(Ett0000s1)};
			(Epp0000) -> [out=-105,in=75] {(Epp0000s1)};
			(Epp0000) -> [out=-105,in=60] {(Epp0000s4)};
			(Ett0000) -> [in=-75,out=105, edge label=2] {(Ett0000s1)};
		};
	}
}
\newcommand{\JustTheta}{\tikz [every new --/.style = {bend left=45},baseline={(0,-0.25)}] \graph [clockwise=4] { a[coordinate] , b/ [boson], c[coordinate], d/[boson]; a -- b/ -- c -- d/ -- a; b -- [bend left=0] d };}
\newcommand{\JustDumbbell}{\tikz [every new --/.style = {bend left=90},baseline={(0,-0.25)}] \graph[branch right]{ a[coordinate] , b/ [boson], c/[boson], d[coordinate]; a -- b -- a; c -- d -- c; b -- [bend left=0] c };}

\tableofcontents{}

\section{Introduction}

String-like objects appear in many quantum field theories, such as
flux tubes in quantum chromodynamics (QCD), vortices such as the Nielsen-Olesen
strings in the 4d Abelian Higgs model\cite{key-9}, and domain walls
in 3d theories such as the Ising model. Their appearance in QCD, as
visible through the spectrum of mesons (and other hadrons), led to
the development of the Veneziano model\cite{key-8} and ultimately
to the development of string theory.

A straight string is a 2d object which breaks the $ISO\left(D-1,1\right)$
symmetry of the $D$-dimensional bulk into an $ISO\left(1,1\right)\times SO\left(D-2\right)$
symmetry group, leading to $\left(D-2\right)$ massless modes of excitation,
known as the Nambu-Goldstone Bosons, or NGBs. These massless excitations
define the low-energy behavior of the string, and we can compute their
energy levels expanded in powers of $1/L$, where $L$ is the length
of the string.

Naively, one might think that the actions computed for string-like
objects in different QFTs are dependent on the underlying theory.
However, as reviewed by Aharony and Komargodski\cite{key-10}, the
first few terms in the expansion - up to and including order of $1/L^{5}$
- are universal, and only the higher order terms are dependent on
the theory. This was shown in 3 different formalisms:
\begin{enumerate}
\item The general case, in which there is no gauge fixing, and allowed terms
in the action must preserve both Lorentz symmetry and diffeomorphism.
\item The unitary (``static'') gauge in which the parameterization of
the world-sheet of the string (``diffeomorphism'') is fixed and
the Lorentz group is broken manifestly. In this formalism, the action
can be expanded by the number of derivatives - corresponding to the
$1/L$ expansion of the energy levels - constrained by Lorentz symmetries.
In this formalism, it was shown that for $D>3$ classical Lorentz
invariance allows a six-derivative term, but its presence modifies
the form of the generators (while higher-derivative allowed terms
do not); and then quantum considerations show that its value is actually
fixed.
\item The orthogonal (``conformal'') gauge in which diffeomorphism is
fixed up to conformal transformations and Lorentz symmetry is maintained.
In this formalism, the action is constrained by conformal invariance.
\end{enumerate}
This work aims at generalizing the results of Aharony and Komargodski
to the case of Supersymmetry (SUSY), specifically $D=4\ N=1$ SUSY.
In a supersymmetric theory, a string may break $D=4\ N=1$ SUSY either
completely, or partially into $D=2,\ N=\left(2,0\right)$, as was
shown by Hughes and Polchinski\cite{key-11}. The breaking of SUSY
generators adds massless fermionic modes of excitation, known as Goldstinos.
The action can then be written as a functional of the NGBs and Goldstinos,
and expanded as in the fully bosonic case by the number of derivatives.
For the two cases of complete and partial breaking of SUSY, a complete
classification of action terms has yet to be made. In the scope of
this work we will only explore the case of complete SUSY breaking,
which is relevant in particular for confining strings in supersymmetric
Yang-Mills theory, and it is the main goal of this work to classify
action terms for this case. As a final step, we will calculate the
form of the energy level correction for a closed string on a circle,
arising from the lowest order new term we find, so that our results
can be verified by lattice simulations at some later point.

The outline of this paper is as follows. In the next section we review
well established results, as well as notations and definitions we
will use, and eventually a graphical approach, originally presented
by Gliozzi and Meineri \cite{key-2}, to find invariant actions for
bosonic effective strings. In section 3 we extend this approach to
include Goldstinos, and in section 4 we use the extended approach
to find invariant actions for SUSY breaking effective strings, including
a new term at order $1/L^{5}$. In section 5 we formulate prohibition
rules which show that our list of invariant actions is indeed exhaustive,
and in section 6 we derive the energy corrections that follow from
our new term. Finally we discuss our results and draw some conclusions.

\section{Review}

\subsection{Bosonic effective strings}

Consider some gapped $D$-dimensional quantum field theory with a
string-like field configuration, so that its width is much smaller
than its length. Such a configuration could be either open, closed,
infinite or semi-infinite. We define this configuration by the space-time
coordinates of its worldsheet $X^{\mu}\left(\sigma^{0},\sigma^{1}\right)$,
where $\sigma^{0},\sigma^{1}$ are some parameterization of the worldsheet
and $\mu=0,\dots,D-1$. The physics can't depend on the parameterization.
The effective string action is the low energy action of the massless
modes on the worldsheet 
\begin{equation}
S=T\intop d^{2}\sigma\ML\left[X^{\mu}\left(\sigma^{0},\sigma^{1}\right)\right]
\end{equation}
where $T$ is the string tension. This general formalism is the first
case referred to in the introduction.

The static gauge is where we fix $\sigma^{0}=X^{0}$ and $\sigma^{1}=X^{1}$.
When working in this gauge we will denote these $\xi^{0},\xi^{1}$
to avoid ambiguity. In this gauge the NGBs are given by the transverse
coordinates $X^{i}$ for $i=2,\dots,D-1$. In this formalism effective
string action is
\begin{equation}
S=T\intop d^{2}\xi\ML\left(\d_{a}X^{i},\d_{a}\d_{b}X^{i},\dots\right)\label{eq:action}
\end{equation}
where $a,b=0,1$. There is no $X^{i}$ dependence with no derivatives
due to translational invariance. For simplicity, we will work mainly
in this formalism, and generalize our results whenever possible. We
will use letters from the beginning of the Latin alphabet such as
$a,b,c,d,\dots$ to denote the worldsheet indices $0,1$, and letters
from the middle of the Latin alphabet such as $i,j,k,\dots$ to denote
the transverse indices $2,\dots,D-1$. 

The gauge choice (\ref{eq:action}) breaks the space-time symmetry
$ISO\left(D-1,1\right)$ by choosing a preferred direction in space.
$ISO\left(D-1,1\right)$ is the Poincaré group which is the group
that preserves the Minkowski metric which we define as $\eta_{\mu\nu}=\text{diag}\left(-1,1,\dots,1\right)$.
It is generated by 
\begin{align}
J_{\mu\nu} & =i\left(x_{\mu}\d_{\nu}-x_{\nu}\d_{\mu}\right)\\
P_{\mu} & =i\d_{\mu}
\end{align}
It is broken into $ISO\left(1,1\right)\times SO\left(D-2\right)$,
where $ISO\left(1,1\right)$ is the symmetry on the worldsheet, which
preserves the metric $\eta_{ab}$, and is generated by $J_{ab}$ and
$P_{a}$; And $SO\left(D-2\right)$ is the symmetry of rotations around
the string and is generated by $J_{ij}$. The remaining generators
$P_{i}$ and $J_{ai}$ are broken. By acting with $J_{ai}$ on the
fields $X^{j}$ we get 
\begin{equation}
\delta X^{j}=i\epsilon^{ai}\left[J_{ai},X^{j}\right]=-\epsilon^{ai}\delta^{ij}\xi_{a}-\epsilon^{ai}X^{i}\d_{a}X^{j}\label{eq:dx}
\end{equation}
which is a non-linear realization of these generators.

When working in the static gauge, we will often work in light-cone
coordinates 
\begin{equation}
\xi^{\pm}=\xi^{0}\pm\xi^{1}
\end{equation}

A well known result in String theory is that the action of a string
is proportional to the area of its worldsheet. This result can be
expressed using the embedded metric on the string 
\begin{equation}
g_{ab}=\eta_{\mu\nu}\d_{a}X^{\mu}\d_{b}X^{\nu}
\end{equation}
which in static gauge can be expressed as 
\begin{equation}
g_{ab}=\eta_{ab}+\d_{a}X^{i}\d_{b}X^{i}\equiv\eta_{ab}+h_{ab},
\end{equation}
and the Nambu-Goto (NG) action equal to the area of the worldsheet
\begin{equation}
S_{NG}=-T\intop d^{2}\sigma\sqrt{-\det\left(g_{ab}\right)}.\label{eq:ng}
\end{equation}

The NG action is highly non-linear. In the context of the effective
string, we can work with it by expanding in terms of derivatives $\d$
around the flat string solution of the static gauge. The determinant
is given by 
\begin{align}
-\det\left(g_{ab}\right) & =-\det\left(\eta_{ab}+h_{ab}\right)=1+\eta^{ab}h_{ab}-\det\left(h_{ab}\right)=\nonumber \\
 & =1+\d_{a}X^{i}\d^{a}X^{i}-\frac{1}{2}\d_{a}X^{i}\d^{b}X^{i}\d_{b}X^{i}\d^{a}X^{i}+\frac{1}{2}\left(\d_{a}X^{i}\d^{a}X^{i}\right)^{2}
\end{align}
and we get 
\begin{equation}
S_{NG}=-T\intop d^{2}\sigma\left(1+\frac{1}{2}\d_{a}X^{i}\d^{a}X^{i}-\frac{1}{4}\d_{a}X^{i}\d^{b}X^{i}\d_{b}X^{i}\d^{a}X^{i}+\frac{1}{8}\left(\d_{a}X^{i}\d^{a}X^{i}\right)^{2}+\MO\left(\d^{6}\right)\right)
\end{equation}

This expansion is meaningful under the assumption of a long string
of length scale $L$. We can then define a small dimensionless parameter
$\left(\sqrt{T}L\right)^{-1}$ and expand the energy levels of the
string in terms of this parameter. This expansion will take the form
\begin{equation}
E_{n}=TL+\frac{a_{n}^{\left(1\right)}}{L}+\frac{a_{n}^{\left(2\right)}}{TL^{3}}+\frac{a_{n}^{\left(3\right)}}{T^{2}L^{5}}+\dots
\end{equation}

Where the term at order $L^{-k}$ corresponds to the terms in the
action at order $\d^{k+1}$. For the NG action, there is a known exact
result for the energy levels of closed strings with no worldsheet
momentum \cite{key-1}
\begin{equation}
E_{n}=TL\sqrt{1+\frac{8\pi}{TL^{2}}\left(n-\frac{D-2}{24}\right)}
\end{equation}

\subsection{Supersymmetry}

Supersymmetry (SUSY) is an extension of the Poincaré algebra to include
fermionic generators. The simplest $\left(\MN=1\right)$ super-Poincaré
generators can be written in $D=4$ as a single Majorana spinor
\begin{equation}
Q=\begin{pmatrix}Q_{1} & Q_{2} & \overline{Q}_{\dot{2}} & -\overline{Q}_{\dot{1}}\end{pmatrix}^{T}
\end{equation}
with the following algebra 
\begin{align}
\left\{ Q,\overline{Q}\right\}  & =-2i\gamma^{\mu}P_{\mu}\nonumber \\
\left[Q,P\right] & =0\\
\left[Q,J_{\mu\nu}\right] & =i\sigma_{\mu\nu}Q\nonumber 
\end{align}
where $\gamma^{\mu}$ are the Dirac gamma matrices satisfying $\left\{ \gamma^{\mu},\gamma^{\nu}\right\} =2\eta^{\mu\nu}$,
$\overline{Q}$ is obtained from $Q$ using the charge conjugation
operator $C=i\gamma^{0}\gamma^{2}$ such that 
\begin{equation}
\overline{Q}=-Q^{T}C=\begin{pmatrix}-Q_{2} & Q_{1} & \overline{Q}_{\dot{1}} & \overline{Q}_{\dot{2}}\end{pmatrix}
\end{equation}
and 
\begin{equation}
\sigma_{\mu\nu}=\frac{i}{4}\left[\gamma_{\mu},\gamma_{\nu}\right].
\end{equation}
We will generally use letters from the beginning of the Greek alphabet
to denote the Majorana spinor indices such as in $Q_{\alpha},\gamma_{\alpha\beta}^{\mu}$
where $\alpha,\beta,\dots=1,2,3,4$. This symmetry can be realized
by introducing a new anti-commuting space-time set of coordinates
$\theta_{\alpha}$, such that 
\begin{equation}
\left\{ \d_{\alpha},\theta_{\beta}\right\} =\delta_{\alpha\beta},\ \d_{\alpha}=\frac{\d}{\d\theta_{\alpha}}
\end{equation}
We will take this to be a Majorana spinor, such that 
\begin{align}
\theta & =\begin{pmatrix}\theta_{1} & \theta_{2} & \overline{\theta}_{\dot{2}} & -\overline{\theta}_{\dot{1}}\end{pmatrix}^{T}\\
\overline{\theta} & =-\theta^{T}C=\begin{pmatrix}-\theta_{2} & \theta_{1} & \overline{\theta}_{\dot{1}} & \overline{\theta}_{\dot{2}}\end{pmatrix}.
\end{align}
Then we can express the super-Poincaré generators in the superspace
$\left\{ x^{\mu},\theta_{\alpha}\right\} $ 
\begin{eqnarray}
J_{\mu\nu} & = & i\left(X_{\mu}\d_{\nu}-X_{\nu}\d_{\mu}\right)+\theta_{\alpha}\left(\sigma_{\mu\nu}\right)_{\alpha\beta}\d_{\beta}\\
Q_{\alpha} & = & -i\overline{\d}_{\alpha}+\gamma_{\alpha\beta}^{\mu}\theta_{\beta}\d_{\mu}\\
P_{\mu} & = & i\d_{\mu}
\end{eqnarray}

\subsection{Fermionic effective strings}

Much like the breaking of commuting symmetry operators results in
the introduction of massless Nambu-Goldstone bosons, Akulov and Volkov
showed \cite{key-4} that the breaking of anti-commuting generators
introduces massless fermions, which were later termed Goldstinos.
In an $\MN=1,\ D=4$ bulk, a string may break either all, or half
of the 4 SUSY generators. Clearly the generators which square to translations
transverse to the string must be broken. In this work we will focus
on the case were all generators are broken. As in the bosonic case,
the coordinates which correspond to the broken generators become a
field configuration which we will denote with the massless Majorana
spinor $\psi_{\alpha}$, and the effective string action is the action
of the massless modes on the worldsheet 
\begin{equation}
S=T\intop d^{2}\sigma\ML\left[X^{\mu}\left(\sigma^{0},\sigma^{1}\right),\psi_{\alpha}\left(\sigma^{0},\sigma^{1}\right)\right]
\end{equation}
The generalization of the Nambu-Goto action (\ref{eq:ng}) to the
supersymmetric case is obtained by replacing 
\begin{equation}
\d_{a}X^{\mu}\r\Pi_{a}^{\mu}\equiv\d_{a}X^{\mu}-i\overline{\psi}\gamma^{\mu}\d_{a}\psi
\end{equation}
to get the Akulov-Volkov action 
\begin{equation}
S_{AV}=-T\intop d^{2}\sigma\sqrt{-\det\left(\eta_{\mu\nu}\Pi_{a}^{\mu}\Pi_{b}^{\nu}\right)},
\end{equation}
When expanding this, dimensional analysis shows that terms of the
form $\d^{k}X^{m}\psi^{2n}$ contribute at order $L^{-k-n+1}$, so
we will denote the free term $i\overline{\psi}\gamma^{\mu}\d_{a}\psi\sim\MO\left(\d^{2}\right)$
and the rest of the terms accordingly. The AV action can then be expanded
as 
\begin{equation}
S_{AV}=-T\intop d^{2}\sigma\left(1+\frac{1}{2}\d_{a}X^{i}\d^{a}X^{i}-\frac{1}{2}\left(\Delta_{+}^{2\dot{2}}+\Delta_{-}^{1\dot{1}}\right)+\MO\left(\d^{3}\right)\right)
\end{equation}
where 
\begin{equation}
\Delta_{a}^{\alpha\dot{\alpha}}\equiv i\overline{\psi}_{\dot{\alpha}}\d_{a}\psi_{\alpha}-i\psi_{\alpha}\d_{a}\overline{\psi}_{\dot{\alpha}}
\end{equation}
This implies the equations of motion 
\begin{equation}
\d_{-}\psi_{1}+\MO\left(\d^{3}\right)=\d_{-}\overline{\psi}_{\dot{1}}+\MO\left(\d^{3}\right)=\d_{+}\psi_{2}+\MO\left(\d^{3}\right)=\d_{+}\overline{\psi}_{\dot{2}}+\MO\left(\d^{3}\right)=0\label{eq:eom}
\end{equation}

\subsection{Classification of the action of bosonic strings}

In their 2013 review of bosonic effective strings, Aharony and Komargodski
(AK) classify the action terms by their scale (which they refer to
as weight). The scale of a term is its dimension of $length^{-1}$,
such that $\d_{a}X^{\mu}$ has scale 0, $\d_{a}\d_{b}X^{\mu}$ has
scale 1, and so on. Translational invariance guarantees that all terms
in the action have non-negative scale. For bosonic strings, $ISO\left(1,1\right)\times SO\left(D-2\right)$
and parity invariance (that we assume) guarantee that all terms have
even scale. AK then show that there is a unique invariant action at
scale zero, which is the Nambu-Goto action (\ref{eq:ng}). At scale
2, AK find a single term which is invariant up to a term proportional
to the EOM, with 6 derivatives and 4 fields 
\begin{equation}
\ML_{6,4}=-32c_{4}\left(\d_{+}^{2}X^{i}\d_{-}^{2}X^{i}\right)\left(\d_{+}X^{j}\d_{-}X^{j}\right)+\dots,\label{eq:c4}
\end{equation}

which can be shown to be forbidden quantum mechanically since it modifies
the algebra of Lorentz transformations \cite{key-1}, which can lead
to anomalies (terms whose variation is proportional to the EOM can
be made invariant by changing the transformation rule, but this can
modify the algebra). The next allowed terms are of scale 4, and have
at least 8 derivatives. The existence of 8 derivatives implies that
those terms contribute to the $1/L$ expansion of the energy levels
at order of at least $1/L^{7}$, so that the coefficients up to and
including order of $1/L^{5}$ are universal. Aharony and Klinghoffer
\cite{key-5} calculated how the first few terms of the NG action
appear in the energy level expansion, as well as the effect of the
$\ML_{6,4}$ term.

\subsection{Gliozzi-Meineri (GM) approach for classifying bosonic string action
terms}

In their 2013 Paper\cite{key-2}, Gliozzi and Meineri (GM) present
a useful graphical approach to finding invariant terms for the action
of a bosonic string. They associate terms with graphs, where the vertices
are the fields $X^{i}$ and their derivatives, and the edges represent
contractions over indices. Since we have 2 types of indices - worldsheet
indices denoted $a,b,c,\dots$ and transverse indices denoted by $i,j,k,\dots$
- we also have 2 types of edges. Worldsheet indices will be represented
by solid lines, and transverse indices will be represented by wavy
lines. The term $\d_{a}X^{i}$ will be represented by a circular node
(slightly changing GM notation) with 2 open edges 

\begin{equation}
\d_{a}X^{i}=\bosonvertex
\end{equation}

So that scale 0 terms can be represented as sums and products of ring
graphs, so for example $\d_{a}X^{i}\d^{a}X^{i}$, $\d_{a}X^{i}\d^{a}X^{j}\d_{b}X^{j}\d^{b}X^{i}$
and a ring with 2n $\d X$'s will be represented and denoted as 
\begin{equation}
\Aring{2},\Aring{4},\Aring{2n}
\end{equation}
 correspondingly. General terms in the action are products of such
rings. GM write the broken infinitesimal Lorentz transformations in
a covariant form 
\begin{align}
\delta X^{i} & =-\epsilon^{aj}\delta^{ij}\xi_{a}-\epsilon^{aj}X^{j}\d_{a}X^{i}\\
\delta\left(\d_{b}X^{i}\right) & =-\epsilon^{aj}\delta^{ij}\eta_{ab}-\epsilon^{aj}\d_{b}X^{j}\d_{a}X^{i}-\epsilon^{aj}X^{j}\d_{a}\d_{b}X^{i}\label{eq:blah}
\end{align}
Eq. (\ref{eq:blah}) can be expressed graphically as
\begin{equation}
\delta\bosonvertex=-\bosontransa-\bosontransb-\bosontumorb
\end{equation}

where the solid circles represent the transformation parameter $\epsilon^{aj}$,
the vertex $X$ represents $X^{j}$, and the vertex which is connected
to 3 edges is simply $\d_{a}\d_{b}X^{i}$. Using this transformation
rule, one can transform the ring $A_{2n}$ which has $2n$ vertices
of the form $\d_{a}X^{i}$ (we will refer to these as boson vertices)
and express it graphically as 
\begin{align}
\Aring{2n} & \r-2n\cdot\Aring{2nl}-2n\cdot\Aring{2n2l}-2n\cdot\Aring{tumor}\label{eq:ringvar}
\end{align}

We can cancel the first two terms in the variation by summing rings
such that one variation from the ring $A_{2n}$ will cancel the other
from the ring $A_{2n+2}$. This gives a recursion relation for the
coefficients of the rings 
\begin{equation}
\left(2n+2\right)a_{2n+2}=-2na_{2n}\ra a_{2n}=\left(-1\right)^{n+1}\frac{1}{n}a_{2}
\end{equation}
Summing this series we get 
\begin{align}
\sum_{n=1}^{\infty}a_{2n}A_{2n} & =a_{2}\sum_{n=1}^{\infty}\left(-1\right)^{n+1}\frac{1}{n}\Tr\left[\left(\d_{a}X^{i}\d_{b}X^{i}\right)^{n}\eta^{bc}\right]=a_{2}\Tr\left[\log\left(\left(\eta_{ab}+\d_{a}X\cdot\d_{b}X\right)\eta^{bc}\right)\right]=\nonumber \\
 & =a_{2}\log\left[-\det\left(\eta_{ab}+h_{ab}\right)\right]=a_{2}\log\left(-g\right)
\end{align}
where $g=\det\left(g_{ab}\right)$. This summation cancels all variations
which come out of those two terms except for the variation 
\begin{equation}
\Aring{2l}
\end{equation}
 We can now consider a sum of terms of the form $b_{n}\left[\log\left(-g\right)\right]^{n}$.
The $n$th order in this sum contains products of $n$ rings, and
in fact every $n$ ring term is contained in it. Looking at the third
term in the variation (\ref{eq:ringvar}), we see that it has a ``tumor''
stemming from the ring. Such a tumor can be handled using integration
by parts of the derivative from which the tumor stems. This will move
the tumor around the ring, so that we get 
\begin{equation}
2n\cdot\Aring{tumor}=-\Aring{2n}\cdot\Aring{2l}+\text{total derivative}
\end{equation}

So for a product of $n$ rings we can cancel this variation using
the surviving $A_{2}$ variation from a product of $n+1$ rings. For
this cancellation we require 
\begin{equation}
b_{n+1}=\frac{1}{2\left(n+1\right)}b_{n}\ra b_{n}=\frac{1}{2^{n}n!}b_{0}
\end{equation}
and we get a unique invariant scale 0 Lagrangian
\begin{equation}
\ML_{0}=b_{0}\sum_{n=0}^{\infty}\frac{1}{n!}\left[\frac{1}{2}\log\left(-g\right)\right]^{n}=b_{0}\sqrt{-g}
\end{equation}

Which is exactly the NG Lagrangian. GM extend this approach for higher
scaling. They obtain two scale 2 invariants 
\begin{align}
I_{1} & =\sqrt{-g}\d_{ab}^{2}X_{i}\d_{cd}^{2}X_{j}t^{ij}g^{ac}g^{bd}\\
I_{2} & =\sqrt{-g}\d_{ab}^{2}X_{i}\d_{cd}^{2}X_{j}t^{ij}g^{ab}g^{cd}
\end{align}
where 
\begin{equation}
g^{ab}=\eta^{ab}-\eta^{ac}h_{cd}\eta^{db}+\eta^{ac}h_{cd}\eta^{de}h_{ef}\eta^{fb}-\dots
\end{equation}
is the matrix inverse of $g_{ab}$, and 
\begin{equation}
t^{ij}=\delta^{ij}-\d_{a}X^{i}\d_{b}X^{j}g^{ab}.
\end{equation}
However, looking at the invariants $I_{1},I_{2}$ one may observe
that $I_{1}-I_{2}=\sqrt{-g}R$ where $R$ is the 2D Ricci scalar.
This is a total derivative so it does not contribute to the action.
Also, the first terms of $I_{2}$, up to eight derivatives, are proportional
to the free EOM and hence are vanishing at the six-derivative order.
This shows there are no contributing invariants at the six-derivative
order. This approach does not find the term (\ref{eq:c4}) since this
term is only invariant up to the EOM. 

GM proceed to apply this method to find higher scale invariants which
will be discussed in chapter 4.

\section{Extending the GM approach to include Goldstinos}

To extend the GM approach to include Goldstinos, we need to look at
the broken supersymmetry transformations on the string. The broken
generators are 
\begin{eqnarray}
J_{ai} & = & i\left(X_{a}\d_{i}-X_{i}\d_{a}\right)+\psi^{\alpha}\left(\sigma_{ai}\right)_{\alpha}^{\,\beta}\d_{\beta}\\
Q_{\alpha} & = & -i\overline{\d}_{\alpha}+\gamma_{\alpha\beta}^{\mu}\psi^{\beta}\d_{\mu}
\end{eqnarray}
so that the transformations can be written as 
\begin{eqnarray}
\delta X^{j} & = & -\epsilon^{ai}\delta^{ij}\xi_{a}-\epsilon^{ai}X^{i}\d_{a}X^{j}+i\overline{\theta}^{\alpha}\gamma_{\alpha\beta}^{j}\psi^{\beta}+i\overline{\theta}^{\alpha}\gamma_{\alpha\beta}^{a}\psi^{\beta}\d_{a}X^{j}\\
\delta\psi^{\beta} & = & i\epsilon^{ai}\psi^{\alpha}\left(\sigma_{ai}\right)_{\gamma}^{\,\beta}-\epsilon^{ai}X_{i}\d_{a}\psi^{\beta}+\overline{\theta}^{\alpha}C_{\alpha}^{\,\beta}+i\overline{\theta}^{\alpha}\gamma_{\alpha\gamma}^{a}\psi^{\gamma}\d_{a}\psi^{\beta}\\
\delta\overline{\psi}^{\beta} & = & \delta\psi^{\delta}C_{\delta}^{\,\beta}=-i\epsilon^{ai}\overline{\psi}^{\alpha}\left(\sigma_{ai}\right)_{\alpha}^{\,\beta}-\epsilon^{ai}X_{i}\d_{a}\psi^{\delta}C_{\delta}^{\,\beta}+\overline{\theta}^{\beta}+i\overline{\theta}^{\alpha}\gamma_{\alpha\gamma}^{a}\psi^{\gamma}\d_{a}\psi^{\delta}C_{\delta}^{\,\beta}
\end{eqnarray}
We can write any fermionic effective string action using the following
vertices 
\begin{align}
\d_{a}X^{i} & =\bosonvertex\\
\d_{a}\psi^{\alpha} & =\dvertex\\
\overline{\psi}^{\alpha}\gamma_{\alpha\beta}^{b} & =\gammavertexb\\
\overline{\psi}^{\alpha}\gamma_{\alpha\beta}^{i} & =\gammavertexa
\end{align}
and their derivatives. In the above we used springs to express spinor
indices. The transformation laws of these vertices can be written
as 
\begin{align}
\delta\d_{b}X^{j} & =-\epsilon^{ai}\delta^{ij}\eta_{ab}-\epsilon^{ai}\d_{b}X^{i}\d_{a}X^{j}-\epsilon^{ai}X^{i}\d_{a}\d_{b}X^{j}+i\overline{\theta}^{\alpha}\gamma_{\alpha\beta}^{j}\d_{b}\psi^{\beta}+i\overline{\theta}^{\alpha}\gamma_{\alpha\beta}^{a}\d_{b}\psi^{\beta}\d_{a}X^{j}+i\overline{\theta}^{\alpha}\gamma_{\alpha\beta}^{a}\psi^{\beta}\d_{a}\d_{b}X^{j}\\
\delta\d_{b}\psi^{\beta} & =i\epsilon^{ai}\left(\sigma_{ai}\right)_{\gamma}^{\,\beta}\d_{b}\psi^{\alpha}-\epsilon^{ai}\d_{b}X_{i}\d_{a}\psi^{\beta}-\epsilon^{ai}X_{i}\d_{a}\d_{b}\psi^{\beta}+i\overline{\theta}^{\alpha}\gamma_{\alpha\gamma}^{a}\d_{b}\psi^{\gamma}\d_{a}\psi^{\beta}+i\overline{\theta}^{\alpha}\gamma_{\alpha\gamma}^{a}\psi^{\gamma}\d_{a}\d_{b}\psi^{\beta}\label{eq:sig1}\\
\delta\overline{\psi}^{\beta}\gamma_{\beta\gamma}^{b} & =-i\epsilon^{ai}\overline{\psi}^{\alpha}\gamma_{\alpha\beta}^{b}\left(\sigma_{ai}\right)_{\,\gamma}^{\beta}-\epsilon^{ai}\overline{\psi}^{\alpha}\gamma_{\alpha\gamma}^{i}\delta_{a}^{b}-\epsilon^{ai}X_{i}\d_{a}\psi^{\beta}\left(C\gamma^{b}\right)_{\beta\gamma}+\overline{\theta}^{\alpha}\gamma_{\alpha\gamma}^{b}+i\overline{\theta}^{\alpha}\gamma_{\alpha\delta}^{a}\psi^{\delta}\d_{a}\psi^{\beta}\left(C\gamma^{b}\right)_{\beta\gamma}\label{eq:sig2}\\
\delta\overline{\psi}^{\beta}\gamma_{\beta\gamma}^{j} & =-i\epsilon^{ai}\overline{\psi}^{\alpha}\gamma_{\alpha\beta}^{j}\left(\sigma_{ai}\right)_{\,\gamma}^{\beta}+\epsilon^{ai}\overline{\psi}^{\alpha}\gamma_{\alpha\gamma}^{a}\delta_{i}^{j}-\epsilon^{ai}X_{i}\d_{a}\psi^{\beta}\left(C\gamma^{j}\right)_{\beta\gamma}+\overline{\theta}^{\alpha}\gamma_{\alpha\gamma}^{j}+i\overline{\theta}^{\alpha}\gamma_{\alpha\delta}^{a}\psi^{\delta}\d_{a}\psi^{\beta}\left(C\gamma^{b}\right)_{\beta\gamma}\label{eq:sig3}
\end{align}
or in graphical representation 
\begin{align}
\delta\bosonvertex= & -\bosontransa &  & -\bosontransb\nonumber \\
 & +i\bosontransc &  & +i\bosontransd\\
 & +i\bosontumora &  & -\bosontumorb\nonumber \\
\delta\dvertex= & \hphantom{+}i\dtransa &  & -\dtransb\nonumber \\
 & +i\dtumora &  & -\dtumorb\\
 & +\sigma\text{-term}\nonumber \\
\delta\gammavertexb= & \hphantom{+}\gammatransba &  & -\gammatransbb\nonumber \\
 & +i\gammatumorba &  & -\gammatumorbb\\
 & +\sigma\text{-term}\nonumber \\
\delta\gammavertexa= & \hphantom{+}\gammatransaa &  & +\gammatransab\nonumber \\
 & i\gammatumoraa &  & -\gammatumorab\\
 & +\sigma\text{-term}\nonumber 
\end{align}
where the $\sigma$-terms are different terms which involves $\sigma_{ai}$
matrices, we used solid circles to represent transformation parameters,
such that 
\begin{align}
\epsilon^{ai} & =\bosontransa\\
\overline{\theta}^{\alpha}\gamma_{\alpha\beta}^{a} & =\gammatransba\\
\overline{\theta}^{\alpha}\gamma_{\alpha\beta}^{i} & =\gammatransaa
\end{align}
and we introduced 3-legged vertices and single legged vertices to
represent double derivatives, the matrices $C\gamma$ and either $X$
or $\psi$. We will refer to diagrams containing single legged vertices
as ``tumor diagrams''.

\section{Finding invariant terms in the unitary gauge}

To find invariant terms, we will begin by eliminating the $\sigma$-terms
and tumors from the transformations. To eliminate $\sigma$-terms,
we note that we must only look at fermion bilinears. Noting that fermion
bilinears with no derivatives will generate a Goldstino mass term
which we know is forbidden, we can construct the following bilinears
at scale zero
\begin{align}
i\overline{\psi}^{\alpha}\gamma_{\alpha\beta}^{i}\d_{b}\psi^{\beta} & =i\fullfermiona\equiv\fermionvertexa\\
i\overline{\psi}^{\alpha}\gamma_{\alpha\beta}^{a}\d_{b}\psi^{\beta} & =i\fullfermionb\equiv\fermionvertexb
\end{align}
These eliminate the $\sigma$-terms which appear both in the variations
of the $\gamma$ vertex and the $\d$ vertex with opposite signs as
can be seen from (\ref{eq:sig1}), (\ref{eq:sig2}) and (\ref{eq:sig3}).
We also define transformation terms 
\begin{align}
i\overline{\theta}^{\alpha}\gamma_{\alpha\beta}^{i}\d_{b}\psi^{\beta} & =i\fulltransa\equiv\fermiontransaa\\
i\overline{\theta}^{\alpha}\gamma_{\alpha\beta}^{a}\d_{b}\psi^{\beta} & =i\fulltransb\equiv\fermiontransba
\end{align}
so that the transformations laws become 
\begin{align}
\delta\fermionvertexa= & \fermiontransaa+\fermiontransab\nonumber \\
 & +\fermiontransac-\fermiontransad\\
 & +X-\text{tumor}+\psi-\text{tumor}\nonumber \\
\delta\fermionvertexb= & \fermiontransba+\fermiontransbb\nonumber \\
 & -\fermiontransbc-\fermiontransbd\\
 & +X-\text{tumor}+\psi-\text{tumor}\nonumber \\
\delta\bosonvertex= & \fermiontransaa+\bosontranse\nonumber \\
 & -\bosontransa-\bosontransb\\
 & +X-\text{tumor}+\psi-\text{tumor}\nonumber 
\end{align}
Note that the variations with a solid boson vertex are due to Lorentz
transformations, and the variations with a solid fermion vertex are
due to SUSY transformations. As in the bosonic case, we can dispose
of tumors through integration by parts, at the price of enlarging
the number of disconnected pieces of a term by 1, where the added
disconnected piece for the $X$ and $\psi$ -tumors are 
\begin{equation}
\Aring{2l},\Bring{0s}
\end{equation}
 correspondingly. We will use this fact to examine fully connected
terms, ignoring tumors, and then reinstate the tumors to sum up terms
with multiple disconnected pieces.

\subsection{Scale 0}

Since the vertices defined above all have scale zero and two legs,
we can build scale zero invariants from them using rings, similarly
to what we have seen in the bosonic case. As in the bosonic case,
we will start with a single ring, and for each (non-tumor) term in
its variation find a new ring which can cancel it, and then repeat
this process with any new rings we find, until all terms are canceled.
We will consider a general ring which has $n$ worldsheet edges $\left(n\geq1\right)$,
and cut all of them. The possible terms we could have between worldsheet
edges and their variations are 
\begin{align}
\delta\doubleboson= & -\doublebosontransa-\doublebosontransb+\nonumber \\
 & +\doublebosontransc+\doublebosontransd+\text{transposed}\\
\delta\fermionvertexb= & \fermiontransba+\fermiontransbb+\nonumber \\
 & -\fermiontransbc-\fermiontransbd\\
\delta\bosonfermion= & \bosonfermiontransa-\bosonfermiontransb+\nonumber \\
 & +\bosonfermiontransc+\bosonfermiontransd+\nonumber \\
 & +\bosonfermiontranse+\bosonfermiontransf+\nonumber \\
 & -\bosonfermiontransg-\bosonfermiontransh\\
\delta\doublefermiona= & \doublefermionatransa+\doublefermionatransb+\nonumber \\
 & +\doublefermionatransc-\doublefermionatransd+\text{transposed}
\end{align}
we will separate these to variations which preserve $n$, and variations
which take $n\r n+1$. The variations which preserve $n$ are 
\begin{align}
\delta_{n}\doubleboson= & -\doublebosontransa+\doublebosontransc+\text{transposed}\\
\delta_{n}\fermionvertexb= & \fermiontransba-\fermiontransbc\\
\delta_{n}\bosonfermion= & \bosonfermiontransa-\bosonfermiontransb+\nonumber \\
 & +\bosonfermiontransd\\
\delta_{n}\doublefermiona= & \doublefermionatransa+\text{transposed}
\end{align}
we can cancel most of these by looking at the combination vertex 
\begin{align}
\combinationvertex & \equiv\doubleboson-\bosonfermion-\fermionboson\label{eq:terms}\\
 & -\fermionvertexb-\fermionvertexc+\doublefermiona\nonumber 
\end{align}
and the ring $A_{n}$ which is just $n$ combination vertices connected
to a ring. The combination vertex leaves only the variations 
\begin{equation}
\delta_{n}\combinationvertex=-\doublebosontransa-\fermiontransba+\text{transposed}\label{eq:trans}
\end{equation}

Now, looking at the variations that take $n\r n+1$ we have 
\begin{align}
\delta_{n+1}\doubleboson= & -\doublebosontransb+\doublebosontransd+\text{transposed}\\
\delta_{n+1}\fermionvertexb= & \fermiontransbb-\fermiontransbd\\
\delta_{n+1}\bosonfermion= & \bosonfermiontransc+\bosonfermiontranse+\nonumber \\
 & +\bosonfermiontransf-\bosonfermiontransg+\nonumber \\
 & -\bosonfermiontransh\\
\delta_{n+1}\doublefermiona= & \doublefermionatransb+\doublefermionatransc+\nonumber \\
 & -\doublefermionatransd+\text{transposed}
\end{align}
these include almost all combinations of terms from (\ref{eq:terms})
and transformations from (\ref{eq:trans}), which means we can cancel
most of the $n\r n+1$ transformations of rings with $n$ terms using
$n$ preserving transformations of rings with $n+1$ terms, exactly
as we did in the boson case, taking 
\begin{equation}
\left(2n+2\right)a_{n+1}=-2na_{n}\ra a_{n}=\left(-1\right)^{n+1}\frac{1}{n}a_{1}
\end{equation}
as the coefficient of the ring $A_{n}$. Note that as in the boson
case there is no need to cancel the $n$ preserving transformations
for $A_{1}$ since it is a total derivative. This leaves us with 2
yet to be canceled terms:
\begin{itemize}
\item A single combination not represented in the $n\r n+1$ transformations
$\doublefermionbtransa$
\item A single $n\r n+1$ transformation which cannot be expressed as such
a combination $\doublefermionatransc$
\end{itemize}
To fix the first problem, we take note that the only term that can
produce this transformation is $\doublefermionb$. No other term can
cancel it. To avoid this problem we will exclude it completely, by
adding a canceling term into the definition of the combination vertex
\begin{align}
\combinationvertex & =\doubleboson-\bosonfermion-\fermionboson\label{eq:terms-1}\\
 & -\fermionvertexb-\fermionvertexc+\doublefermiona+\alpha\doublefermionb\nonumber 
\end{align}
One can check that in order to cancel the $\doublefermionbtransa$
transformation from the $\doublefermionb$ in $A_{n}$ with that transformation
from the combinations of $\fermionvertexb$ and $\fermionvertexc$
in $A_{n+1}$, taking into account their respectable coefficients,
we should take $\alpha=1$. We now need to also include the variations
of $\doublefermionb$ which are 
\begin{align}
\delta\doublefermionb & =\doublefermionbtransa+\doublefermionbtransb+\nonumber \\
 & -\doublefermionbtransd-\doublefermionbtransc
\end{align}
The first 3 transformations here are automatically dropped since all
terms including $\doublefermionb$ were excluded. We are left with
the last transformation, but this is exactly the last transformation
we could not cancel before, and it is now canceled! This means we
have constructed an invariant using $A_{n}$ rings in exactly the
same way we have in the boson case, with the switch 
\begin{equation}
\doubleboson\r\combinationvertex
\end{equation}
switching back from the diagrammatic notation, this means 
\begin{align}
\d_{a}X^{i}\d_{b}X^{i}\r & \d_{a}X^{i}\d_{b}X^{i}-i\overline{\psi}\gamma_{a}\d_{b}\psi-i\overline{\psi}\gamma_{b}\d_{a}\psi-i\overline{\psi}\gamma^{i}\d_{a}\psi\d_{b}X^{i}-i\overline{\psi}\gamma^{i}\d_{b}\psi\d_{a}X^{i}+\nonumber \\
 & -\left(\overline{\psi}\gamma^{i}\d_{a}\psi\right)\overline{\psi}\gamma^{i}\d_{b}\psi-\eta_{cd}\left(\overline{\psi}\gamma^{c}\d_{a}\psi\right)\overline{\psi}\gamma^{d}\d_{b}\psi
\end{align}
(where the spinor indices are contracted between adjacent $\overline{\psi}'s$
and $\psi$'s), or alternatively 
\begin{eqnarray}
\d_{a}X^{\mu}\d_{b}X_{\mu} & = & \eta_{ab}+\d_{a}X^{i}\d_{b}X^{i}\r\nonumber \\
 & \r & \eta_{ab}+\d_{a}X^{i}\d_{b}X^{i}-i\overline{\psi}\gamma_{a}\d_{b}\psi-i\overline{\psi}\gamma_{b}\d_{a}\psi-i\overline{\psi}\gamma^{i}\d_{a}\psi\d_{b}X^{i}-i\overline{\psi}\gamma^{i}\d_{b}\psi\d_{a}X^{i}+\nonumber \\
 &  & -\left(\overline{\psi}\gamma^{i}\d_{a}\psi\right)\overline{\psi}\gamma^{i}\d_{b}\psi-\eta_{cd}\left(\overline{\psi}\gamma^{c}\d_{a}\psi\right)\overline{\psi}\gamma^{d}\d_{b}\psi=\\
 & = & \d_{a}X^{\mu}\d_{b}X_{\mu}-\d_{a}X^{\mu}i\overline{\psi}\gamma_{\mu}\d_{b}\psi-i\overline{\psi}\gamma^{\mu}\d_{a}\psi\d_{b}X_{\mu}-\left(\overline{\psi}\gamma^{\mu}\d_{a}\psi\right)\overline{\psi}\gamma_{\mu}\d_{b}\psi=\nonumber \\
 & = & \left(\d_{a}X^{\mu}-i\overline{\psi}\gamma^{\mu}\d_{a}\psi\right)\left(\d_{b}X_{\mu}-i\overline{\psi}\gamma_{\mu}\d_{b}\psi\right)\nonumber 
\end{eqnarray}
and we get the invariant scale zero action
\begin{equation}
S_{0}=-c_{0}\intop d^{2}\xi\sqrt{-\det\left[\left(\d_{a}X^{\mu}-i\overline{\psi}\gamma^{\mu}\d_{a}\psi\right)\left(\d_{b}X_{\mu}-i\overline{\psi}\gamma_{\mu}\d_{b}\psi\right)\right]}=-c_{0}\intop d^{2}\xi\sqrt{-g}
\end{equation}
which is exactly the Akulov-Volkov action with 
\begin{align}
g= & \det g_{ab}\\
g_{ab}= & \left(\d_{a}X^{\mu}-i\overline{\psi}\gamma^{\mu}\d_{a}\psi\right)\left(\d_{b}X_{\mu}-i\overline{\psi}\gamma_{\mu}\d_{b}\psi\right)=\eta_{ab}+h_{ab}\\
h_{ab}= & \d_{a}X^{i}\d_{b}X^{i}-i\overline{\psi}\gamma_{a}\d_{b}\psi-i\overline{\psi}\gamma_{b}\d_{a}\psi-i\overline{\psi}\gamma^{i}\d_{a}\psi\d_{b}X^{i}-i\overline{\psi}\gamma^{i}\d_{b}\psi\d_{a}X^{i}+\nonumber \\
 & -\left(\overline{\psi}\gamma^{i}\d_{a}\psi\right)\overline{\psi}\gamma^{i}\d_{b}\psi-\eta_{cd}\left(\overline{\psi}\gamma^{c}\d_{a}\psi\right)\overline{\psi}\gamma^{d}\d_{b}\psi
\end{align}
and we have $g^{ab}$ the matrix inverse of $g_{ab}$
\begin{equation}
g^{ab}=\eta^{ab}-\eta^{ac}h_{cd}\eta^{db}+\eta^{ac}h_{cd}\eta^{de}h_{ef}\eta^{fb}-\dots\label{eq:g}
\end{equation}
We can see that this method is exhaustive since up to the overall
$c_{0}$ it fixes the coefficients of all possible terms.

\subsection{Scale 1}

In order to find a scale one invariant action, we first list all possible
independent scale one vertices, which are
\begin{equation}
\d_{a}\d_{b}X^{i},\ \overline{\psi}\gamma^{i}\d_{a}\d_{b}\psi,\ \overline{\psi}\gamma^{c}\d_{a}\d_{b}\psi,\ \d_{a}\overline{\psi}\d_{b}\psi\label{eq:scale1}
\end{equation}
Since we can only include one such vertex in our action, all 3-legged
vertices are excluded, and the only one we can use is $\d_{a}\overline{\psi}\d_{b}\psi$
in ring topology, where all other terms are scale zero. However this
vertex is antisymmetric in the indices $\left(a,b\right)$, while
the rest of the ring is symmetric, and so the scale 1 action is dropped.

\subsection{Scale 2}

There are several ways to construct scale two invariants: either with
two scale one vertices, or with a single scale two vertex. The scale
one vertices are listed in (\ref{eq:scale1}). and we can either use
3-leeged vertices in ``$\Theta$'' or ``dumbbell'' topologies
as shown below, or two copies of the 2-legged vertex $\d_{a}\overline{\psi}\d_{b}\psi$
in a ring topology. 
\begin{align*}
\JustTheta & \quad & \JustDumbbell\\
\Theta\text{ topology} &  & \text{Dumbbell topology}
\end{align*}
The possible independent scale two vertices are 
\begin{equation}
\d_{a}\d_{b}\d_{c}X^{i},\ \d_{a}\overline{\psi}\d_{b}\d_{c}\psi,\ \overline{\psi}\gamma^{i}\d_{a}\d_{b}\d_{c}\psi,\ \overline{\psi}\gamma^{d}\d_{a}\d_{b}\d_{c}\psi
\end{equation}
Excluding the 3-legged vertex $\d_{a}\overline{\psi}\d_{b}\d_{c}\psi$
we are left with three 4-legged vertices which can be used in an ``$8$''
topology.

\subsubsection{Ring topology}

A ring topology invariant can be obtained by placing two $\d_{a}\overline{\psi}\d_{b}\psi$
vertices in a ring. We will first give this vertex a diagrammatic
representation 
\begin{equation}
i\d_{a}\overline{\psi}\d_{b}\psi=\dfdfvertex
\end{equation}
The variation for this vertex is 
\begin{align}
\delta\dfdfvertex= & \dfdftransa+\dfdftransb+\nonumber \\
 & -\dfdftransc-\dfdftransd+\label{eq:dfdfvar}\\
 & +\text{tumors}\nonumber 
\end{align}

Looking back at our calculation for the scale zero term, we see immediately
that this vertex has no $n$ preserving variations, and that its $n\r n+1$
variations fit right into our cancellation scheme for $\combinationvertex$
vertices, without allowing for the excluded $\doublefermionb$. Looking
at the ring $B_{k\ell}$, $\left(k+\ell=n-2\geq0\right)$ which is
\begin{equation}
\Scaleoneringkl=\d^{a}\overline{\psi}\d_{b}\psi\left(h^{k}\right)_{\,c}^{b}\d^{c}\overline{\psi}\d_{d}\psi\left(h^{\ell}\right)_{\,a}^{d}
\end{equation}
where $\left(h^{k}\right)_{\,c}^{b}$ is the matrix $h_{\,c}^{b}=\eta^{ba}h_{ac}$
taken to the $k$th power, we see that its variation can be canceled
by the variations of $B_{k+1,\ell}$ and $B_{k,\ell+1}$. Taking the
sum so that the variations cancel we have
\begin{align}
 & \sum_{k,\ell=0}^{\infty}\left(-1\right)^{k+\ell}\d^{a}\overline{\psi}\d_{b}\psi\left(h^{k}\right)_{\,c}^{b}\d^{c}\overline{\psi}\d_{d}\psi\left(h^{\ell}\right)_{\,a}^{d}=\nonumber \\
= & \d^{a}\overline{\psi}\d_{b}\psi\sum_{k=0}^{\infty}\left(\left(-h\right)^{k}\right)_{\,c}^{b}\d^{c}\overline{\psi}\d_{d}\psi\sum_{\ell=0}^{\infty}\left(\left(-h\right)^{\ell}\right)_{\,a}^{d}=\\
= & \d^{a}\overline{\psi}\d_{b}\psi\left(\left(\identity+h\right)^{-1}\right)_{\,c}^{b}\d^{c}\overline{\psi}\d_{d}\psi\left(\left(\identity+h\right)^{-1}\right)_{\,a}^{d}=\nonumber \\
= & \d_{a}\overline{\psi}\d_{b}\psi g^{bc}\d_{c}\overline{\psi}\d_{d}\psi g^{da}\nonumber 
\end{align}
Taking tumors into account means that this term must be multiplied
by the scale zero invariant $\sqrt{-g}$, and we get the scale two
ring invariant 
\begin{equation}
\ML_{2}^{\text{ring}}=c_{2}\sqrt{-g}\d_{a}\overline{\psi}\d_{b}\psi\d_{c}\overline{\psi}\d_{d}\psi g^{bc}g^{da}\label{eq:Lring}
\end{equation}
we can generalize this in a similar manner to what Gliozzi and Meineri
did to obtain high scaling invariants. To do so, we define \emph{seed}
graphs, which are minimal connected graphs in the sense that they
cannot be reduced to a non-trivial graph by erasing scale zero chains
and connecting their edges together, and have no fermion vertices
which can be reduced to boson vertices with the same scale and leg
structure. The scale two ring topology seed graph is 
\begin{equation}
\ScaleTwoSeed{Ring}
\end{equation}
Given a seed graph, for each worldsheet edge $\eta^{ab}$, if the
vertices connected to it have a variation structure like the one in
(\ref{eq:dfdfvar}), we can replace $\eta^{ab}\r g^{ab}$ to eliminate
the non-tumor variations, and multiply by $\sqrt{-g}$ to eliminate
tumors.

\subsubsection{$\Theta$ and dumbbell topologies}

These invariants are created using two 3-legged scale one vertices,
which are $\d_{a}\d_{b}X^{i},\ i\overline{\psi}\gamma^{i}\d_{a}\d_{b}\psi,\ i\overline{\psi}\gamma^{c}\d_{a}\d_{b}\psi$.
We will use the following graphical representations for them 
\begin{equation}
\ScaleOneVertex{bos}{},\ScaleOneVertex{fera}{},\ScaleOneVertex{ferb}{}
\end{equation}

and their variations 
\begin{align}
\delta\ScaleOneVertex{bos}{}= & \ScaleOneVertex{bos}{trans2}+\ScaleOneVertex{bos}{trans3}-\ScaleOneVertex{bos}{trans6}-\ScaleOneVertex{bos}{trans7}+\nonumber \\
 & -\ScaleOneVertex{bos}{trans1}+\ScaleOneVertex{bos}{trans4}+\ScaleOneVertex{bos}{trans5}+\text{tumors}\\
\delta\ScaleOneVertex{fera}{}= & \ScaleOneVertex{fera}{trans2}+\ScaleOneVertex{fera}{trans3}-\ScaleOneVertex{fera}{trans6}-\ScaleOneVertex{fera}{trans7}+\nonumber \\
 & -\ScaleOneVertex{fera}{trans1}+\ScaleOneVertex{fera}{trans8}+\ScaleOneVertex{fera}{trans4}+\ScaleOneVertex{fera}{trans5}+\text{tumors}\\
\delta\ScaleOneVertex{ferb}{}= & \ScaleOneVertex{ferb}{trans2}+\ScaleOneVertex{ferb}{trans3}-\ScaleOneVertex{ferb}{trans6}-\ScaleOneVertex{ferb}{trans7}+\nonumber \\
 & -\ScaleOneVertex{ferb}{trans1}-\ScaleOneVertex{ferb}{trans8}+\ScaleOneVertex{ferb}{trans4}+\ScaleOneVertex{ferb}{trans5}+\text{tumors}
\end{align}

Where in the last vertex it's important to note that we use an isosceles
triangle, so that the two derivative legs behave differently than
the $\gamma^{a}$ leg. Looking at the first line in every variation,
we see the exact same structure we saw for the ring topology, so these
variations can be eliminated by replacing $\eta^{ab}\r g^{ab}$ on
seed graphs in which worldsheet legs of scale one vertices are connected
together. The seed graphs we can construct using these vertices are
\begin{itemize}
\item $\Theta$ topology 
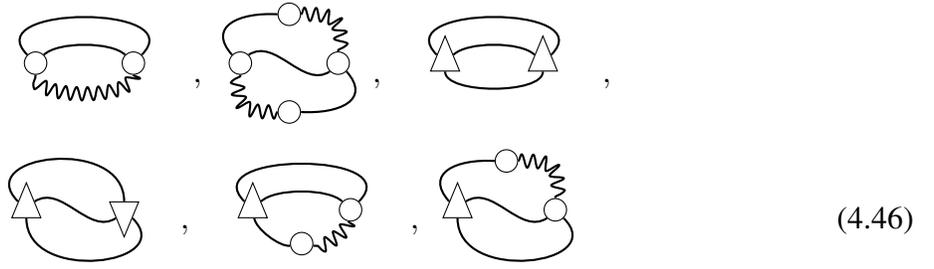
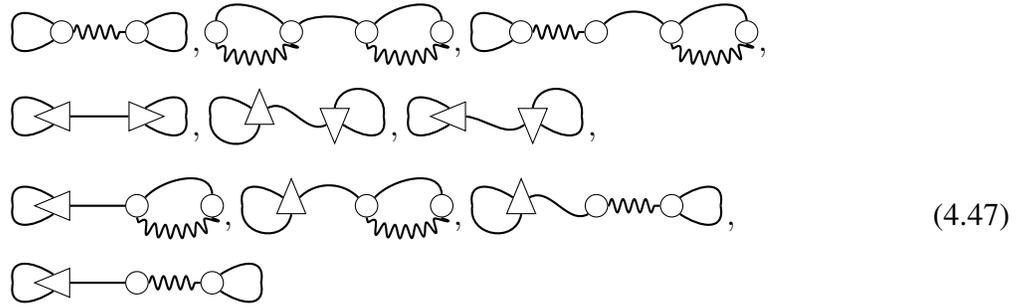
\begin{align}
\ScaleTwoSeed{Theta1},\ScaleTwoSeed{Theta2},\ScaleTwoSeed{Theta3},\nonumber \\
\ScaleTwoSeed{Theta4},\ScaleTwoSeed{Theta5},\ScaleTwoSeed{Theta6}
\end{align}
\item Dumbbell topology 
\begin{align}
 & \ScaleTwoSeed{Dumbbell1},\ScaleTwoSeed{Dumbbell2},\ScaleTwoSeed{Dumbbell3},\nonumber \\
 & \ScaleTwoSeed{Dumbbell4},\ScaleTwoSeed{Dumbbell5},\ScaleTwoSeed{Dumbbell6},\nonumber \\
 & \ScaleTwoSeed{Dumbbell7},\ScaleTwoSeed{Dumbbell9},\ScaleTwoSeed{Dumbbell10},\\
 & \ScaleTwoSeed{Dumbbell8}\nonumber 
\end{align}
\end{itemize}
Most of these seed graphs can be easily eliminated as candidates for
invariants, since they have variations which cannot be canceled, such
as 
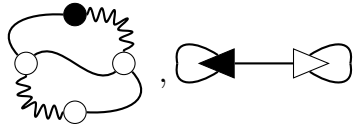
\[
\ScaleTwoSeed{forbidden1},\ScaleTwoSeed{forbidden2}
\]
In fact, it is easy to check that each of these graphs with either
a single scale-0 boson not connected to the $\gamma$ leg of a scale-2
fermion, or a scale-2 fermion not connected by its $\gamma$ leg to
a scale-0 boson, will have such a variation. This leaves us with the
following seed graphs 
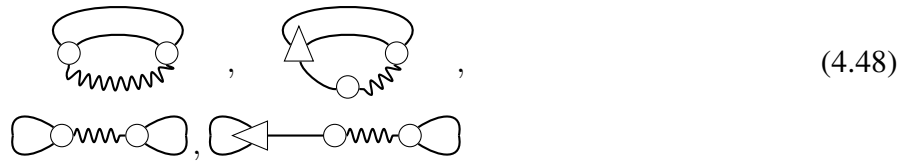
\begin{align}
\ScaleTwoSeed{Theta1},\ScaleTwoSeed{Theta5},\label{eq:seeds}\\
\ScaleTwoSeed{Dumbbell1},\ScaleTwoSeed{Dumbbell8}\nonumber 
\end{align}
However, these seed graphs are not independent, since both graphs
in each line of (\ref{eq:seeds}) appear in the same cancellation
flow. We can define a new scale-2 fermion-boson vertex 
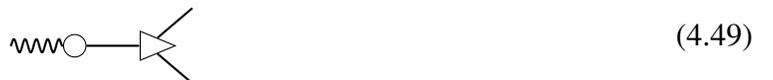
\begin{equation}
\scaleonechainc
\end{equation}
and include such vertices as fermion vertices in our definition for
seed graphs, leaving us with exactly 2 seed graphs we can create an
invariant out of 
\begin{equation}
\ScaleTwoSeed{Theta1},\ScaleTwoSeed{Dumbbell1}
\end{equation}
These are in fact exactly the same seed graphs GM used to create scale-2
invariants in the bosonic case. To make the invariants in our case,
we need to eliminate all the variations. The variations on the derivative
edges are eliminated by the exchange of the embedded metric. To eliminate
the variations on the transverse edge, we look at all the legal chains
between the scale one vertex and the first transverse edge. Those
legal chains have to be made out of $\fermionvertexb$ terms, which
must be directed into the scale one vertex, since back-to-back fermions
connected by a worldsheet vertex are forbidden. So we have the following
chains 
\begin{equation}
\scaleonechaina,\scaleonechainb,\scaleonechaine,\scaleonechainf\qquad n\geq0
\end{equation}

We can take a sum of these so that all variations which appear between
the scale one vertex and the first transverse edge are canceled, by
defining the scale one combination vertex
\begin{align}
\ScaleOneVertex{comb}{} & =\scaleonechaina-\scaleonechainb-\sum_{n=0}^{\infty}\left(-1\right)^{n}\left(\scaleonechaine-\scaleonechainf\right)=\nonumber \\
 & =\d_{a}\d_{b}X^{i}-i\overline{\psi}\gamma^{i}\d_{a}\d_{b}\psi-\left(\d_{c}X^{i}-i\overline{\psi}\gamma^{i}\d_{c}\psi\right)\left(\delta_{\,d}^{c}+i\overline{\psi}\gamma^{c}\d_{d}\psi\right)^{-1}i\overline{\psi}\gamma^{d}\d_{a}\d_{b}\psi\equiv C_{ab}^{i}
\end{align}

Replacing the scale one boson vertices with these combination vertices,
we get that there are still variations on the transverse edge. To
eliminate these we sum up all the ways the two combination vertices
can connect. This is either directly through the transverse edge,
or the transverse edge can be terminated on both ends either by a
boson vertex or a fermion vertex, and these connect through a scale
zero chain with worldsheet edges on each side. This is equivalent
to making a similar replacement to the one GM do for the boson case
\begin{equation}
\delta^{ij}\r t^{ij}=\delta^{ij}-\left(\d_{a}X^{i}-i\overline{\psi}\gamma^{i}\d_{a}\psi\right)g^{ab}\left(\d_{b}X^{j}-i\overline{\psi}\gamma^{j}\d_{b}\psi\right)\label{eq:t}
\end{equation}
Thus eliminating the rest of the variations. To conclude, scale two
invariants are obtained by looking at scale two seed graphs and performing
the following moves
\begin{enumerate}
\item Replacing $\eta^{ab}\r g^{ab}$ on worldsheet edges
\item Replacing the scale one vertices $\scaleonechaina\r\ScaleOneVertex{comb}{}$
\item Replacing $\delta^{ij}\r t^{ij}$ on transverse edges
\end{enumerate}
Where in these topologies we get the invariants 
\begin{align}
I_{1} & =\sqrt{-g}C_{ab}^{i}t^{ij}C_{cd}^{j}g^{ac}g^{bd}\\
I_{2} & =\sqrt{-g}C_{ab}^{i}t^{ij}C_{cd}^{j}g^{ab}g^{cd}
\end{align}
Here, as in the bosonic case, $I_{2}$ is proportional to the EOM
up to $\MO\left(\d^{6}\right)$. We have not checked if $I_{1}-I_{2}$
is a total derivative to any order, but at least up to $\MO\left(\d^{6}\right)$
it is. Thus, there are no new corrections to the fermionic string
energy levels up to this order.

\subsubsection{$8$ topology}

Creating an $8$ topology invariant requires using a single 4-legged
scale two vertex, which can be either $\d_{a}\d_{b}\d_{c}X^{i}$,
$\overline{\psi}\gamma^{i}\d_{a}\d_{b}\d_{c}\psi$ or $\overline{\psi}\gamma^{d}\d_{a}\d_{b}\d_{c}\psi$.
The available seed graphs are 
\begin{equation}
\ScaleTwoSeed{Eight1},\ScaleTwoSeed{Eight2}
\end{equation}
where the third, with the vertex $\overline{\psi}\gamma^{i}\d_{a}\d_{b}\d_{c}\psi$,
is included in the chain obtained from $\d_{a}\d_{b}\d_{c}X^{i}$.
Both graphs have variations which cannot be canceled 
\begin{equation}
\ScaleTwoSeed{forbidden3},\ScaleTwoSeed{forbidden4}
\end{equation}
so an $8$ topology invariant is excluded.

\subsection{Higher scaling}

The number of invariants proliferates rapidly as the scaling increases,
since the number of different vertices available and the number of
different topologies both increase. We can generalize the vertices
we have introduce into three types of higher scaling vertices 
\begin{itemize}
\item $\d^{n}X^{i}$ at scaling $n-1$ and with $n+1$ legs
\item $\overline{\psi}\gamma^{a/i}\d^{n}\psi$ at scaling $n-1$ and with
$n+1$ legs
\item $\d^{m}\overline{\psi}\d^{n}\psi$ at scaling $m+n-1$ and with $m+n$
legs
\end{itemize}
From their transformation laws, it is easy to see that the first two
types are highly related. In fact, for any bosonic invariant that
we can create using just the first type of vertices, we can generate
a corresponding supersymmetric invariant by replacing the high scaling
bosonic vertices with the appropriate combination vertices like we
did in the last section 
\begin{equation}
\d_{a_{1}\cdots a_{n}}^{n}X^{i}\r\d_{a_{1}\cdots a_{n}}^{n}X^{i}-i\overline{\psi}\gamma^{i}\d_{a_{1}\cdots a_{n}}^{n}\psi-\left(\d_{c}X^{i}-i\overline{\psi}\gamma^{i}\d_{c}\psi\right)\left(\delta_{\,d}^{c}+i\overline{\psi}\gamma^{c}\d_{d}\psi\right)^{-1}i\overline{\psi}\gamma^{d}\d_{a_{1}\cdots a_{n}}^{n}\psi\equiv C_{a_{1}\cdots a_{n}}^{i}\label{eq:C}
\end{equation}

GM formulate the generation of higher scaling bosonic invariants by
looking at the variation of the scaling $n-1$ vertex $\left(n>1\right)$
\begin{align}
\delta\left(\d_{a_{1}\cdots a_{n}}^{n}X^{i}\right) & =-\epsilon^{bj}\paren{\d_{b}X^{i}\d_{a_{1}\cdots a_{n}}^{n}X^{j}+\sum_{k}\d_{a_{k}}X^{j}\d_{ba_{1}\cdots a_{k-1}a_{k+1}\cdots a_{n}}^{n}X^{i}+}\nonumber \\
 & \qquad\ \thesis{+\sum_{k,l}\d_{a_{k}a_{l}}^{2}X^{j}\d_{ba_{1}\cdots a_{k-1}a_{k+1}\cdots a_{l-1}a_{l+1}\cdots\cdots a_{n}}^{n-1}X^{i}+\dots}
\end{align}
Where the first two terms add a scale zero vertex on each on the legs,
and are canceled by the moves $\eta^{ab}\r g^{ab},\delta^{ij}\r t^{ij}$
as we have seen in the previous chapter. The third term has a scale
$n-2$ vertex connected to a scale 1 vertex so it can only be canceled
by terms containing such vertices, the fourth has a scale $n-3$ vertex
connected to a scale 3 vertex and so on. We can cancel these terms
by defining a sort of covariant derivative. GM define this for the
scale 2 term 
\begin{equation}
\d_{abc}^{3}X^{i}\r\del_{abc}^{3}X^{i}=\d_{abc}^{3}X^{i}-\left(\d_{ab}^{2}X^{j}\d_{d}X^{j}\d_{ec}^{2}X^{i}g^{de}+\text{cyclic permutations of }abc\right)
\end{equation}
so that 
\begin{equation}
\delta\left(\del_{abc}^{3}X^{i}\right)=-\epsilon^{bj}\left(\d_{b}X^{i}\d_{a_{1}\cdots a_{n}}^{n}X^{j}+\sum_{k}\d_{a_{k}}X^{j}\d_{ba_{1}\cdots a_{k-1}a_{k+1}\cdots a_{n}}^{n}X^{i}\right)
\end{equation}
which can be generalized to the $n$-th derivative with 
\begin{align}
\del_{a_{1}\cdots a_{n}}^{n}X^{i} & =\d_{a_{1}\cdots a_{n}}^{n}X^{i}-\left(\d_{a_{1}\cdots a_{n-1}}^{n-1}X^{j}\d_{b}X^{j}\d_{ca_{n}}^{2}X^{i}g^{bc}+\text{cyclic permutations of }a_{1}\dots a_{n}\right)+\nonumber \\
 & -\left(\d_{a_{1}\cdots a_{n-2}}^{n-2}X^{j}\d_{b}X^{j}\d_{ca_{n-1}a_{n}}^{3}X^{i}g^{bc}+\text{cyclic permutations of }a_{1}\dots a_{n}\right)+\dots
\end{align}
We can generalize this for the supersymmetric case by noting that
\begin{align}
\delta\d_{a_{1}\cdots a_{n}}^{n}X^{i} & =-\epsilon^{bj}\paren{\d_{b}X^{i}\d_{a_{1}\cdots a_{n}}^{n}X^{j}+\sum_{k}\d_{a_{k}}X^{j}\d_{ba_{1}\cdots a_{k-1}a_{k+1}\cdots a_{n}}^{n}X^{i}+}\nonumber \\
 & \qquad\ \thesis{+\sum_{k,l}\d_{a_{k}a_{l}}^{2}X^{j}\d_{ba_{1}\cdots a_{k-1}a_{k+1}\cdots a_{l-1}a_{l+1}\cdots\cdots a_{n}}^{n-1}X^{i}+\dots}\nonumber \\
 & +i\overline{\theta}\paren{\gamma^{i}\d_{a_{1}\cdots a_{n}}^{n}\psi+\gamma^{b}\sum_{k}\d_{a_{k}}\psi\d_{ba_{1}\cdots a_{k-1}a_{k+1}\cdots a_{n}}^{n}X^{i}+}\label{eq:d1}\\
 & \qquad\thesis{+\gamma^{b}\sum_{k,l}\d_{a_{k}a_{l}}^{2}\psi\d_{ba_{1}\cdots a_{k-1}a_{k+1}\cdots a_{l-1}a_{l+1}\cdots\cdots a_{n}}^{n-1}X^{i}+\dots}\nonumber 
\end{align}
where the first two terms of the $\theta$ variation are canceled
by the move $\d_{a_{1}\cdots a_{n}}^{n}X^{i}\r C_{a_{1}\cdots a_{n}}^{i}$,
and the following terms can be canceled by generalizing the above
covariant derivative to the supersymmetric case such that 
\begin{align}
\del_{a_{1}\cdots a_{n}}^{n}X^{i} & =\d_{a_{1}\cdots a_{n}}^{n}X^{i}-\left(\d_{a_{1}\cdots a_{n-1}}^{n-1}X^{j}\d_{b}X^{j}\d_{ca_{n}}^{2}X^{i}g^{bc}+\text{cyclic permutations of }a_{1}\dots a_{n}\right)+\nonumber \\
 & -\left(\d_{a_{1}\cdots a_{n-2}}^{n-2}X^{j}\d_{b}X^{j}\d_{ca_{n-1}a_{n}}^{3}X^{i}g^{bc}+\text{cyclic permutations of }a_{1}\dots a_{n}\right)+\dots\nonumber \\
 & -\left(i\overline{\psi}\gamma_{b}\d_{ca_{n}}^{2}\psi\d_{a_{1}\cdots a_{n-2}}^{n-2}X^{i}g^{bc}+\text{cyclic permutations of }a_{1}\dots a_{n}\right)+\\
 & -\left(i\overline{\psi}\gamma_{b}\d_{ca_{n-1}a_{n}}^{3}\psi\d_{a_{1}\cdots a_{n-3}}^{n-3}X^{i}g^{bc}+\text{cyclic permutations of }a_{1}\dots a_{n}\right)+\dots\nonumber 
\end{align}
and similarly for derivatives acting on fermions where 
\begin{align}
\delta\left(i\overline{\psi}\gamma^{i}\d_{a_{1}\cdots a_{n}}^{n}\psi\right) & =-i\epsilon^{bj}\paren{\overline{\psi}\gamma^{i}\d_{b}\psi\d_{a_{1}\cdots a_{n}}^{n}X^{j}+\sum_{k}\d_{a_{k}}X^{j}\overline{\psi}\gamma^{i}\d_{ba_{1}\cdots a_{k-1}a_{k+1}\cdots a_{n}}^{n}\psi+}\nonumber \\
 & \qquad\ \thesis{+\sum_{k,l}\d_{a_{k}a_{l}}^{2}X^{j}\overline{\psi}\gamma^{i}\d_{ba_{1}\cdots a_{k-1}a_{k+1}\cdots a_{l-1}a_{l+1}\cdots\cdots a_{n}}^{n-1}\psi+\dots}+\nonumber \\
 & +i\overline{\theta}\paren{\gamma^{i}\d_{a_{1}\cdots a_{n}}^{n}\psi+\gamma^{b}\sum_{k}\d_{a_{k}}\psi i\overline{\psi}\gamma^{i}\d_{ba_{1}\cdots a_{k-1}a_{k+1}\cdots a_{n}}^{n}\psi+}\nonumber \\
 & \qquad\thesis{+\gamma^{b}\sum_{k,l}\d_{a_{k}a_{l}}^{2}\psi i\overline{\psi}\gamma^{i}\d_{ba_{1}\cdots a_{k-1}a_{k+1}\cdots a_{l-1}a_{l+1}\cdots\cdots a_{n}}^{n-1}\psi+\dots}
\end{align}
and a similar expression when replacing the transverse index $i$
with a worldsheet index $c$. We get 
\begin{align}
\del_{a_{1}\cdots a_{n}}^{n}\psi & =\d_{a_{1}\cdots a_{n}}^{n}\psi-\left(\d_{ba_{1}\cdots a_{n-2}}^{n-1}\psi\d_{c}X^{i}\d_{a_{n-1}a_{n}}^{2}X^{i}g^{bc}+\text{cyclic permutations of }a_{1}\dots a_{n}\right)+\nonumber \\
 & -\left(\d_{ba_{1}\cdots a_{n-3}}^{n-2}\psi\d_{c}X^{i}\d_{a_{n-2}a_{n-1}a_{n}}^{3}X^{i}g^{bc}+\text{cyclic permutations of }a_{1}\dots a_{n}\right)+\dots\nonumber \\
 & -\left(\d_{ba_{1}\cdots a_{n-2}}^{n-1}\psi i\overline{\psi}\gamma_{c}\d_{a_{n-1}a_{n}}^{2}\psi g^{bc}+\text{cyclic permutations of }a_{1}\dots a_{n}\right)+\label{eq:d2}\\
 & -\left(\d_{ba_{1}\cdots a_{n-3}}^{n-2}\psi i\overline{\psi}\gamma_{c}\d_{a_{n-2}a_{n-1}a_{n}}^{3}\psi g^{bc}+\text{cyclic permutations of }a_{1}\dots a_{n}\right)+\dots\nonumber 
\end{align}

So we can get invariants by taking any bosonic seed graph, and acting
on it with the following moves 
\begin{enumerate}
\item Replacing $\eta^{ab}\r g^{ab}$ on worldsheet edges,
\item Replacing the bosonic vertices with combination vertices for $n\geq2$,
$\d_{a_{1}\cdots a_{n}}^{n}X^{i}\r D_{a_{1}\cdots a_{n}}^{i}$,
\item Replacing $\delta^{ij}\r t^{ij}$ on transverse edges,
\end{enumerate}
where $D_{a_{1}\cdots a_{n}}^{i}$ is a combination vertex with higher
derivatives replaced with covariant derivatives 
\begin{equation}
D_{a_{1}\cdots a_{n}}^{i}=\del_{a_{1}\cdots a_{n}}^{n}X^{i}-i\overline{\psi}\gamma^{i}\del_{a_{1}\cdots a_{n}}^{n}\psi-\left(\d_{c}X^{i}-i\overline{\psi}\gamma^{i}\d_{c}\psi\right)\left(\delta_{\,d}^{c}+i\overline{\psi}\gamma^{c}\d_{d}\psi\right)^{-1}i\overline{\psi}\gamma^{d}\del_{a_{1}\cdots a_{n}}^{n}\psi,
\end{equation}

GM generate two scale 4 invariants 
\begin{align}
I_{3} & =\sqrt{-g}t^{ij}t^{kl}\d_{ab}^{2}X^{i}\d_{cd}^{2}X^{j}\d_{ef}^{2}X^{k}\d_{gh}^{2}X^{l}g^{ha}g^{bc}g^{de}g^{fg}\\
I_{4} & =\sqrt{-g}\del_{abc}^{3}X^{i}\del_{efg}^{3}X^{j}t^{ij}g^{ae}g^{bf}g^{cg}
\end{align}
which we can use to generate the supersymmetric invariants 
\begin{align}
I_{3} & =\sqrt{-g}t^{ij}t^{kl}D_{ab}^{i}D_{cd}^{j}D_{ef}^{k}D_{gh}^{l}g^{ha}g^{bc}g^{de}g^{fg}\\
I_{4} & =\sqrt{-g}D_{abc}^{i}D_{efg}^{j}t^{ij}g^{ae}g^{bf}g^{cg}
\end{align}

We are left with the term $\d^{m}\overline{\psi}\d^{n}\psi$, which
is the only vertex which gives us non-trivial supersymmetric invariants.
First we note that it is antisymmetric, which means any invariant
we generate must have an even number of these vertices. Second, we
note that we can use the above argument for the variation of $\d^{n}\psi$
to show that given a seed graph which contains such vertices, the
above moves are sufficient to generate an invariant, along with replacing
$\d^{m}\overline{\psi}\d^{n}\psi\r\del^{m}\overline{\psi}\del^{n}\psi$.
We will define a seed graph at scaling higher than zero as a graph
containing only boson and $\d^{m}\overline{\psi}\d^{n}\psi$ vertices,
which does not contain scale zero vertices. The procedure for generating
invariants at scale $n$ will then be 
\begin{enumerate}
\item Draw all seed graphs at this scale
\item Perform the above moves to generate an invariant
\end{enumerate}
We can then generate two non-trivial scale 4 invariants 
\begin{align}
I_{5} & =\sqrt{-g}\d_{a}\overline{\psi}\d_{bc}^{2}\psi\d_{d}\overline{\psi}\d_{ef}^{2}\psi g^{ad}g^{be}g^{cf}\\
I_{6} & =\sqrt{-g}\d_{a}\overline{\psi}\d_{b}\psi\d_{cd}^{2}\overline{\psi}\d_{ef}^{2}\psi g^{ad}g^{be}g^{cf}
\end{align}
as well as many higher scaling invariants. As in the scale zero case,
this method is exhaustive since up to the overall multiplicative constant
it fixes the coefficients of all possible terms.

\section{Exhaustiveness of the seed terms}

To show that our list of invariants is exhaustive, we will formulate
prohibition rules on the form the seed terms are allowed to take.
To do so, we first define the lowering and raising variations under
symmetry generators $Q$ and J, such that 
\begin{align}
\delta_{Q}\mathcal{L}_{f} & =\mathcal{L}_{f-1}^{<}+\ML_{f+1}^{>}\\
\delta_{J}\mathcal{L}_{d} & =\mathcal{L}_{d-1}^{<}+\mathcal{L}_{d,d+1}^{>}
\end{align}

where $f$ is the number of fermions and $d$ is the number of derivatives
in the term. 

In the case of supercharges the lowering part $\mathcal{L}_{f-1}^{<}$
is obtained by removal of one bare fermion (fermion without any derivatives
acting on it); in the Lorentz transformation case to obtain the lowering
part $\mathcal{L}_{d-1}^{<}$ we need to erase one $\partial X$ term
from the initial $\mathcal{L}_{d}$. Let's call the bare fermion $\psi$
and $\partial X$ - the lowering factor. If and only if a seed term
contains at least one lowering factor, its lowering variation isn't
zero. 

\subsection{First prohibition rule}

\textbf{Claim: }If the seed term has two or more lowering factors
it cannot be made invariant. 

Note that we cannot add any terms with a higher number of fermions
or derivatives to cancel the lowering variation. Hence, the lowering
variation should be either zero or a total derivative. 

\textbf{Proof:} Let's start from the reverse. Assume we have the term
with two lowering factors. Now, let's try to make its lowering variation
a total derivative. For simplicity of notations let's assume that
the two lowering factors are two bare fermions (the other two cases:
one bare $\psi$ and one $\partial X$ or two $\partial X$, are equivalent
to this one).

Consider the most general form of the term with two bare fermions,
where we have explicitly emphasized two derivatives we're going to
use to make the lowering variation a total derivative 
\begin{equation}
\mathcal{L}_{\psi\psi}=\psi^{\alpha}\psi^{\beta}\partial_{a}f\partial_{b}gh,
\end{equation}
where $f$, $g$ and $h$ are any combinations of the fields and their
derivatives. Consider its lowering variations under $Q^{\alpha}$and
$Q^{\beta}$ 

\begin{align}
\delta_{Q^{\alpha}}^{<}\mathcal{L}_{\psi\psi} & =\psi^{\beta}\partial_{a}f\partial_{b}gh,\nonumber \\
\delta_{Q^{\beta}}^{<}\mathcal{L}_{\psi\psi} & =\psi^{\alpha}\partial_{a}f\partial_{b}gh.
\end{align}

Let's make the first variation $\delta_{Q^{\alpha}}^{<}\mathcal{L}_{\psi\psi}$
a total derivative $\d_{a}\left(\psi^{\beta}f\partial_{b}gh\right)$.
To do so we need to add to the initial term $\mathcal{L}_{\psi\psi}$
other terms

\begin{equation}
\mathcal{L}_{a}^{\alpha}=\psi^{\alpha}(\partial_{a}\psi^{\beta}f\partial_{b}gh+\psi^{\beta}f\partial_{ab}^{2}gh+\psi^{\beta}f\partial_{b}g\partial_{a}h).
\end{equation}

Generalizing this, we can make the variation $\delta_{Q^{\alpha}}^{<}\mathcal{L}_{\psi\psi}$
or $\delta_{Q^{\beta}}^{<}\mathcal{L}_{\psi\psi}$ a total derivative
with respect to $\partial_{a}$ or $\partial_{b}$ by adding one of
the four terms $\mathcal{L}_{a(b)}^{\alpha(\beta)}$ to $\mathcal{L}_{\psi\psi}$,
such that 
\begin{equation}
\delta_{Q^{\alpha}}^{<}\left(\mathcal{L}_{\psi\psi}+\mathcal{L}_{a(b)}^{\alpha}\right)=\partial_{a(b)}N_{a(b)}^{\alpha},\qquad\delta_{Q^{\beta}}^{<}\left(\mathcal{L}_{\psi\psi}+\mathcal{L}_{a(b)}^{\beta}\right)=\partial_{a(b)}N_{a(b)}^{\beta}
\end{equation}
where $N_{a}^{\alpha}=\psi^{\beta}f\partial_{b}gh$ is $\mathcal{L}_{\psi\psi}$without
$\psi^{\alpha}$ and $\partial_{a}.$

It is crucial that we satisfy both of the variations simultaneously.
Naively, one would add to $\mathcal{L}_{\psi\psi}$ one of the combinations
$\mathcal{L}_{a}^{\alpha}+\mathcal{L}_{a}^{\beta}$ or $\mathcal{L}_{a}^{\alpha}+\mathcal{L}_{b}^{\beta}$
or $\mathcal{L}_{b}^{\alpha}+\mathcal{L}_{a}^{\beta}$ or $\mathcal{L}_{b}^{\alpha}+\mathcal{L}_{b}^{\beta}$.
However, there is a problem that $\delta_{Q^{\alpha}}^{<}\mathcal{L}_{a(b)}^{\beta}\neq0$
and $\delta_{Q^{\beta}}^{<}\mathcal{L}_{a(b)}^{\alpha}\neq0$. Let
us consider the combination $\mathcal{L_{\psi\psi}}+\mathcal{L}_{a}^{\alpha}+\mathcal{L}_{a}^{\beta}$.
When we act on this term with $\delta_{Q^{\alpha}}^{<}$ we get 

\begin{equation}
\delta_{Q^{\alpha}}^{<}(\mathcal{L_{\psi\psi}}+\mathcal{L}_{a}^{\alpha}+\mathcal{L}_{a}^{\beta})=\partial_{a}N_{a}^{\alpha}+\delta_{Q^{\alpha}}^{<}\mathcal{L}_{a}^{\beta}\neq\partial_{a}F.
\end{equation}

But we can notice that $\psi^{\beta}\delta_{Q^{\beta}}^{<}\mathcal{L}_{a}^{\alpha}=\psi^{\alpha}\delta_{Q^{\alpha}}^{<}\mathcal{L}_{a}^{\beta}$.
So we can subtract it from the action and verify that the variation
indeed gives a total derivative 

\begin{align}
\delta_{Q^{\alpha}}^{<}(\mathcal{L_{\psi\psi}}+\mathcal{L}_{a}^{\alpha}+\mathcal{L}_{a}^{\beta}-\psi^{\alpha}\delta_{Q^{\alpha}}^{<}\mathcal{L}_{a}^{\beta}) & =\partial_{a}N_{a}^{\alpha}+\delta_{Q^{\alpha}}^{<}\mathcal{L}_{a}^{\beta}-\delta_{Q^{\alpha}}^{<}\mathcal{L}_{a}^{\beta}=\partial_{a}N_{a}^{\alpha},\\
\delta_{Q^{\beta}}^{<}(\mathcal{L_{\psi\psi}}+\mathcal{L}_{a}^{\alpha}+\mathcal{L}_{a}^{\beta}-\psi^{\beta}\delta_{Q^{\beta}}^{<}\mathcal{L}_{a}^{\alpha}) & =\partial_{a}N_{a}^{\beta}+\delta_{Q^{\beta}}^{<}\mathcal{L}_{a}^{\alpha}-\delta_{Q^{\beta}}^{<}\mathcal{L}_{a}^{\alpha}=\partial_{a}N_{a}^{\beta}.
\end{align}

So, both of the variations $\delta_{Q^{\alpha}}^{<}$ and $\delta_{Q^{\beta}}^{<}$
of $\mathcal{L_{\psi\psi}}+\mathcal{L}_{a}^{\alpha}+\mathcal{L}_{a}^{\beta}-\psi^{\alpha}\delta_{Q^{\alpha}}^{<}\mathcal{L}_{a}^{\beta}$
are total derivatives. However, this term by itself is a total derivative.
Also, $\mathcal{L}_{\psi\psi}+\mathcal{L}_{a}^{\alpha}+\mathcal{L}_{b}^{\beta}$
and $\mathcal{L_{\psi\psi}}+\mathcal{L}_{b}^{\alpha}+\mathcal{L}_{a}^{\beta}$
don't work since $\psi^{\beta}\delta_{Q^{\beta}}^{<}\mathcal{L}_{a}^{\alpha}\neq\psi^{\alpha}\delta_{Q^{\alpha}}^{<}\mathcal{L}_{b}^{\beta}$.
This concludes the proof. 

This prohibition rule applied to two bare $\psi$'s illuminates the
fact that Goldstinos cannot have a mass term.

\subsection{Second prohibition rule}

\textbf{Claim: }Any seed term which contains a factor of $\psi_{\alpha}\partial_{a}\psi_{\alpha}\sim\psi\sigma^{bi}\partial_{a}\psi$
cannot generate an invariant chain.

\textbf{proof: }The most general form of such terms is 
\begin{equation}
\ML_{\psi\d\psi}=\psi_{\alpha}\partial_{a}\psi_{\alpha}\partial_{b}hf
\end{equation}
where $f$ should not contain bare $\psi_{\alpha}$. Applying $Q_{\alpha}$to
this term we find in the leading order 

\begin{equation}
\delta_{Q_{\alpha}}^{<}\ML_{\psi\d\psi}=\partial_{a}\psi_{\alpha}\partial_{b}hf
\end{equation}

We can not make this variation a total derivative with respect to
$\partial_{a}$. Because then we should add $\psi_{\alpha}\psi_{\alpha}(\partial_{ab}^{2}hf+\partial_{a}h\partial_{bf})$
which is identically zero. We can try to make this variation a total
derivative with respect to $\partial_{b}$, $\d_{b}\left(\partial_{a}\psi_{\alpha}hf\right)$
by adding 

\begin{equation}
\psi_{\alpha}\partial_{a}\psi_{\alpha}\partial_{b}hf+\psi_{\alpha}\partial_{ab}^{2}\psi_{\alpha}hf+\psi_{\alpha}\partial_{a}\psi_{\alpha}h\partial_{b}f
\end{equation}

This expression can be rewritten after integration by parts of the
middle term and then switching the order of fermions as: 

\begin{equation}
\partial_{a}\psi_{\alpha}\partial_{b}\psi_{\alpha}hf
\end{equation}

This term is either zero, if $a=b$, or is proportional to the EOM
(\ref{eq:eom}), if $a\ne b$, since for any value of $\alpha=1,\dot{1},2,\dot{2}$
one of the terms $\d_{a}\psi_{\alpha}$ or $\d_{a}\psi_{\alpha}$
will be proportional to the EOM. This concludes the proof.

\subsection{Exhausting seeds up to scale 2}

We first write all possible irreducible terms up to scale 2 with up
to 3 indices 
\begin{align}
\text{Scale -1: } & \bar{\psi}\psi,\ \bar{\psi}\gamma_{\mu}\psi,\ \bar{\psi}\sigma^{\mu\nu}\psi\\
\text{Scale 0: } & \bar{\psi}\partial_{a}\psi,\ \bar{\psi}\gamma_{\mu}\partial_{a}\psi,\ \bar{\psi}\sigma^{\mu\nu}\partial_{a}\psi,\ \partial_{a}X^{i}\\
\text{Scale 1: } & \partial_{a}\bar{\psi}\partial_{b}\psi,\ \partial_{a}\bar{\psi}\gamma^{\mu}\partial_{b}\psi,\ \partial_{a}\bar{\psi}\sigma^{\mu\nu}\partial_{b}\psi,\ \partial_{ab}X^{i}\\
\text{Scale 2: } & \partial_{a}\bar{\psi}\partial_{bc}\psi
\end{align}

Where we have ignored terms which can be eliminated using integration
by parts. For example, on scale 1 we can write the term $\bar{\psi}\partial_{a}\partial_{b}\psi$,
but when we consider $\psi\partial_{a}\partial_{b}\psi F$ for any
$F$, we can integrate by parts to get $-\partial_{a}\bar{\psi}\partial_{b}\psi F-\bar{\psi}\partial_{b}\psi\partial_{a}F$,
so it is enough to consider the scale 1 term $\partial_{a}\bar{\psi}\partial_{b}\psi$
and scale 0 term $\psi\partial_{b}\psi$.

Any possible term can be obtained by multiplying some combination
of irreducible terms. Note that we are uninterested in purely bosonic
terms, since they were considered in previous papers and it was shown
that the first allowed term appears at higher orders, and that all
seeds containing a scale -1 irreducible term are eliminated by the
first prohibition rule. 

Moreover, the second prohibition rule tells us that terms $\bar{\psi}\sigma^{ai}\partial_{b}\psi\sim\psi_{\alpha}\partial\psi_{\alpha}$
are prohibited. So we should use only $\bar{\psi}\sigma^{ab}\partial_{c}\psi$
or $\bar{\psi}\sigma^{ij}\partial_{c}\psi$ instead of $\bar{\psi}\sigma^{\mu\nu}\partial_{c}\psi$.
Then immediately we can forget about $\bar{\psi}\sigma^{ij}\partial_{c}\psi$
because to contract transverse indices $i$ and $j$ we will need
to go higher orders. Finally, we ignore terms which are proportional
to the EOM and can be eliminated by field redefinitions 
\begin{align}
\gamma^{a}\partial_{a}\psi & =0\quad\partial^{2}\psi=0\\
\partial^{2}X_{i} & =0
\end{align}

and for the fermions these can be written in light-cone coordinates
as 
\begin{equation}
\partial_{-}\psi_{1}=\partial_{-}\bar{\psi}_{1}=\partial_{+}\psi_{2}=\partial_{+}\bar{\psi}_{2}=0
\end{equation}

Scale 0 terms can be obtained by multiplying irreducible scale 0 terms
or scale -1 and scale 1. However, as discussed above such terms with
fermions are subjected to prohibition rules or proportional to the
EOM. Scale 1 terms can be obtained by contracting the indices of a
scale 1 irreducible term either with itself, or with the indices of
scale 0 terms. This gives the following terms 
\begin{align}
 & (\bar{\psi}\partial_{a}\psi)(\partial^{b}\bar{\psi}\gamma^{a}\partial_{b}\psi),\ (\bar{\psi}\gamma^{a}\partial^{b}\psi)(\partial_{a}\bar{\psi}\partial_{b}\psi),\ (\bar{\psi}\gamma_{a}\partial_{b}\psi)(\partial^{c}\bar{\psi}\sigma^{ab}\partial_{c}\psi),\nonumber \\
 & (\bar{\psi}\sigma^{ab}\partial^{c}\psi)(\partial_{a}\bar{\psi}\gamma_{c}\partial_{b}\psi),\ (\bar{\psi}\sigma^{ab}\partial^{c}\psi)(\partial_{a}\bar{\psi}\gamma_{c}\partial_{b}\psi)
\end{align}

All of which are proportional to the EOM, as can be seen by writing
them in light-cone coordinates.

Scale two terms can be obtained from contraction of scale 2 irreducible
with scale 0 irreducible or as contraction of two scale 1 irreducible
terms. We get 
\begin{align}
1\times1: & (\partial_{a}\bar{\psi}\partial_{b}\psi)(\partial^{a}\bar{\psi}\partial^{b}\psi),\ (\partial_{a}\bar{\psi}\partial_{b}\psi)(\partial_{c}\bar{\psi}\sigma^{ab}\partial^{c}\psi),\ (\partial^{a}\bar{\psi}\gamma^{b}\partial^{c}\psi)(\partial_{a}\bar{\psi}\gamma_{b}\partial_{c}\psi),\nonumber \\
 & (\partial^{a}\bar{\psi}\gamma^{b}\partial_{a}\psi)(\partial^{c}\bar{\psi}\gamma_{b}\partial_{c}\psi),\ (\partial^{a}\bar{\psi}\gamma^{b}\partial^{c}\psi)(\partial_{b}\bar{\psi}\gamma_{c}\partial_{a}\psi),\ (\partial^{a}\bar{\psi}\gamma^{i}\partial^{b}\psi)\partial_{ab}X_{i},\nonumber \\
 & (\partial^{c}\bar{\psi}\sigma^{ab}\partial_{c}\psi)(\partial^{d}\bar{\psi}\sigma_{ab}\partial_{d}\psi),\ (\partial_{c}\bar{\psi}\sigma^{ab}\partial_{d}\psi)(\partial_{a}\bar{\psi}\sigma^{cd}\partial_{b}\psi),\ (\partial^{a}\bar{\psi}\sigma^{bc}\partial^{d}\psi)(\partial_{a}\bar{\psi}\sigma^{bc}\partial_{d}\psi),\nonumber \\
 & (\partial^{a}\bar{\psi}\sigma^{bc}\partial^{d}\psi)(\partial_{b}\bar{\psi}\sigma^{cd}\partial_{a}\psi)\\
2\times0: & (\partial^{a}\bar{\psi}\partial_{ab}\psi)(\bar{\psi}\partial_{b}\psi),\ (\partial_{a}\bar{\psi}\partial_{bc}\psi)(\bar{\psi}\sigma^{ab}\partial_{c}\psi)
\end{align}

After some manipulation we see that among all of these terms only
two are independent 
\begin{align}
 & (\partial_{a}\bar{\psi}\partial_{b}\psi)(\partial^{a}\bar{\psi}\partial^{b}\psi),\label{eq:yay}\\
 & (\partial^{a}\bar{\psi}\gamma^{i}\partial^{b}\psi)\partial_{ab}X_{i}\label{eq:nay}
\end{align}

However using integration by parts, one can see that (\ref{eq:nay})
is proportional to the EOM, leaving us with (\ref{eq:yay}) as the
only non-purely bosonic seed up to scale 2. This is the term we found
in (\ref{eq:Lring}) above.

\section{Energy Correction of the $\protect\d\psi\protect\d\psi\protect\d\psi\protect\d\psi$
term}

The most interesting lowest scale result we have arrived at in the
analysis of the previous chapters is the term 
\begin{equation}
\ML_{2}^{\text{ring}}=c_{2}\sqrt{-g}\d_{a}\overline{\psi}\d_{b}\psi\d_{c}\overline{\psi}\d_{d}\psi g^{bc}g^{da}
\end{equation}
In order to make this result testable we would like to see how it
affects the energy levels of the Akulov-Volkov string at large $L$.
To do so we will consider this term in the static gauge, and in the
lowest order in derivative expansion 
\begin{align}
\ML_{2}^{\text{ring}} & =c_{2}\d_{a}\overline{\psi}\d_{b}\psi\d_{c}\overline{\psi}\d_{d}\psi\eta^{bc}\eta^{da}+\MO\left(\d^{5}\right)=\nonumber \\
 & =4c_{2}\d_{+}\overline{\psi}_{\dot{1}}\d_{-}\overline{\psi}_{\dot{2}}\d_{+}\psi_{1}\d_{-}\psi_{2}+\MO\left(\d^{5}\right)
\end{align}
We will treat this as a perturbation for the free part of the AV action
\begin{align}
\ML_{AV,free} & =\frac{T}{2}\left(i\overline{\psi}_{\dot{2}}\d_{+}\psi_{2}+i\psi_{2}\d_{+}\overline{\psi}_{\dot{2}}+i\overline{\psi}_{\dot{1}}\d_{-}\psi_{1}+i\psi_{1}\d_{-}\overline{\psi}_{\dot{1}}\right)\\
\ML & =\ML_{AV,free}+\ML_{2}^{\text{ring}}
\end{align}
Since the leading order perturbation is purely fermionic, the boson
field is completely free and we can ignore it for this derivation.
We define the conjugate momenta
\begin{align}
\overline{\Pi} & =\frac{\delta\ML}{\delta\left(\d_{0}\psi\right)}=\begin{pmatrix}\frac{\delta\ML}{\delta\left(\d_{0}\psi_{1}\right)}\\
\frac{\delta\ML}{\delta\left(\d_{0}\psi_{2}\right)}\\
\frac{\delta\ML}{\delta\left(\d_{0}\overline{\psi}_{\dot{2}}\right)}\\
\frac{\delta\ML}{\delta\left(-\d_{0}\overline{\psi}_{\dot{1}}\right)}
\end{pmatrix}^{T}=\begin{pmatrix}-\frac{1}{2}Ti\overline{\psi}_{\dot{1}}-2c_{2}\d_{+}\overline{\psi}_{\dot{1}}\d_{-}\overline{\psi}_{\dot{2}}\d_{-}\psi_{2}\\
-\frac{1}{2}Ti\overline{\psi}_{\dot{2}}+2c_{2}\d_{+}\overline{\psi}_{\dot{1}}\d_{-}\overline{\psi}_{\dot{2}}\d_{+}\psi_{1}\\
\frac{1}{2}Ti\psi_{2}+2c_{2}\d_{+}\overline{\psi}_{\dot{1}}\d_{+}\psi_{1}\d_{-}\psi_{2}\\
-\frac{1}{2}Ti\psi_{1}+2c_{2}\d_{-}\overline{\psi}_{\dot{2}}\d_{+}\psi_{1}\d_{-}\psi_{2}
\end{pmatrix}
\end{align}

so that the Hamiltonian is 
\begin{align}
\MH & =\intop_{0}^{2\pi R}d\sigma\left(\overline{\Pi}\d_{0}\psi-\ML\right)=\nonumber \\
 & =\intop_{0}^{2\pi R}d\sigma\left(\overline{\Pi}\d_{1}\psi\right)-\frac{1}{4}c_{2}\intop_{0}^{2\pi R}d\sigma\left(\d_{1}\overline{\psi}_{\dot{1}}\d_{1}\overline{\psi}_{\dot{2}}\d_{1}\psi_{1}\d_{1}\psi_{2}+\MO\left(\d^{5}\right)\right)=\nonumber \\
 & =\MH_{free}+\MH_{4}
\end{align}
where we use $L=2\pi R$ and have plugged in the equations of motion
(\ref{eq:eom}). We look at the Fourier expansion of the fields on
a closed string in the NS sector at $\tau=0$ (for the R sector take
$n\in\BZ$ instead of $r\in\BZ+\frac{1}{2}$) 
\begin{align}
\psi_{1}\left(\sigma\right) & =\sqrt{\frac{2}{\pi RT}}\sum_{r\in\BZ+\frac{1}{2}}b_{r}^{1}e^{ir\frac{\sigma}{R}},\quad\overline{\Pi}_{1}\left(\sigma\right)=-i\sqrt{\frac{T}{2\pi R}}\sum_{r\in\BZ+\frac{1}{2}}b_{r}^{1}e^{ir\frac{\sigma}{R}}\\
\psi_{2}\left(\sigma\right) & =\sqrt{\frac{2}{\pi RT}}\sum_{r\in\BZ+\frac{1}{2}}b_{r}^{2}e^{ir\frac{\sigma}{R}},\quad\overline{\Pi}_{2}\left(\sigma\right)=-i\sqrt{\frac{T}{2\pi R}}\sum_{r\in\BZ+\frac{1}{2}}b_{r}^{2}e^{ir\frac{\sigma}{R}}\\
\overline{\psi}_{\dot{1}}\left(\sigma\right) & =\sqrt{\frac{2}{\pi RT}}\sum_{r\in\BZ+\frac{1}{2}}b_{r}^{\dot{1}}e^{ir\frac{\sigma}{R}},\quad\Pi_{1}\left(\sigma\right)=-i\sqrt{\frac{T}{2\pi R}}\sum_{r\in\BZ+\frac{1}{2}}b_{r}^{\dot{1}}e^{ir\frac{\sigma}{R}}\\
\overline{\psi}_{\dot{2}}\left(\sigma\right) & =\sqrt{\frac{2}{\pi RT}}\sum_{r\in\BZ+\frac{1}{2}}b_{r}^{\dot{2}}e^{ir\frac{\sigma}{R}},\quad\Pi_{2}\left(\sigma\right)=-i\sqrt{\frac{T}{2\pi R}}\sum_{r\in\BZ+\frac{1}{2}}b_{r}^{\dot{2}}e^{ir\frac{\sigma}{R}}
\end{align}
where $\sigma\in\left[0,2\pi R\right]$ and $\left\{ b_{r}^{s},b_{r^{\tg}}^{s^{\tg}}\right\} =\delta_{r+r^{\tg}}\delta_{ss^{\tg}},\ s=1,2,\dot{1},\dot{2}$.
Note that $\sigma$ is really periodic only under $\sigma\r\sigma+4\pi R$
in the NS-NS sector since 
\begin{equation}
\psi\left(\sigma\right)=-\psi\left(\sigma+2\pi R\right).
\end{equation}
The commutator is

\begin{align}
\left\{ \psi_{s}\left(\sigma\right),\overline{\Pi}_{s^{\tg}}\left(\sigma^{\tg}\right)\right\}  & =-i\delta_{ss^{\tg}}\left[2\delta\left(\frac{\sigma-\sigma^{\tg}}{4\pi R}\right)-\delta\left(\frac{\sigma-\sigma^{\tg}}{2\pi R}\right)\right]
\end{align}

where $\delta\left(x\right)$ is non-zero for all $x\in\BZ$. Plugging
this into the free (fermion) Hamiltonian we get 
\begin{align}
\MH_{free} & =\frac{2}{R}\sum_{s=1,2,\dot{1},\dot{2}}\left(\sum_{r\in\BN+\frac{1}{2}}r\left(b_{-r}^{s}b_{r}^{s}-b_{r}^{s}b_{-r}^{s}\right)\right)
\end{align}
This Hamiltonian is Weyl ordered, meaning that products of fermionic
operators appear in the form 
\begin{equation}
\frac{1}{k!}\sum_{\left(p_{1},\dots p_{k}\right)\in\text{perms}\left(k\right)}\left(-1\right)^{s\left(p_{1},\dots p_{k}\right)}b_{r_{p_{1}}}\cdots b_{r_{p_{k}}}
\end{equation}
where $s\left(p_{1},\dots p_{k}\right)$ is the parity of the permutation
$\left(p_{1},\dots p_{k}\right)$. We now take the normal ordering
to get 
\begin{align}
\MH_{free} & =\frac{4}{R}\sum_{s=1,2,\dot{1},\dot{2}}\left(\sum_{r\in\BN+\frac{1}{2}}rb_{-r}^{s}b_{r}^{s}-\frac{1}{48}\right)
\end{align}
where we used zeta function regularization to take sums of the form
\begin{align}
\sum_{r=\frac{1}{2}}^{\infty}r^{k} & =\sum_{n=1}^{\infty}\left(\frac{2n+1}{2}\right)^{k}=-\frac{2^{k}-1}{2^{k}}\zeta\left(-k\right)=\begin{cases}
\frac{1}{24} & k=1\\
0 & k=2,4\\
-\frac{7}{8*120} & k=3
\end{cases}
\end{align}

The perturbation Hamiltonian is 
\begin{align}
\MH_{4} & =\frac{c_{2}}{4}\intop_{0}^{2\pi R}d\sigma\d_{1}\overline{\psi}_{\dot{1}}\d_{1}\overline{\psi}_{\dot{2}}\d_{1}\psi_{2}\d_{1}\psi_{1}=\nonumber \\
 & =\frac{c_{2}}{\pi^{2}T^{2}R^{6}}\intop_{0}^{2\pi R}d\sigma\sum_{r_{1},r_{2},r_{3},r_{4}}b_{r_{1}}^{\dot{1}}r_{1}e^{ir_{1}\frac{\sigma}{R}}b_{r_{2}}^{\dot{2}}r_{2}e^{ir_{2}\frac{\sigma}{R}}b_{r_{3}}^{2}r_{3}e^{ir_{3}\frac{\sigma}{R}}b_{r_{4}}^{1}r_{4}e^{ir_{4}\frac{\sigma}{R}}=\nonumber \\
 & =\frac{c_{2}}{\pi^{2}T^{2}R^{6}}\sum_{r_{1},r_{2},r_{3},r_{4}}r_{1}r_{2}r_{3}r_{4}b_{r_{1}}^{\dot{1}}b_{r_{2}}^{\dot{2}}b_{r_{3}}^{2}b_{r_{4}}^{1}\intop_{0}^{2\pi R}d\sigma e^{-2\pi i\left(r_{1}+r_{3}+r_{2}+r_{4}\right)\frac{\sigma}{2\pi R}}=\nonumber \\
 & =\frac{2c_{2}}{\pi T^{2}R^{5}}\sum_{n\in\BZ}\sum_{r,r^{\tg}}r\left(r+n\right)r^{\tg}\left(r^{\tg}-n\right)b_{-r}^{\dot{1}}b_{r+n}^{1}b_{-r^{\tg}}^{\dot{2}}b_{r^{\tg}-n}^{2}
\end{align}

This clearly annihilates the ground state. The simplest of its eigenstates
with non-zero eigenvalues are

\begin{equation}
\ket{\psi}=\frac{1}{\sqrt{2}}\left(\ket{1_{\frac{1}{2}}^{1},1_{\frac{1}{2}}^{2}}\pm\ket{1_{\frac{1}{2}}^{\dot{1}},1_{\frac{1}{2}}^{\dot{2}}}\right)
\end{equation}
since 
\begin{align*}
\MH_{4}\ket{\psi} & =\frac{2c_{2}}{\pi T^{2}R^{5}}\sum_{n\in\BZ}\sum_{r,r^{\tg}}r\left(r+n\right)r^{\tg}\left(r^{\tg}-n\right)b_{-r}^{\dot{1}}b_{r+n}^{1}b_{-r^{\tg}}^{\dot{2}}b_{r^{\tg}-n}^{2}\frac{1}{\sqrt{2}}\left(\ket{1_{\frac{1}{2}}^{1},1_{\frac{1}{2}}^{2}}\pm\ket{1_{\frac{1}{2}}^{\dot{1}},1_{\frac{1}{2}}^{\dot{2}}}\right)=\\
 & =\frac{2c_{2}}{\pi T^{2}R^{5}}\sum_{n\in\BZ}\sum_{r,r^{\tg}}r\left(r+n\right)r^{\tg}\left(r^{\tg}-n\right)b_{-r}^{\dot{1}}b_{r+n}^{1}b_{-r^{\tg}}^{\dot{2}}b_{r^{\tg}-n}^{2}\frac{1}{\sqrt{2}}\left(b_{-\frac{1}{2}}^{1}b_{-\frac{1}{2}}^{2}\pm b_{-\frac{1}{2}}^{\dot{1}}b_{-\frac{1}{2}}^{\dot{2}}\right)\ket 0=\\
 & =\frac{c_{2}}{8\pi T^{2}R^{5}}\frac{1}{\sqrt{2}}\left(b_{-\frac{1}{2}}^{\dot{1}}b_{\frac{1}{2}}^{1}b_{-\frac{1}{2}}^{\dot{2}}b_{\frac{1}{2}}^{2}b_{-\frac{1}{2}}^{1}b_{-\frac{1}{2}}^{2}\pm b_{\frac{1}{2}}^{\dot{1}}b_{-\frac{1}{2}}^{1}b_{\frac{1}{2}}^{\dot{2}}b_{-\frac{1}{2}}^{2}b_{-\frac{1}{2}}^{\dot{1}}b_{-\frac{1}{2}}^{\dot{2}}\right)\ket 0=\\
 & =\frac{c_{2}}{8\pi T^{2}R^{5}}\frac{1}{\sqrt{2}}\left(b_{-\frac{1}{2}}^{\dot{1}}b_{-\frac{1}{2}}^{\dot{2}}\pm b_{-\frac{1}{2}}^{1}b_{-\frac{1}{2}}^{2}\right)\ket 0=\pm\frac{c_{2}}{8\pi T^{2}R^{5}}\ket{\psi}
\end{align*}
 which gives rise to the energy correction
\begin{equation}
\Delta E_{NS}=\pm\frac{c_{2}}{8\pi T^{2}R^{5}}
\end{equation}
Where other eigenstates will give rise to more complicated energy
corrections at this order in $1/R$. We can repeat this analysis for
the Ramond sector with $\ket{\psi}=\frac{1}{\sqrt{2}}\left(\ket{1_{1}^{1},1_{1}^{2}}\pm\ket{1_{1}^{\dot{1}},1_{1}^{\dot{2}}}\right)$
to get 
\begin{align}
\MH_{free} & =\frac{4}{R}\sum_{s=1,2,\dot{1},\dot{2}}\left(\sum_{n\in\BN}nb_{-n}^{s}b_{n}^{s}+\frac{1}{24}\right)\\
\Delta E_{R} & =\pm\frac{2c_{2}}{\pi T^{2}R^{5}}
\end{align}

\section{Discussion and conclusions}

In this work we present a general method to generate invariant actions
for effective strings which break $D=4\ N=1$ SUSY. We have shown
that this method recreates known results, as well as producing new
ones. Our method does not generate terms which are only invariant
up to the equations of motion, which may be related to anomalies as
in the bosonic case, but seems to be exhaustive otherwise. We can
summarize our method as taking a seed term - a minimal term of Goldstone
bosons and Goldstinos which is invariant under the non-broken $ISO\left(1,1\right)\cross SO\left(D-2\right)$,
and performing 4 simple moves: (a) replacing the Minkowski metric
$\eta^{ab}$ with the worldsheet metric $g^{ab}$ as defined in (\ref{eq:g}),
(b) replacing $n\geq2$ scaling boson vertices with the combination
vertex $C_{a_{1}\cdots a_{n}}^{i}$ as defined in (\ref{eq:C}), (c)
replacing $n\geq3$ derivatives with covariant derivatives as defined
in (\ref{eq:d1}), (\ref{eq:d2}) and (d) replacing the Euclidean
transverse metric $\delta_{ij}$ with the transverse metric $t_{ij}$
as defined in (\ref{eq:t}). This method clearly shows that every
known bosonic invariant has a supersymmetric counterpart, as well
as the existence of new supersymmetric invariants with no bosonic
counterparts, the simplest of which we have termed $\ML_{2}^{\text{ring}}$
and for which we have calculated its energy corrections. As directions
for future research, we can consider repeating this analysis for the
case in which only half of the SUSY generators are broken, such that
the worldsheet theory has $\MN=\left(0,2\right)$ supersymmetry, analyzing
$\ML_{2}^{\text{ring}}$ in the conformal gauge and verifying it does
not contribute to the conformal anomaly and that no other terms are
possible also in that approach, and generalizing this work to other
dimensions and $\MN>1$.

\subsection*{Acknowledgments}

We thank O. Aharony for advising us during this work. Our work was
supported in part by the I-CORE program of the Planning and Budgeting
Committee and the Israel Science Foundation (grant number 1937/12)
and by an Israel Science Foundation center for excellence grant (grant
number 1989/14), as well as the Minerva foundation with funding from
the Federal German Ministry for Education and Research.


\begin{thebibliography}{10}
\bibitem{key-8}S. Fubini and G Veneziano, ``Level Structure of Dual
Resonance Models'', Il Nuovo Cimento 64A (1969) 811

\bibitem{key-9}N.B. Nielsen and P. Olesen, ``Vortex-Line Models
for Dual Strings'', Nucl. Phys. B61 (1973) 45

\bibitem{key-10}O. Aharony and Z. Komargodski, ``The Effective Theory
of Long Strings'' {[}arXiv:1302.6257v2 {[}hep-th{]}{]}

\bibitem{key-11}J. Hughes and J. Polchinski, ``Partially Broken
Global Supersymmetry and the Superstring'', Nucl. Phys. B278 (1986)
147

\bibitem{key-12}O. Aharony and E. Karzbrun, ``On the Effective Action
of Confining Strings'' JHEP 0906, 012 (2009) {[}arXiv:0903.1927 {[}hep-th{]}{]}

\bibitem{key-4}D.V. Volkov and V.P. Akulov, ``Possible Universal
Neutrino Interaction'' JETP Lett., 16, (1972), 367

\bibitem{key-1}S. Dubovsky, R. Flauger and V. Gorbenko, \textquotedblleft Effective
String Theory Revisited\textquotedblright{} JHEP 1209, 044 (2012)
{[}arXiv:1203.1054 {[}hep-th{]}{]}

\bibitem{key-5}O. Aharony and N. Klinghoffer, ``Corrections to Nambu-Goto
energy levels from the effective string action'' JHEP 1012, 058 (2010)
{[}arXiv:1008.2648v2 {[}hep-th{]}{]}

\bibitem{key-2}F. Gliozzi and M. Meineri, ``Lorentz completion of
effective string (and p-brane) action'' JHEP 1208, 056 (2012), {[}arXiv:1207.2912v2
{[}hep-th{]}{]}

\bibitem{key-1}M. Luscher and P. Weisz, \textquotedblleft String
excitation energies in SU(N) gauge theories beyond the free-string
approximation,\textquotedblright{} JHEP 07 (2004) 014, {[}arXiv:0406205
{[}hep-th{]}{]}
\end{thebibliography}
\end{document}